 \documentclass[prl,amsmath,amssymb,twocolumn, showpacs, superscriptaddress,10pt]{revtex4-1} 

\usepackage{amsmath}
\usepackage{hyperref}
\usepackage{graphicx}
\usepackage{amsfonts}
\usepackage{amsthm}
\usepackage{cases}
\usepackage{bm}
\usepackage[utf8]{inputenc}
\usepackage{enumitem}

\usepackage[normalem]{ulem}

\usepackage{color}
\definecolor{Blue}{rgb}{0.00, 0.00, 1.00}
\definecolor{Red}{rgb}{1.00, 0.00, 0.00}
\definecolor{Green}{rgb}{0.00, 0.70, 0.00}
\newcommand{\red}{\color{Red}}

\def\XXint#1#2#3{{\setbox0=\hbox{$#1{#2#3}{\int}$}
     \vcenter{\hbox{$#2#3$}}\kern-.5\wd0}}

\usepackage{ifxetex}
\ifxetex
\usepackage{fontspec}
    \defaultfontfeatures{Ligatures={TeX}}
\usepackage[math-style=ISO,bold-style=ISO]{unicode-math}
\else
    \usepackage{esint}
\fi

\hypersetup{
    colorlinks=true,       
    linkcolor=red,          
    citecolor=blue,        
    filecolor=magenta,      
    urlcolor=cyan           
}

\newcommand{\nn}{\nonumber}
\newcommand{\be}{\begin{equation}}
\newcommand{\ee}{\end{equation}}
\newcommand{\bea}{\begin{eqnarray}}
\newcommand{\eea}{\end{eqnarray}}

\newcommand{\beq}{\begin{equation}}
\newcommand{\eeq}{\end{equation}}
\newcommand{\beqn}{\begin{eqnarray}}
\newcommand{\eeqn}{\end{eqnarray}}

\begin{document}

\title{Interacting, running and tumbling: the active Dyson Brownian motion}

\author{Léo Touzo}
\affiliation{Laboratoire de Physique de l'\'Ecole Normale Sup\'erieure, CNRS, ENS $\&$ PSL University, Sorbonne Universit\'e, Universit\'e de Paris, 75005 Paris, France}
\author{Pierre Le Doussal}
\affiliation{Laboratoire de Physique de l'\'Ecole Normale Sup\'erieure, CNRS, ENS $\&$ PSL University, Sorbonne Universit\'e, Universit\'e de Paris, 75005 Paris, France}
\author{Gr\'egory \surname{Schehr}}
\affiliation{Sorbonne Universit\'e, Laboratoire de Physique Th\'eorique et Hautes Energies, CNRS UMR 7589, 4 Place Jussieu, 75252 Paris Cedex 05, France}
\date{\today}

\begin{abstract}
We introduce and study a model in one dimension of $N$ run-and-tumble particles (RTP)
{which repel each other} logarithmically in the presence of an external quadratic potential. This is an ``active'' version of the well-known Dyson Brownian motion (DBM) where the particles are subjected to a telegraphic noise, with two possible states $\pm$ with velocity $\pm v_0$. We study analytically and numerically two different versions of this model. In model I a particle only interacts with particles in the same state, while in model II all the particles interact with each other. In the large time limit, both models converge to a steady state where the stationary density has a finite support. 
For finite $N$, the stationary density exhibits singularities, which disappear when $N \to +\infty$. In that limit, for model I, using a Dean-Kawasaki approach, we show that the stationary density of $+$ (respectively $-$) particles deviates from the DBM Wigner semi-circular shape, and vanishes with an exponent $3/2$ at one of the edges. 
In model II, the Dean-Kawasaki approach fails but we obtain strong evidence that the density in the large $N$ limit retains a Wigner semi-circular shape.
\end{abstract}


\maketitle

{\bf Introduction}. There is a tremendous current interest in the study of interacting active particles both from the theoretical and experimental point of view~\cite{soft,BechingerRev,Ramaswamy2017,Marchetti2018,Berg2004,Cates2012,TailleurCates}. At variance with a passive
particle, an active particle has a self-propelled motion modelled by a driving ``active" noise, with a finite persistence time. 
A paradigmatic model is the so-called run-and-tumble particle (RTP) -- a motion exhibited by E. Coli bacteria \cite{Berg2004,TailleurCates},
driven by telegraphic noise \cite{HJ95,W02,ML17}. It was also studied in the math literature 
on ``persistent'' random walks~\cite{kac74,Orshinger90}.
Even 
a single RTP 
exhibits interesting properties, e.g. 
in the presence of a trapping potential, the system reaches a non-Boltzmann 
stationary state,
retaining the effect of activity even at late times \cite{HJ95,Solon15, TDV16, DKM19, DD19,3statesBasu,LMS2020}. 


A crucial question is what happens to these stationary states in the presence of interactions between the RTP's. 
Interacting RTP's are known to exhibit remarkable collective effects, such as motility-induced phase separation,  
clustering and jamming even for repulsive interactions and in the absence of alignement \cite{TailleurCates,FM2012,soft,Buttinoni2013,FHM2014,CT2015,Ramaswamy2017,slowman,slowman2,Active_OU,BG2021,CMPT2010,SG2014}. To describe the effects of interactions beyond numerical simulations, hydrodynamic approaches and perturbative exact results have been developed \cite{TailleurCates,KH2018,Agranov2021,Agranov2022,Active_OU}. At present, very few exact results exist, even in one dimension, beyond two interacting RTP's on the line \cite{slowman,slowman2,us_bound_state,Maes_bound_state,nonexistence,MBE2019,KunduGap2020,LMS2019}, and 
harmonic chains \cite{SinghChain2020,PutBerxVanderzande2019}. {Recent exact solutions have also been obtained for some specific many-particle models on a lattice with contact interactions~\cite{Metson2022,MetsonLong,Dandekar2020,Thom2011}}.

In the passive case, a well studied model of $N$ interacting particles in one dimension is the Dyson Brownian motion (DBM)~\cite{mehta_book,forrester_book,bouchaud_book}. In this model, the particles interact via a pairwise logarithmic potential and are subjected to independent white noises. There is a host of exact results in the limit of
large $N$, principally due to the connection between the positions of the particles and the eigenvalues of a random matrix. 
In the presence of a quadratic external potential, these matrices belong to the celebrated Gaussian $\beta$-ensemble~\cite{forrester_book}. 
An important result in that case is that the scaled particle density converges at large time and large $N$ to the Wigner semi-circle density
$\rho_{sc}(x)=\frac{2}{\pi x_e^2} \sqrt{x_e^2 - x^2}$, which has a finite support $[-x_e,x_e]$. 
It is then quite natural to look for an extension of this model to the realm of active matter, where each particle
becomes an RTP. An interesting question is whether exact results can also be obtained in that case, 
and what kind of new phenomena can be expected as compared to the passive DBM model. In particular, let us recall that the stationary density for independent RTP's in a quadratic potential also has a finite support $[-x_+,x_+]$, of the form $\rho_{1}(x) \propto (x_+^2 - x^2)^{\phi}$ 
where the edge exponent $\phi \in ]-1, +\infty[$ can vary continuously \cite{HJ95,TailleurCates,DKM19}.
One can thus ask if, by turning on the interactions, one may eventually interpolate between these two density profiles. 

To address such questions we introduce and study in this paper a model that we call {\it the active DBM}. It is defined by the evolution equation 
for the positions $x_i(t)$ of $N$ particles 
\bea \label{model0} 
 \dot x_i(t) &=& - \lambda x_i(t) +  \frac{2 }{N} \sum_{j \neq i} 
\frac{g_{\sigma_i(t), \sigma_j(t)}}{x_i(t)-x_j(t)} \\
&+&  v_0 \sigma_i(t) + \sqrt{\frac{2 T}{N}} \xi_i(t) \nn \;.
\eea 
Each particle can be in two internal states $\sigma_i(t)=\pm 1$ of velocities respectively $\pm v_0$, and flips its sign
with a constant rate $\gamma$. In addition each particle is submitted to an external potential $V(x)=\frac{\lambda}{2} x^2$
and to a thermal noise at temperature $T/N$, where the $\xi_i(t)$'s are independent standard white noises. 
The particles interact via a repulsive pairwise logarithmic potential (i.e. a $1/x$ force) of strength
$g_{\sigma,\sigma'} \geq 0$ which depends a priori on their internal states. We will focus on two cases 
\begin{eqnarray} \label{def_model}
g_{\sigma,\sigma'}=
\begin{cases}
& g \, \delta_{\sigma,\sigma'} \quad \quad \quad \quad \;\, \rm{(model \; I)} \;,\\ 
& g \quad, \quad \forall (\sigma, \sigma') \quad  \rm{(model \; II)} \, .
\end{cases}
\end{eqnarray}
Model II looks a priori like the most natural extension of the DBM to RTP particles. However, its non-interacting limit $g\to 0$ is singular (see below). On the other hand, model I interpolates naturally between the independent RTPs limit and the usual DBM.

The simplest observables are the densities of each species $\sigma = \pm 1$
\be \label{def_rho}
\rho_\sigma(x,t) = \frac{1}{N} \sum_i \delta_{\sigma_i(t),\sigma} \delta(x-x_i(t)) \;,
\ee 
as well as the total density $\rho_s(x,t)=\rho_+(x,t)+\rho_-(x,t)$, normalized to unity, and 
$\rho_d(x,t)=\rho_+(x,t)-\rho_-(x,t)$. The model \eqref{model0}-\eqref{def_model} is invariant under the symmetry $(x_i,\sigma_i) \to (-x_i,-\sigma_i)$,
which implies that $\rho_-(x,t)=\rho_+(-x,t)$ for all time $t$ if the initial condition is symmetric. 
For model II, Eq. \eqref{model0} with $v_0=0$ 
describes the standard DBM for the Gaussian $\beta$-ensemble
with $\beta=2g/T$ \cite{forrester_book}, leading to the stationary semi-circle density with $x_e=2 \sqrt{g/\lambda}$, which
is independent of $T$. The factors of $N$ in \eqref{model0} are chosen 
so that the support of the stationary density is of $O(1)$, which, 
as we will see below, remains true for the active case.

\begin{figure}
    \centering
    \includegraphics[width=0.5\linewidth, trim={0 5cm 0 2cm}, clip]{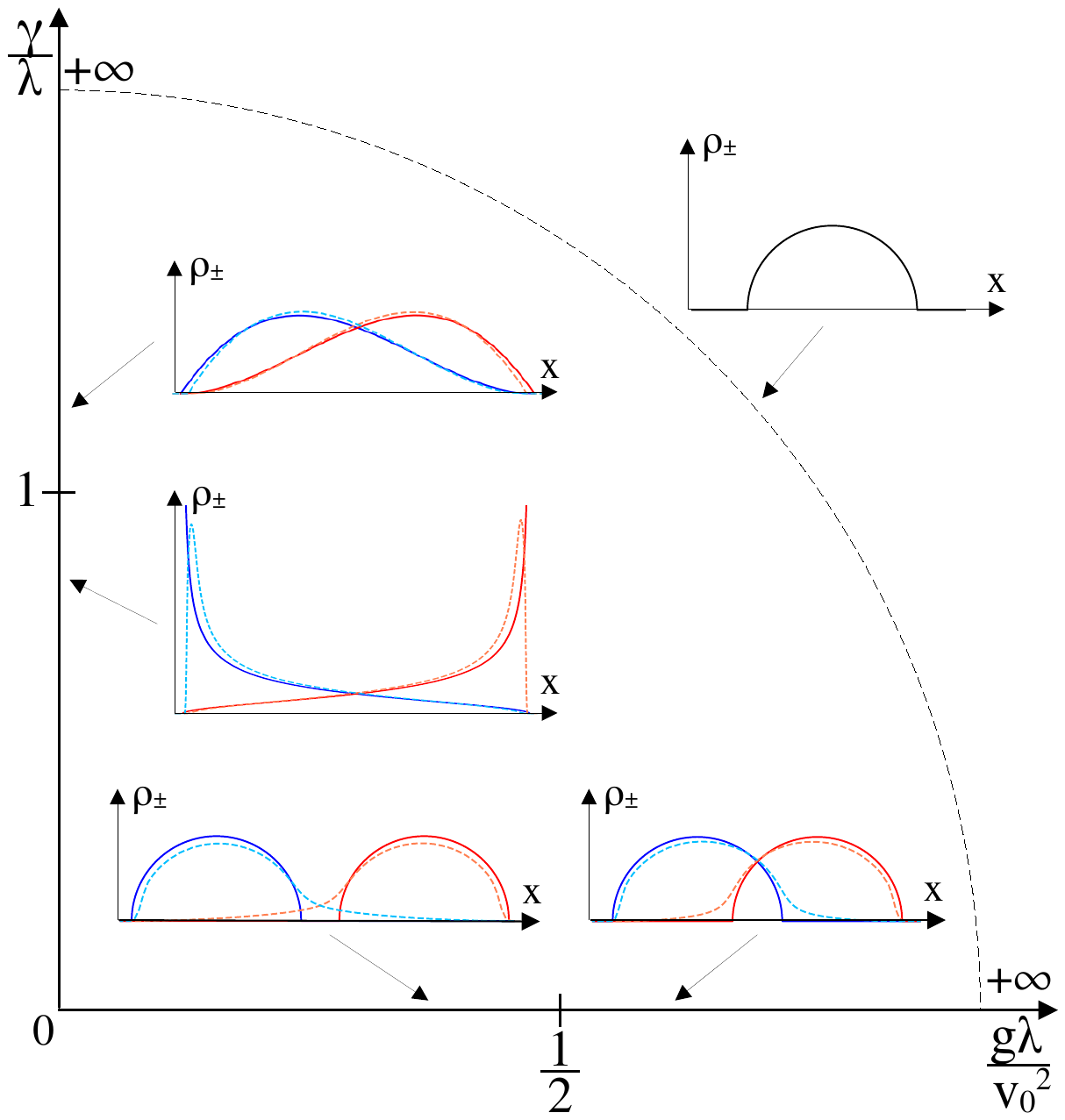}
    \includegraphics[width=\linewidth, trim={0 3.2cm 0 8.2cm}, clip]{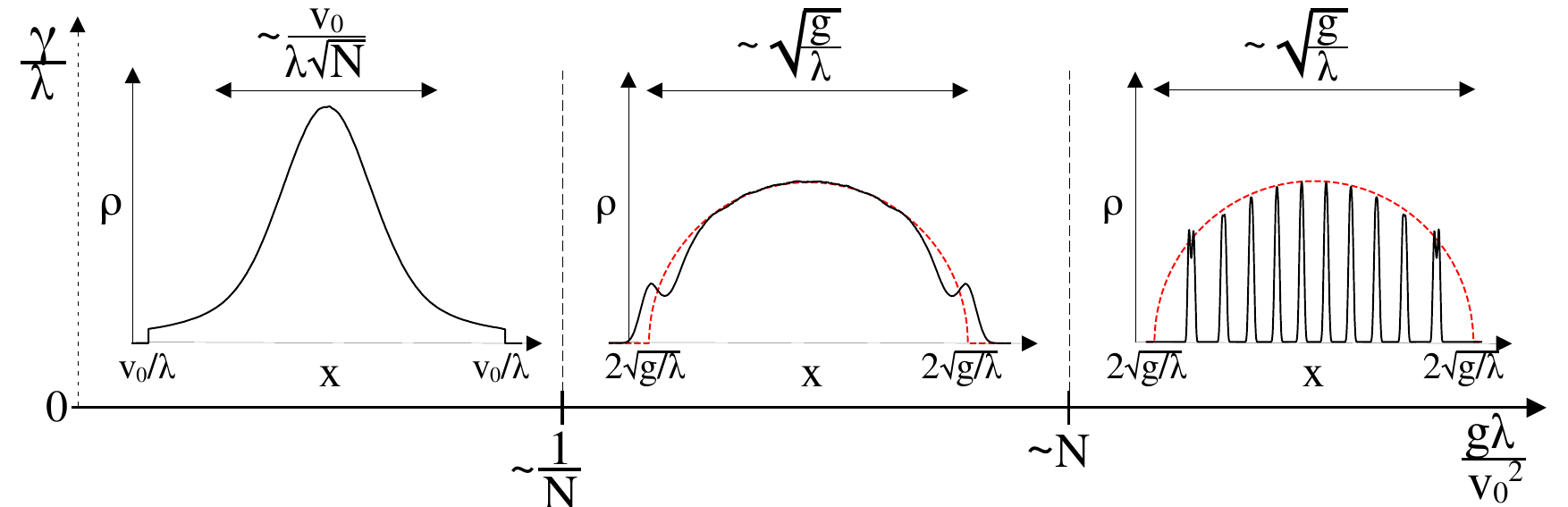}
    \caption{Top: Shape of the particle density in model I in the plane $(g\lambda/v_0^2,\gamma/\lambda)$ in different limits. The density $\rho_+$ is plotted in red and $\rho_-$ in blue. When the two coincide they are plotted in black. The light dashed curves represent the density slightly away from the limit considered. The dashed circular line in the diagram symbolizes infinity. The diffusive limit, which requires a specific scaling between $v_0$ and $\gamma$, is not shown here. Bottom: Shape of the total density $\rho_s$ in model II as a function of the parameter $g\lambda/v_0^2$ showing the different regimes at large $N$. The results were derived in the limit $\gamma \to 0$ but simulations suggest that they are valid beyond this limit. The dashed red line shows the semi-circle. The spatial extension of the density as a function of the model parameters is also shown in the different regimes.}
    \label{phase_diagrams}
\end{figure}

In this paper we study the model defined by Eq. \eqref{model0} for $v_0>0$ and focus on the limit $T \to 0$ with fixed $g$ and $\lambda$, which is the purely active problem with repulsion. By combining analytical tools and numerical solutions of Eq. \eqref{model0}, we obtain results at finite $N$ as well as in the large $N$ limit for various observables. This includes
the stationary limit of the densities defined in \eqref{def_rho}, which have a single support
with edges at $x_{\pm,N}$.  We demonstrate that the two models I and II exhibit
very different behaviors (see Fig. \ref{phase_diagrams}). 
For model I, particles of opposite velocities do not interact and can thus cross, and we find that
the hydrodynamic description, based on the Dean-Kawasaki (DK) approach \cite{Dean,Kawa}, becomes exact at large $N$. This allows 
to show
that in that limit the stationary density of $+$ (respectively $-$) particles deviates from the Wigner semi-circular shape. It vanishes with an exponent $3/2$ at one of the edges and we obtain its dependence
on the parameters (see top panel of Fig. \ref{phase_diagrams}).
In model II, particles cannot cross, and as a result they tend to aggregate into clusters at small $g$.
In this case the hydrodynamic approach at large $N$ fails and we characterize the distribution of the sizes of the clusters as well as the stationary density numerically (see bottom panel of Fig. \ref{phase_diagrams}). 
We obtain strong evidence, e.g., by computing perturbatively the fluctuations of the particle positions,
that the density in the large $N$ limit still retains a Wigner semi-circular shape. 

{\bf General properties of both models}. We start with some general observations which are valid for both model I and model II. First, notice that, in the limit $T \to 0$ that we consider in this paper, there are only two dimensionless parameters $\frac{v_0}{\sqrt{g \lambda}}$ and $\frac{\gamma}{\lambda}$. Hence, from
now on we set $\lambda=1$. For each model there are four interesting limits
discussed below: (i) the passive limit $v_0=0$ (ii) the limit $g \to 0^+$, (iii) the limit where the velocities are frozen $\gamma=0$, and (iv) the diffusive limit $\gamma \to +\infty$, $v_0 \to +\infty$ with $D=\frac{v_0^2}{2\gamma}$ fixed.

For a single particle, $N=1$, we know that the particle density has a finite support $[-v_0,v_0]$ with edges obtained by solving $f(x_{\pm,1})=\pm v_0$ where $f(x) = - x$ is the force due to the harmonic potential \cite{LMS2020}. For $N >1$ the support is still finite $[x_{-,N},x_{+,N}]$,
but it is modified by the interactions. We find the upper edge $x_{+,N}$ by fixing $\sigma_i=+1$ for all particles (hence it is the same for models I and II) and computing the equilibrium positions $\{x^{\rm eq}_i\}_{1\leq i\leq N}$ of the particles. Then $x_{+,N}$ corresponds to the largest $x^{\rm eq}_i$ (and $x_{-,N}=-x_{+,N}$ by symmetry). From Eq. (\ref{model0}), we thus see that we need to solve the following set of equations
\beq \label{eq_pos}
v_0 - x_i + \frac{2g}{N}\sum_{j\neq i}\frac{1}{x_i-x_j}=0 \ \ \ {\rm for} \ i=1, \ \cdots \ , N \;.
\eeq
The solution is found as $x_i=v_0 + \sqrt{\frac{2g}{N}}y_i$ where the $y_i$'s are the zeroes of the Hermite polynomials, $H_N(y)=0$ \cite{Hermite1,Hermite2}. This allows to show that $\lim_{N \to + \infty} x_{\pm,N} = \pm (v_0+2\sqrt{g})$, with $O(N^{-2/3})$ corrections, 
see \cite{SM} for the systematic expansion. However we will see below that for $N=+\infty$ strictly, 
the support is also an interval which we denote $[x_-,x_+]$. This interval must be included in
$[-v_0-2\sqrt{g},v_0+2\sqrt{g}]$ but can be strictly smaller (see \cite{SM}).
Finally note that since for $\gamma>0$, $+$ particles can change into $-$ particles at any time and vice-versa, $\rho_+(x)$ and $\rho_-(x)$ necessarily have the same support. 

\begin{figure}[t]
    \centering
    \includegraphics[width=0.32\linewidth]{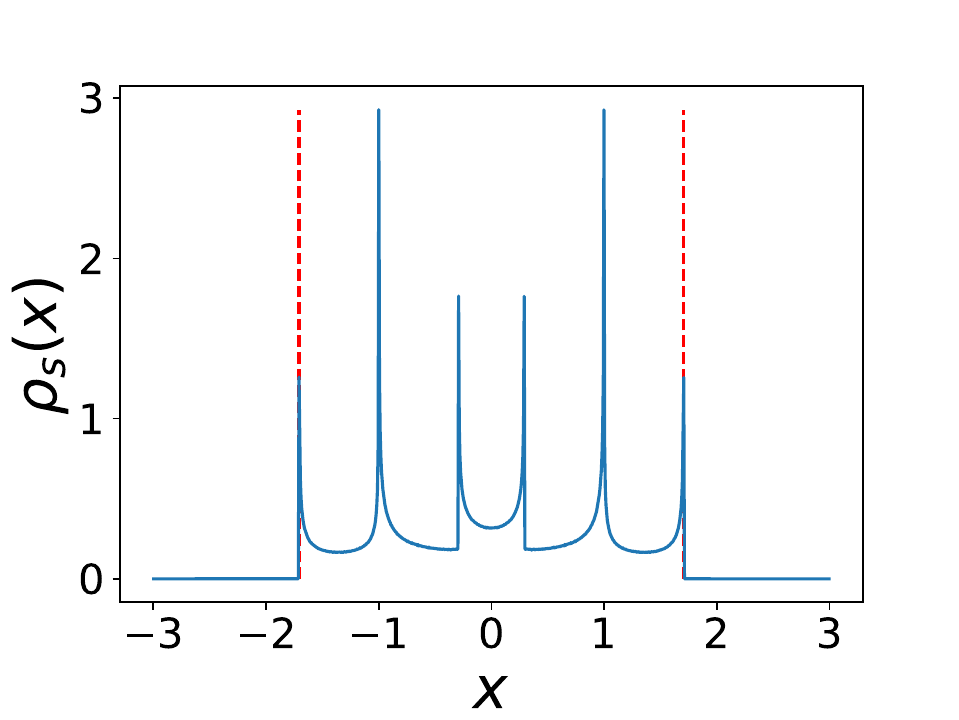}
    \includegraphics[width=0.32\linewidth]{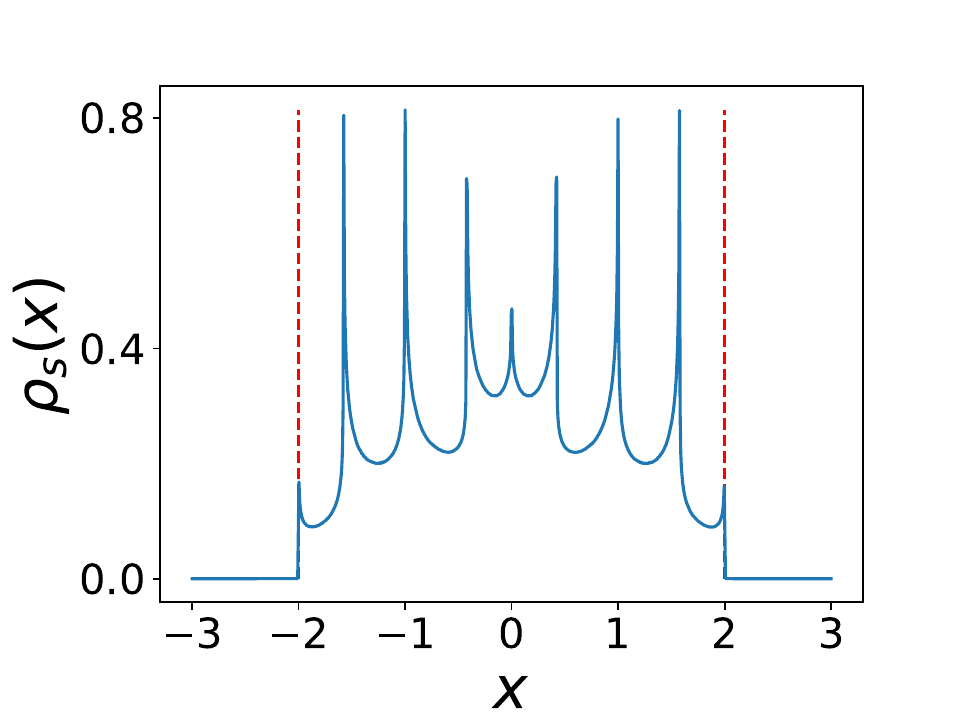}
    \includegraphics[width=0.32\linewidth]{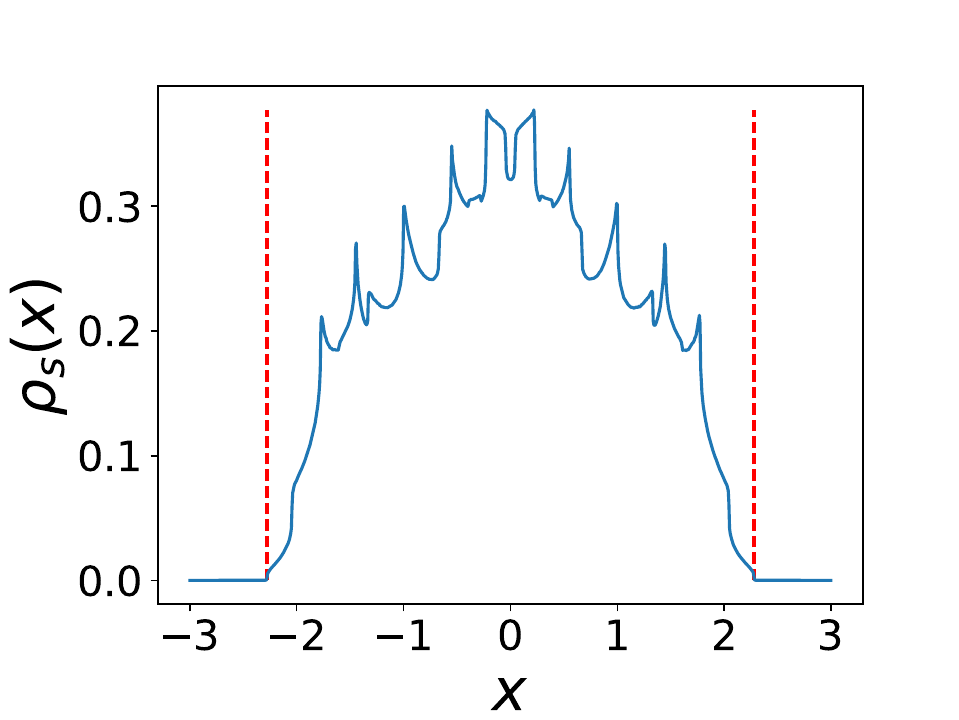}
    \includegraphics[width=0.32\linewidth]{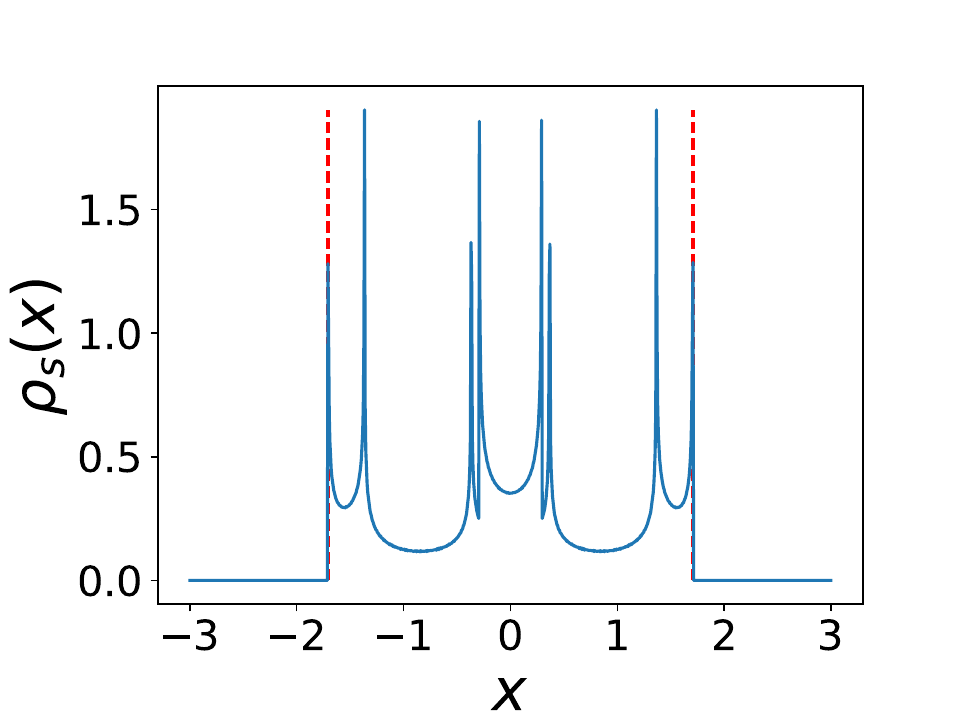}
    \includegraphics[width=0.32\linewidth]{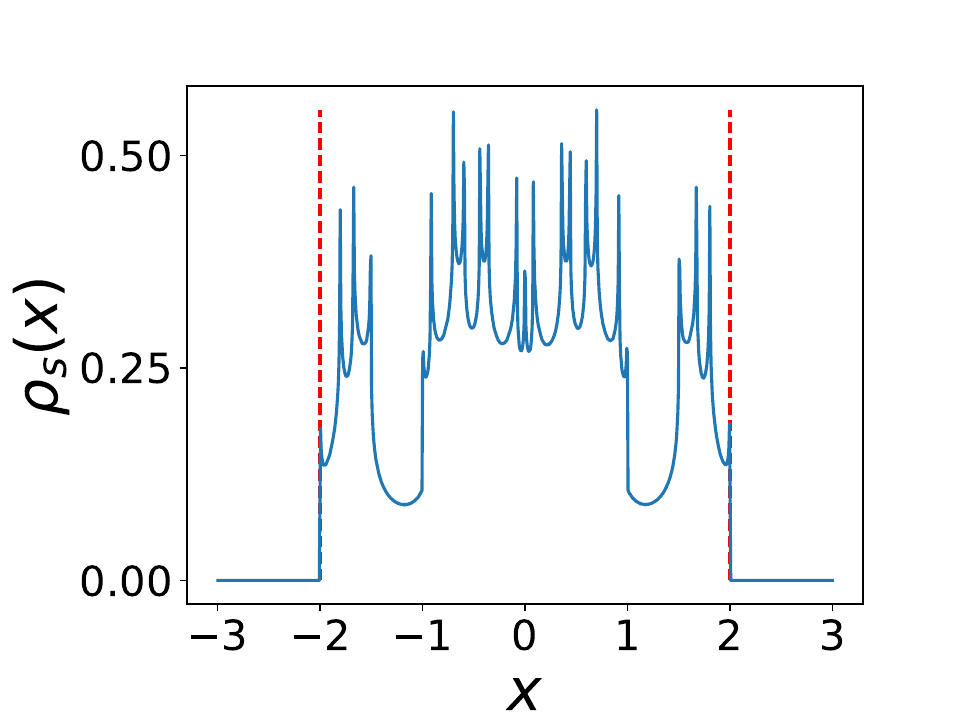}
    \includegraphics[width=0.32\linewidth]{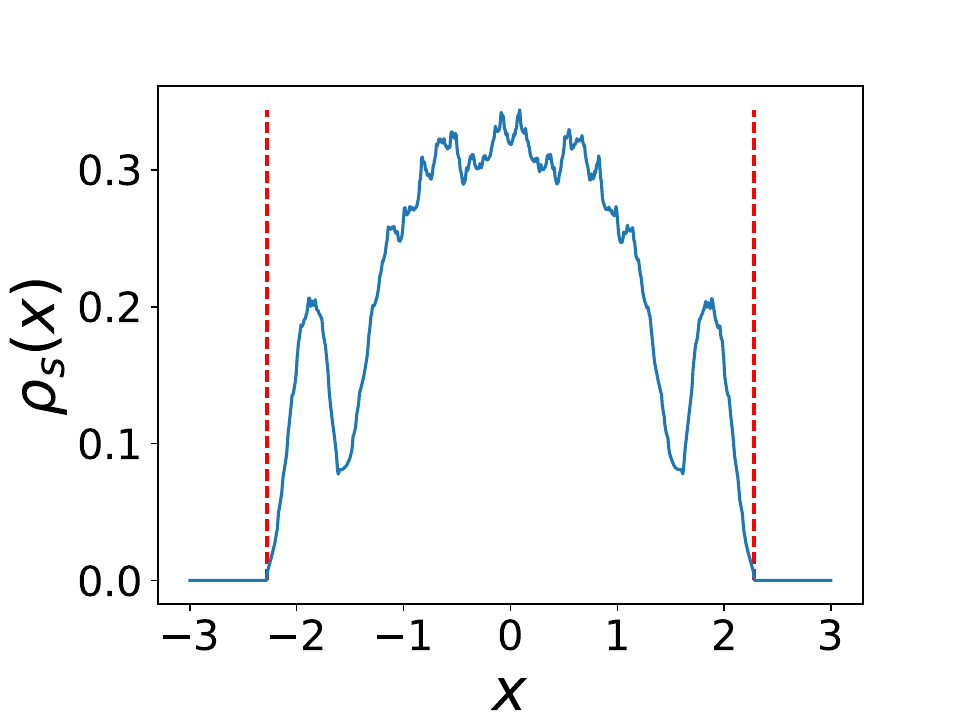}
    \caption{Total particle density $\rho_s(x)$ in model I (top) and II (bottom) for $N=2$, 3 and 5. The other parameters are $g=1$, $v_0=1$ and $\gamma=0.25$. When $N\gamma \leq 1$ we observe singularities in the density. The red lines show the predicted edges of the support.
    }
    \label{finiteN}
\end{figure}

To understand the stationary measure for finite $N$ it is useful to consider first the
case $\gamma=0$, in which case the $\vec \sigma=(\sigma_1,...,\sigma_N)$ are frozen. This allows to identify the "fixed points" of the dynamics, which will
also determine the singularities of the stationary measure for $\gamma>0$. 
For each given $\vec \sigma$ there is a set of fixed points in the space $\vec x=(x_1,...,x_N)$, which correspond to the local minima of the potential
\begin{equation} \label{potential} 
    V_{\vec \sigma}(\vec x) = \frac{1}{2} \sum_i x_i^2 -\frac{2}{N}\sum_{i<j} g_{\sigma_i,\sigma_j} \log |x_i-x_j| - v_0 \sum_i \sigma_i x_i
\end{equation}
which is invariant by a global permutation $\tau$ of the particles $(x_i,\sigma_i) \to (x_{\tau(i)},\sigma_{\tau(i)})$. 
Indeed the potential \eqref{potential} has the property that any fixed point $\partial_t \vec x = - \vec{\nabla} V_{\vec \sigma}(\vec x)=0$
is fully stable hence is a local minimum \cite{SM}. Therefore, starting from a given initial condition there is a unique accessible fixed point.

In model I, for $\gamma=0$ and for any given initial condition, the system ends up forming two groups with the $N_-$ particles with $\sigma=-1$ shifted to the left, 
and the $N_+$ particles with $\sigma=1$ shifted to the right. Within each group the particles keep their initial order. Note that the
two groups can overlap. Up to a permutation $\tau$ of the particles the fixed points are thus labeled by $(N_+,N_-)$,
hence there are $N+1$ of them. The positions of the particles in each group are 
$x_i=\sigma_i v_0 + \sqrt{\frac{2g}{N}}y_i^{{\sigma_i}}$, where the $y_i^{{\pm}}$'s are again the zeroes of the Hermite polynomials $H_{N_+}$ and $H_{N_-}$ respectively, see \cite{SM}, leading in the large $N$ limit to two shifted semi-circles. 

In model II, for $\gamma=0$, since the particles cannot cross it is convenient to assume that the $x_i$'s are ordered (which is always possible up to
a permutation $\tau$). Then 
each $\vec \sigma$ leads to a different fixed point, 
so that in total there are $2^N$ different fixed points. It is difficult to compute them analytically.

These fixed points lead to singularities in the stationary state for $\gamma>0$ which are visible in the stationary particle density $\rho_s(x)$, determined numerically in
Fig. \ref{finiteN}. Each time the velocity
of a particle switches sign, the system flows towards the corresponding fixed point until the next change of sign. The smaller $\gamma$, the longer the particles stay near the fixed point $\vec x$. This leads to an algebraic singularity in $\rho_s(x)$ of the type $|x-x^*|^{N \gamma/\lambda -1}$
around all $x^*$ which coincide with one of the $x_i$'s (see also \cite{footnotelog}). There are 
generically
$N (N+1)$ (model I) and
$N 2^N$ (model II) such singularities in $\rho_s(x)$ 
(some of which may coincide). Note that it agrees with the known result for $N=1$~\cite{HJ95,TailleurCates,DKM19}.
In the large $N$ limit these singularities are washed out in the bulk
and at the edge they result in a shrinking of the support of the density mentioned above. We now study this limit for
each model separately.

{\bf Model 1 and large $N$ limit.} Consider now interactions between the same species only, i.e. the model \eqref{model0} with 
the choice 
$g_{\sigma,\sigma'}= g \, \delta_{\sigma,\sigma'}$ in \eqref{def_model},
for any $T$. To study the limit of large $N$ we extend the DK approach~\cite{Dean,Kawa} to derive evolution equations
for the densities $\rho_\pm(x,t)$ defined in \eqref{def_rho} in presence of the active noise.
As shown in \cite{SM} for the model I, they take the form at large $N$
\bea
\partial_t \rho_\sigma(x,t) &=&  \partial_x \left[ \rho_\sigma(x,t) \left( x - v_0 \sigma    -  2 g
\fint dy \frac{\rho_\sigma(y,t)}{x-y}  \right)\right] \nonumber \\
 &+& \gamma \left(\rho_{-\sigma}(x,t) - \rho_{\sigma}(x,t) \right) + O(1/\sqrt{N})
\label{Dean_eq} \;,
\eea 
for $\sigma = \pm 1$ and where $\fint$ denotes the Cauchy principal value. The correction terms are (i) a $O(1/\sqrt{N})$ random term coming from the active noise, 
(ii) deterministic terms of order $O(1/N)$ (which in the standard DBM case $v_0=0$ can be written
exactly), see \cite{SM} for more details on these terms. As in the case of the DBM one 
introduces the "resolvents", i.e., the Stieltjes transforms of the $\rho_\sigma(x,t)$ 
\be 
G_\sigma(z,t) = \int dx \frac{\rho_\sigma(x,t)}{z-x}  = \frac{1}{N} \sum_i \frac{\delta_{\sigma_i(t),\sigma} }{z-x_i(t)} 
\ee 
for $z$ in the complex plane minus the support of the densities. Their asymptotic behaviors are 
\bea \label{G_largez}
G_\pm(z,t) \simeq_{z \to \infty}   \frac{1}{z}  \int dx \rho_\pm(x,t) =  \frac{p_\pm(t)}{z} 
\eea 
with $p_\sigma(t) =\frac{1}{N} \sum_i \delta_{\sigma_i(t),\sigma}$ and $p_+(t)+p_-(t)=1$. In the limit $N \to +\infty$
one can neglect the $O(1/\sqrt{N})$ terms in \eqref{Dean_eq} and
the time evolution of the density becomes deterministic. This equation can be rewritten as 
a pair of equations for $G_\pm(z,t)$
\beq
\partial_t G_\sigma = \partial_z (-v_0 \sigma G_\sigma +  z G_\sigma {-} g G_\sigma^2) + \gamma G_{-\sigma} - \gamma G_{\sigma} \;.
\label{2species_G}
\eeq
These equations allow to study the stationary densities of the system, which we denote $\rho_\pm(x)$ -- each being normalized
to $\frac{1}{2}$ by symmetry. Setting the time derivatives to zero and introducing
the densities $\rho_s=\rho_++\rho_-$ and $\rho_d=\rho_+-\rho_-$ and their Stieltjes transforms $G_s = G_+ + G_-$ and $G_d = G_+ - G_-$ respectively,
we get from (\ref{2species_G})
\bea
&& 0 = \partial_z (-v_0 G_d +  z G_s  - \frac{g}{2} (G_s^2+G_d^2)) \;, \label{2species_Gs} \\
&& 0 = \partial_z (-v_0 G_s +  z G_d  - g G_s G_d) - 2\gamma G_d \;. \label{2species_Gd}
\eea
The first equation can be integrated, using the large $z$ behaviors in (\ref{G_largez})
\be \label{eq1integrated} 
-v_0 G_d +  z G_s  - \frac{g}{2} (G_s^2+G_d^2) = 1 \;.
\ee 
In the case $v_0=0$, one finds that $G_d=0$ is indeed a solution, as expected, together with
$G_s(z)= \frac{z}{g} (1-\sqrt{1-\frac{2 g}{z^2}})$ which recovers the semi-circle density 
\be
 \rho_s(x) = \frac{1}{\pi} {\rm Im} G_s(x-i 0^+) =  \frac{1}{ \pi g} \sqrt{(2 g - x^2)_+} 
\ee 
for the total density $\rho_s$. It has support over $[-\sqrt{2 g},\sqrt{2 g}]$ which is indeed strictly 
included in $[-v_0-2\sqrt{g},v_0+2\sqrt{g}]$ as discussed above.

We now turn to the case $v_0>0$. In the limit $g \to 0$, one obtains the solution
$G_s(z) = \frac{1}{z} \, _2F_1\left(\frac{1}{2},1;\gamma  +\frac{1}{2};\frac{v_0^2}{z^2}\right)$
and $v_0 G_d(z)= z G_s(z)-1$, which is consistent with
\bea
\rho_s(x) = A \left(1 - \left(\frac{x}{v_0}\right)^2\right)_+^{\gamma-1} \, , \, \,
\rho_d(x)=\frac{x}{v_0} \rho_s(x)
\label{rhosol_g0}
\eea
with $A=\frac{\Gamma \left(\gamma+\frac{1}{2}\right)}{\sqrt{\pi } v_0 \Gamma (\gamma)}$ which recovers, as expected, the solution for a single RTP (i.e., $N=1$) \cite{DKM19}. In the case $g>0$ the coupled equations \eqref{2species_Gd}-\eqref{eq1integrated} are more difficult to solve. Interesting
observables are the integer moments $\langle x^k \rangle_\pm$ of the densities $2 \rho_{\pm}(x)$ (normalized to unity), 
as well as $m^s_k$ and $m^d_k$ of $\rho_s$ and $\rho_d$. They are obtained exactly by recursion from the large $z$ expansion 
$2 G_\pm(z)=\sum_{k=0}^\infty \frac{\langle x^k \rangle_{\pm}}{z^{k+1}}$. From the unicity of the stationary state \cite{unicity}
we have the symmetry $\rho_+(x)=\rho_-(-x)$ leading to 
$m^s_{2p}=\langle x^{2p} \rangle_+=\langle x^{2p} \rangle_-$ as well as 
$m^d_{2p+1}=\langle x^{2p+1} \rangle_+=-\langle x^{2p+1} \rangle_-$,
{and} $m^s_{2p+1}=m_{2p}^d=0$. One finds
\be
 m^d_1 = \langle x {\rangle}_+ = \frac{v_0}{1+2\gamma} ~  , ~
 m^s_2 = \langle x^2 \rangle_+ = \frac{v_0^2}{1+2\gamma} + \frac{g}{2}
\ee 
The higher moments are obtained from the following recursion relation  
\beq \label{moments}
\langle x^k \rangle_+ = \frac{\frac{k}{2}}{\frac{k}{2}+\gamma \delta_{k,{\rm odd}}} (v_0 \langle x^{k-1} \rangle_+ + \frac{g}{2} \sum_{l=0}^{k-2} \langle x^l \rangle_+ \langle x^{k-2-l} \rangle_+) \;
\eeq
with $\delta_{k,{\rm odd}} = 1$ if $k$ is odd and $0$ otherwise. Explicit expressions are given in \cite{SM},
where we also obtain exactly the first three moments for any $N$. 
These predictions are in excellent agreement with our numerical simulations. It is also possible
to obtain predictions for the {\it time dependent} moments (beyond stationarity) from the
large $z$ expansion of Eq. (\ref{2species_G}) and the agreement with numerics is also excellent.

From the recursion \eqref{moments} we compute the moments to a high order. This allows
to determine numerically \cite{flajolet2009}, for $N \to \infty$, (i) the position of the edges $x_+=-x_-$, (ii)
the behavior of the densities near the edges. Over a wide range of parameters $(v_0, g,\gamma)$, we obtain the large $k$ behavior compatible with \beq
\langle x^k \rangle_+ \simeq x_+^k (A k^{-\frac{3}{2}} + (-1)^k B k^{-\alpha-1}) \; \quad , \quad \alpha \approx \frac{3}{2}.
\label{singularity_evenodd}
\eeq
This
indicates that the density $\rho_+(x)$ exhibits two distinct behaviors near the upper and lower edges, i.e., 
$\rho_{+}(x) \sim (x_+-x)^{1/2}$ (as for the semi-circle) and 
$\rho_{+}(x) \sim (x+x_+)^{3/2}$ respectively. The fact that the exponents near $\pm x_+$ differ by unity appears to
be a more general feature also valid for the finite $N$ singularities \cite{SM}. 
The value $\alpha=3/2$ is confirmed by a small $\gamma$ expansion, as we now discuss.

We start from the integrated version of \eqref{2species_G} {(in the stationary state)}
\bea
&& -v_0 G_+(z) +  z (G_+(z) - \frac{1}{2}) - g G_{+}^2(z) \\
&& = \gamma \int_{-\infty}^z [G_+(z') + G_+(-z')] dz'
\label{integrated_G+0}
\eea
where we have used $G_-(z)=-G_+(-z)$ and $G_+(z) \sim \frac{1}{2z}$ when $z \to \pm \infty$ \cite{foot_infinity}.

Let us first discuss the limit $\gamma=0^+$ \cite{foot_frozen}. 
From \eqref{integrated_G+0} one obtains $G_+(z) \simeq \frac{z-v_0}{2} \left(1 - \sqrt{1-\frac{2 g}{(z-v_0)^2}}\right) $
corresponding to a semi-circle density of support $[v_0-\sqrt{2 g},v_0+\sqrt{2 g}]$
\beq
\rho_+(x) = \frac{\sqrt{(2 g-(x-v_0)^2)_+}}{2\pi g}
\eeq
Similarly $\rho_-(z)$ is a semi-circle of support $[-v_0-\sqrt{2 g},-v_0+\sqrt{2 g}]$. 
The total density $\rho_s(x)$ is thus the superposition of two shifted semi-circles
centered at $\pm v_0$.

\begin{figure}[t!]
    \centering
    \includegraphics[width=0.45\linewidth]{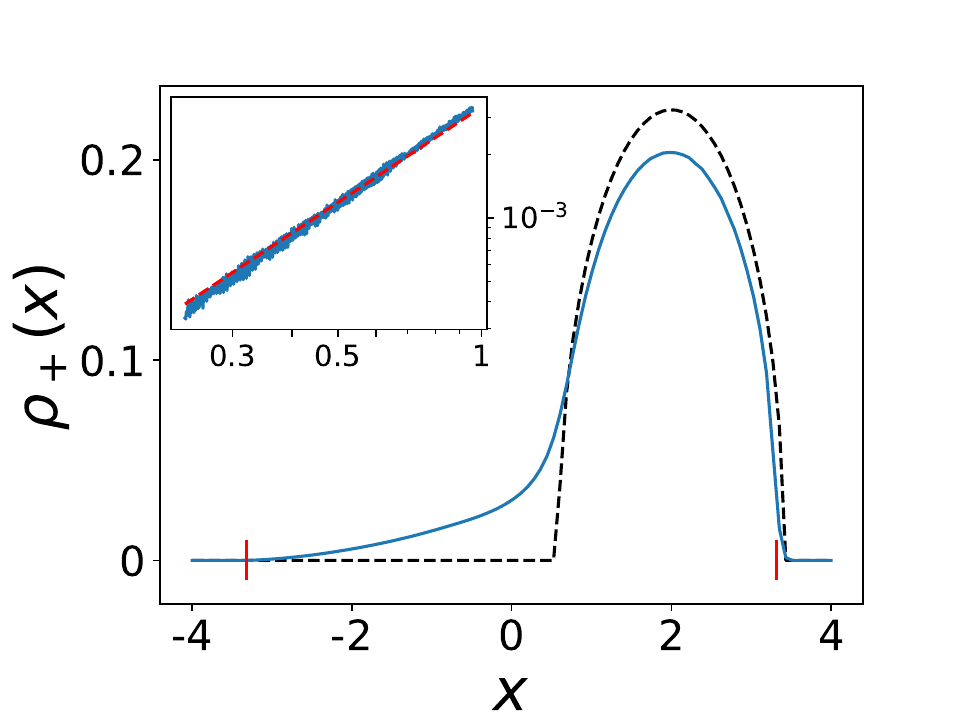}
    \includegraphics[width=0.45\linewidth]{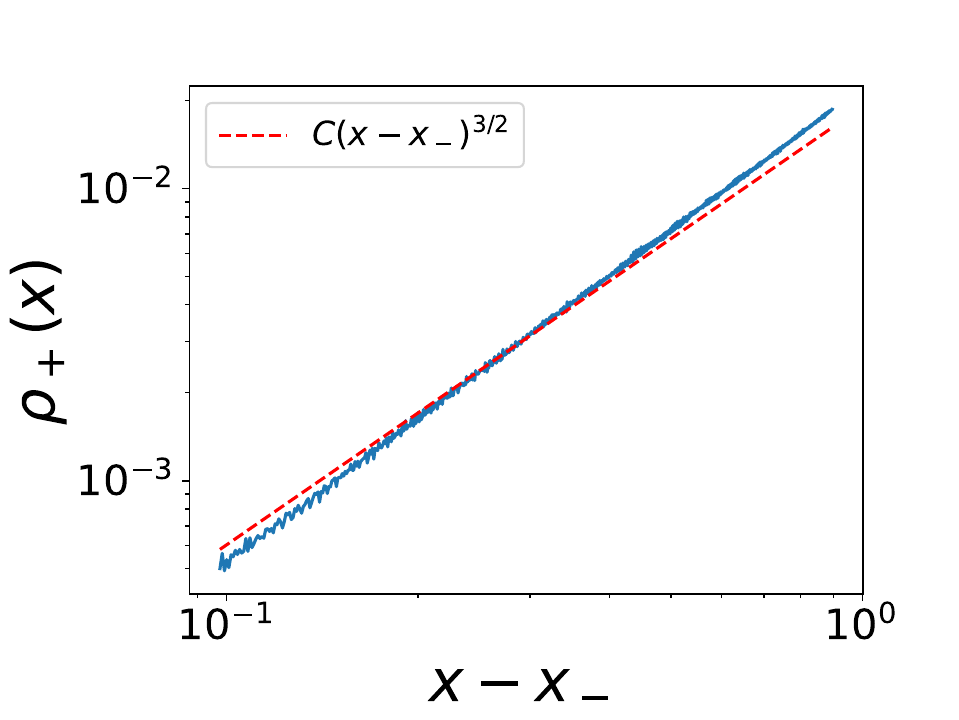}
    \caption{Left: Density $\rho_+(x)$ for $g=1$, $v_0=2$ and $\gamma=0.1$, for $N=100$. The dashed black line shows the limit $\gamma=0^+$ for $N\to +\infty$. The small red lines show the edges $x_{\pm}$ of the support for $N\to +\infty$, computed numerically using \eqref{moments}. Inset: same density close to the left edge in log-log scale. The dashed red line has slope $3/2$. The $3/2$ exponent is observed in a small window between the finite $N$ exponential regime around the edge and the bulk regime. Right: Same plot as the inset for $\gamma=1$ which shows that the $3/2$ exponent is valid beyond the $\gamma \to 0$ limit. 
}
    \label{plot_model2.1}
\end{figure}

It is possible to carry a systematic expansion in powers of $\gamma$ \cite{SM}.
One finds that the semi-circle exponent $1/2$ near $x_+$ survives for $\gamma>0$,
$\rho_+(x) \simeq A\sqrt{x_+-x}$, with a $O(\gamma)$ shift in 
the position of the upper edge $x_+$ 
and in the amplitude $A$, given explicitly in \cite{SM}. 
Interestingly, near the lower edge $x_-=- x_+$
\beq
\rho_+(x) \simeq \frac{\gamma (x-x_-)^{3/2}}{3\pi 2^{1/4} {g^{{3}/4}} \sqrt{{v_0(v_0+\sqrt{2g})}} } \;,
\eeq
which confirms that $\alpha = 3/2$, as anticipated above in Eq. (\ref{singularity_evenodd}) and in agreement with numerics (see Fig. \ref{plot_model2.1}). 

Another solvable limit is the diffusive limit $v_0,\gamma \to +\infty$ with a fixed "effective"
diffusion constant $\frac{v_0^2}{2\gamma}=D$. In that limit the telegraphic noise converges
to Gaussian white noise and it is natural to ask whether model I recovers the physics of the DBM.
From \eqref{2species_Gd} one finds $G_d\simeq\frac{v_0}{2\gamma} \partial_z G_s$ and 
from \eqref{eq1integrated} we obtain the following equation for $G_s$
\beq \label{bouchaud} 
(zG_s-1) -\frac{g}{2} G_s^2 + D \partial_z G_s = 0 \;.
\eeq
If in \eqref{bouchaud} {$D \ll 1$ (e.g., one takes $\gamma$ to infinity while $v_0$ remains finite), see Fig. \ref{phase_diagrams}}, one recovers the semi-circle density
with edge $\pm \sqrt{g}$. On the other hand, 
here $D=O(1)$ which corresponds to a white noise in \eqref{model0} with $T=D N$. Our Eq.
\eqref{bouchaud} is then the same equation as in \cite{BouchaudGuionnet}, 
where they considered the DBM with $\beta=\frac{2c}{N}$ and $c=\frac{g}{2 D}$. 
Performing the change of units, and using the solution given in \cite{BouchaudGuionnet} 
we obtain the total density in our model as
\bea
\rho_s(x) &=& \sqrt{\frac{D}{2\pi}} \frac{1}{\Gamma(1+c)} \frac{1}{|{\cal D}_{-c}(ix)|^2} \\
{\cal D}_{-c}(z) &=& \frac{e^{-\frac{z^2}{4D}}}{\Gamma(c)} \int_0^{+\infty} dx \ e^{-\frac{1}{D} (zx + \frac{x^2}{2})} \left( \frac{x}{\sqrt{D}} \right)^{c-1} \eea
where ${\cal D}_{-c}(z)$ is the parabolic cylinder function. This density
interpolates between the semi-circle for $c \to +\infty$ and the Gaussian for $c = 0$. It is thus interesting to see that the diffusive limit of model I corresponds to the DBM in the special limit $\beta = O(1/N)$, well studied in the context of random matrix theory~\cite{cuenca,dumaz,allez_satya}. Note also that the effective coupling constant is actually $g/2$, instead of $g$, since at any given time each particle interacts only with half of the system.

{\bf Model 2 and large $N$ limit.}
We now turn to the fully interacting version of the model. 
One important difference with model I is that the trajectories of the particles cannot cross, so that they keep the same ordering at all times. This case is more difficult to study analytically and in particular we do not have the equivalent of
the DK approach (its standard version fails, see below). Based on a thorough numerical study, we find a quite interesting feature of the model, {namely that there are three different regimes as $\frac{g}{v_0^2}$ is varied (see Fig. \ref{phase_diagrams}). Let us first focus on the first two regimes.}  
For $\frac{g}{v_0^2} = O(1)$, the stationary density is consistent with a Wigner semi-circle, independently of $\gamma > 0$. On the other hand, for $\frac{g}{v_0^2} \to 0^+$ while keeping the non intersection constraint, the particles tend to form clusters which we characterize numerically. The crossover between the two regimes appears to occur on a scale $\frac{g}{v_0^2} = O(1/N)$ so we surmise, based on our numerical results, that
\bea
&& \rho_s(x)  \sim \frac{1}{\sqrt{g}}  f\left( \frac{x}{\sqrt{g}} \right) \quad ,\quad \frac{g}{v_0^2} \gg \frac{1}{N} \;, \label{fsc}\\
&& \rho_s(x)  \sim \frac{\sqrt{N}}{v_0}  \phi\left( \sqrt{N} \frac{x}{v_0} \right) \quad , \quad \frac{g}{v_0^2} \ll \frac{1}{N} \;, \label{phi}
\eea
where $f(z)$ and $\phi(z)$ are scaling functions which we discuss below. 

\begin{figure}[t!]
    \centering
    \includegraphics[width=0.45\linewidth]{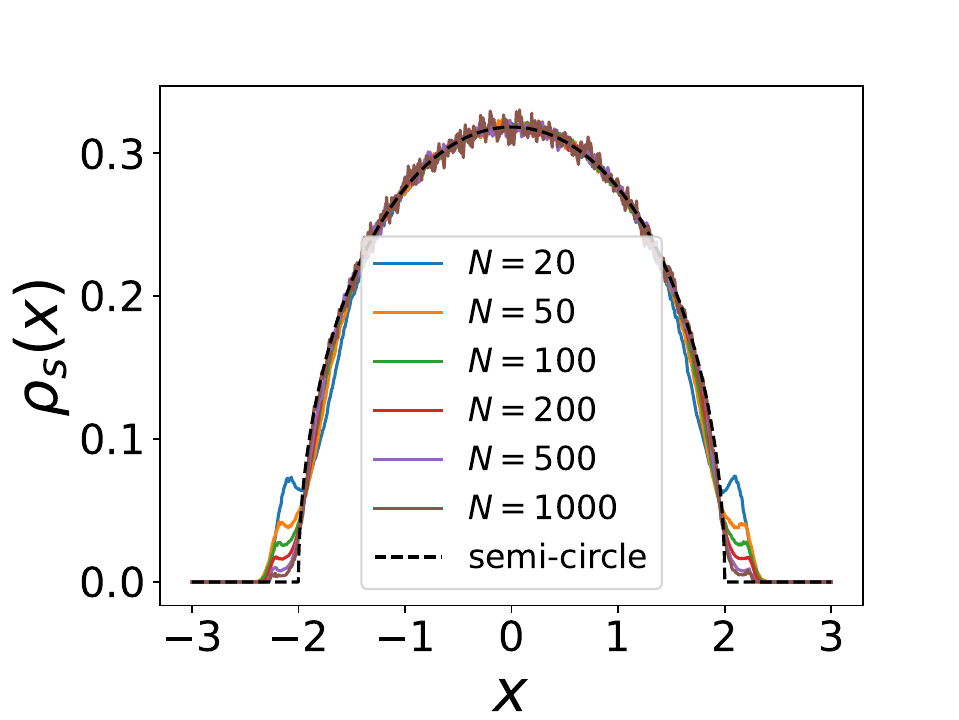}
    \includegraphics[width=0.45\linewidth]{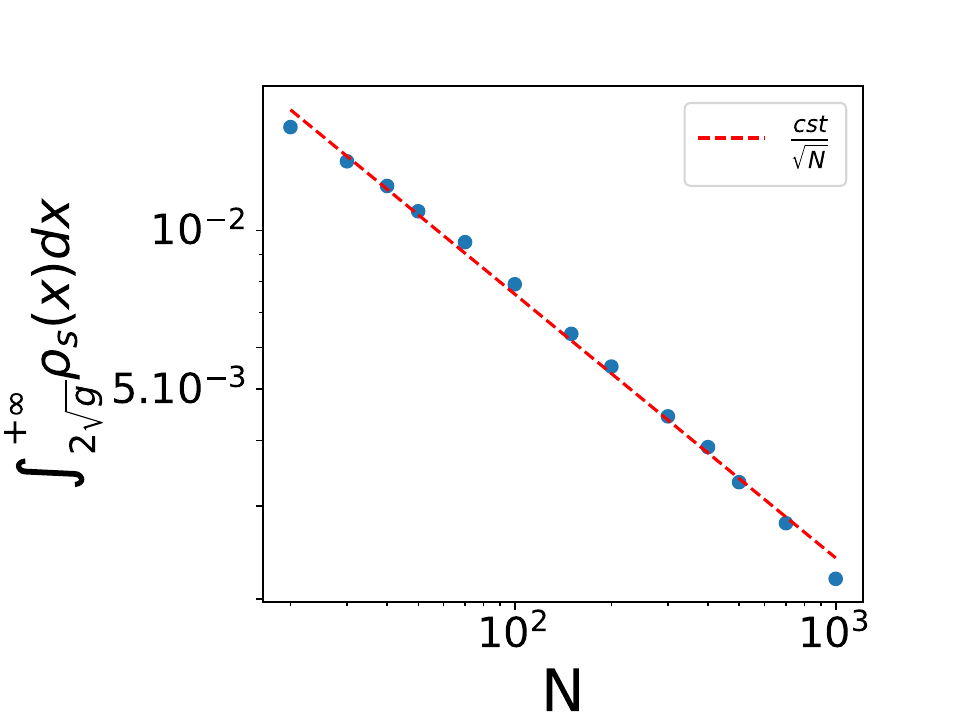}
    \caption{Left: 
    Density of particles for $\lambda=1$, $g=1$, $v_0=1$ and $\gamma=1$ for  different values of $N$. Right: fraction of particles above the right edge of the semi-circle as a function of $N$, for the same set of parameters. It decreases as $1/\sqrt{N}$.}
    \label{plot_model2}
\end{figure}

Let us start with the regime $\frac{g}{v_0^2}=O(1)$. The naive generalization of the DK equation to this case amounts to Eq. \eqref{Dean_eq} 
where one replaces \cite{SM} $2 g
\fint dy \frac{\rho_\sigma(y,t)}{x-y}$ by $2 g
\fint dy \frac{\rho_s(y,t)}{x-y}$ since each particle interacts identically with both species (we recall
that $\rho_s=\rho_+ +\rho_-$). However this equation does not hold even at large $N$ as we have carefully checked numerically.
To understand this we consider for both models the exact equation satisfied by $p_\sigma(x,t)=\langle \rho_\sigma(x,t) \rangle$ 
where $\langle \dots \rangle$ denotes the average over the different histories $\sigma(t)$. This equation 
is not closed but involves, in its interaction term, the pair correlation $p^{(2)}_{\sigma,\sigma}(x,y,t)$ 
for model I and $p^{(2)}_{\sigma,s}(x,y,t)$ for model II (see \cite{SM} for details). For model I, we have checked numerically
that at large $N$ this pair correlation can be replaced by its factorized form $p_\sigma(x,t) p_\sigma(y,t)$ leading to a closed equation for
$p_\sigma(x,t)$ which coincides with the DK equation. By contrast, for model II, we find that replacing the pair 
correlation by its factorized form $p_\sigma(x,t) p_s(y,t)$ is inconsistent with the numerics, even at large $N$~\cite{SM}. Remarkably, we find that, in that case, the density $\rho_s(x)$ converges to a Wigner semi-circle with support on $[-2\sqrt{g}, 2 \sqrt{g}]$, i.e., it takes the scaling form as in \eqref{fsc} with $f(z)=f_{sc}(z) = \frac{1}{2\pi}\sqrt{4-z^2}$, independently of $\gamma$. This convergence is shown in Fig. \ref{plot_model2} (left) for $g=1$ where we also
see that the density $\rho_s(x)$ exhibits "wings" outside the semi-circle, with total weight 
scaling as $N^{-1/2}$, see Fig. \ref{plot_model2} (right). 
In addition we find that $\rho_d(x)=\rho_+(x)-\rho_-(x)$ vanishes at large $N$.

\begin{figure}[t!]
    \centering
    \includegraphics[width=0.32\linewidth]{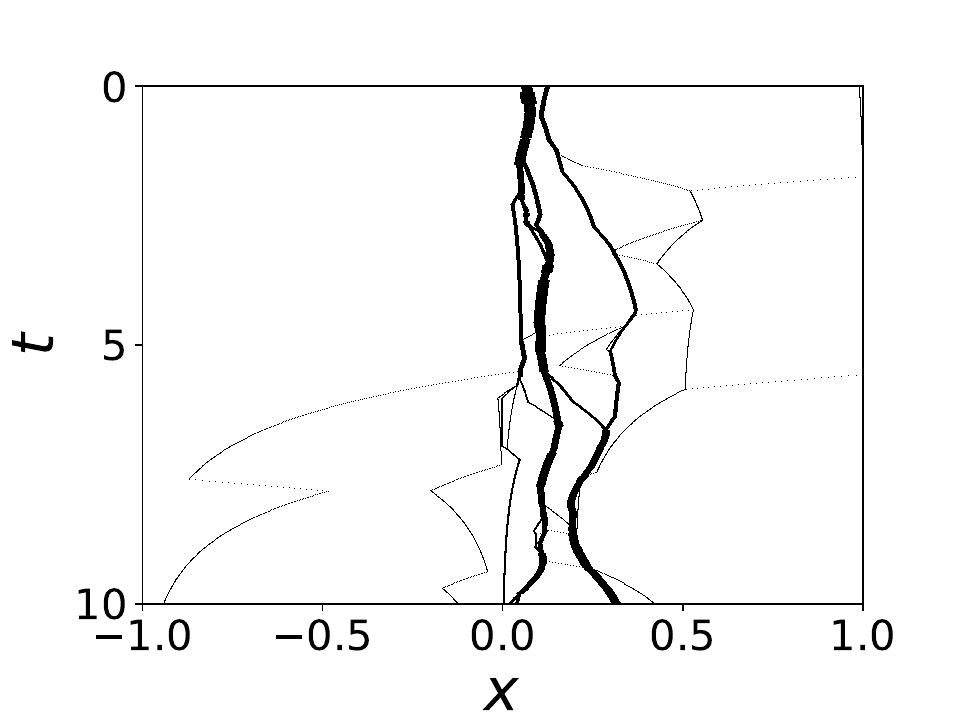}
    \includegraphics[width=0.32\linewidth]{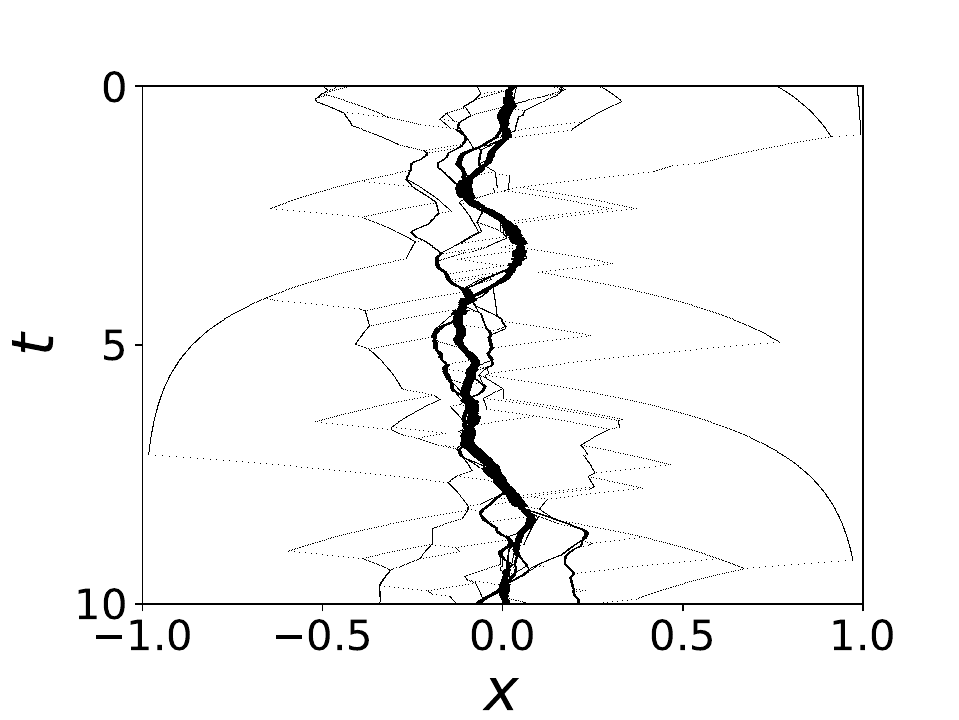}
    \includegraphics[width=0.32\linewidth]{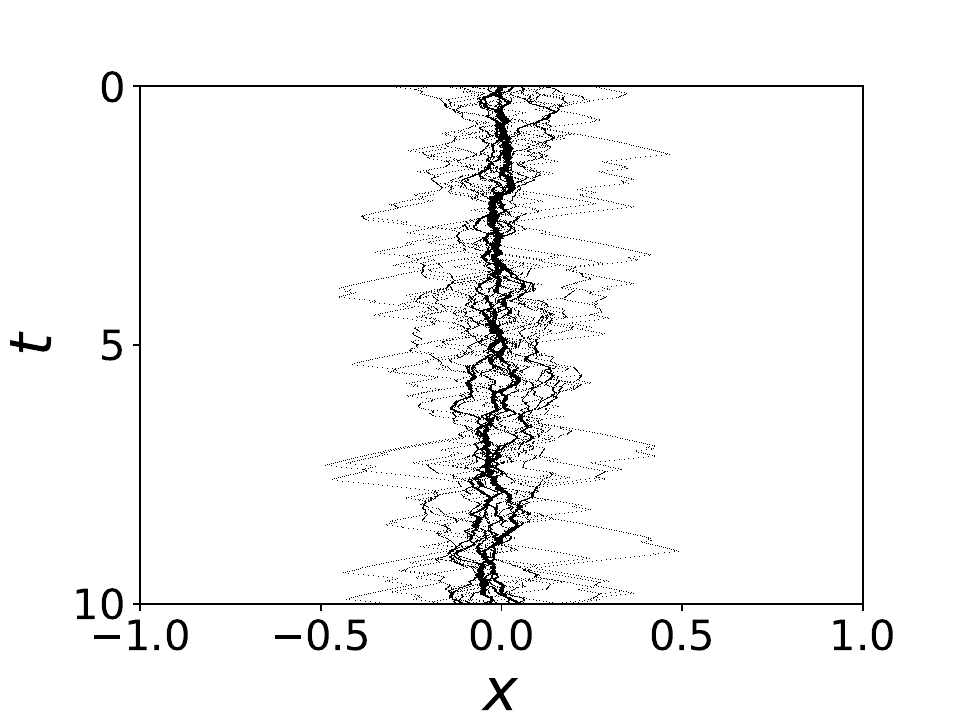}
    \caption{Time evolution of the positions of $N=100$ particles for the effective model $g=0^+$ of model II, for $\lambda=1$, $v_0=1$ and $\gamma=0.1$, $1$ and $10$ from left to right.}
    \label{spacetime_g0}
\end{figure}

\begin{figure}[t!]
    \centering
    \includegraphics[width=0.32\linewidth]{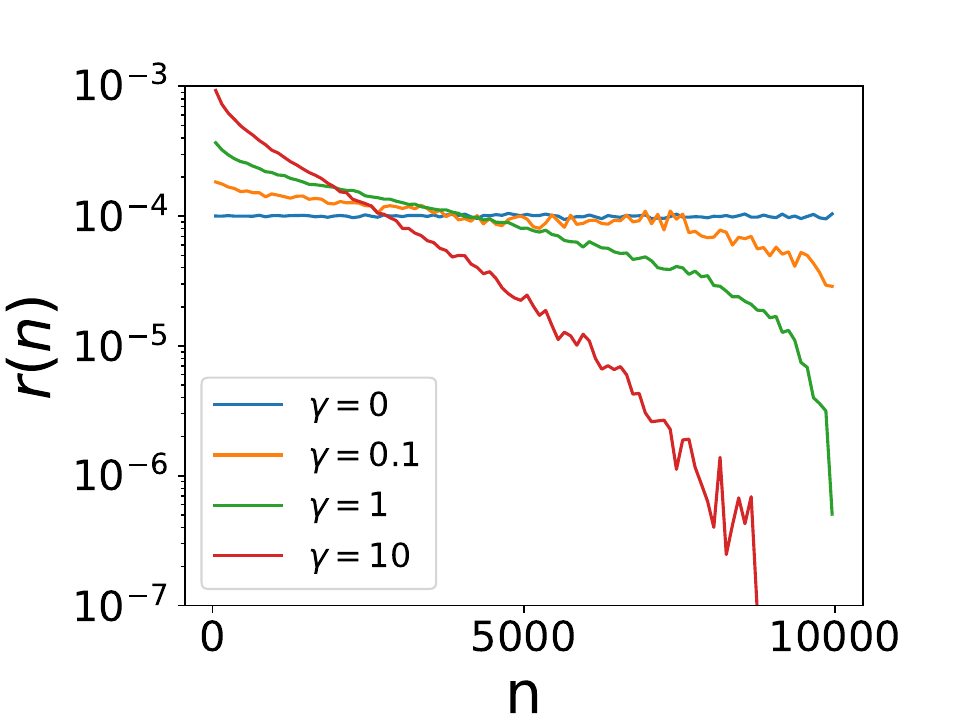}
    \includegraphics[width=0.32\linewidth]{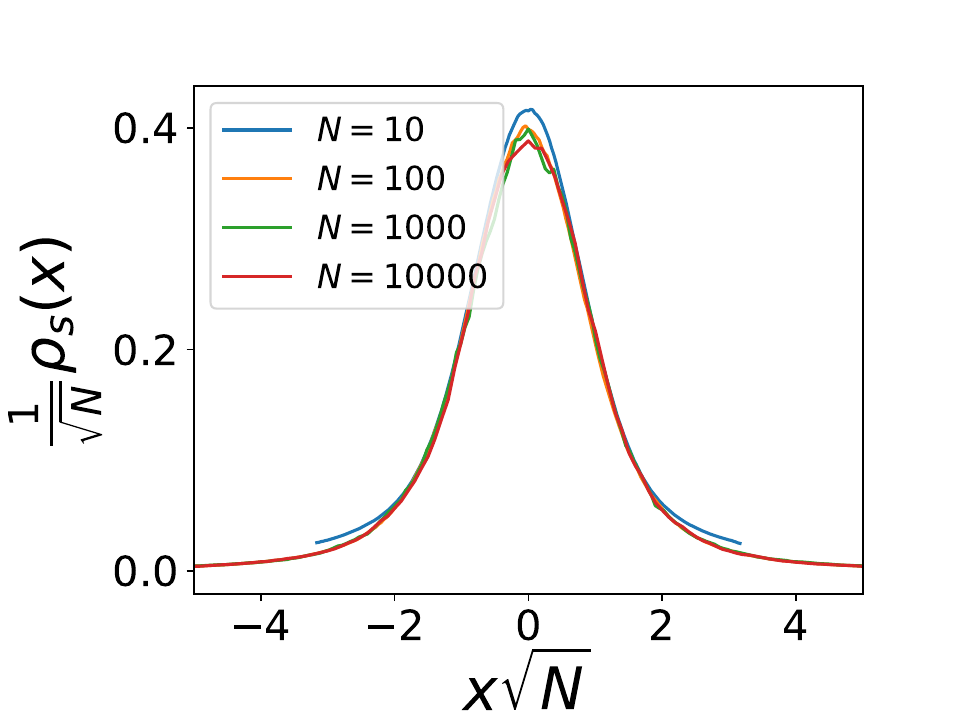}
    \includegraphics[width=0.32\linewidth]{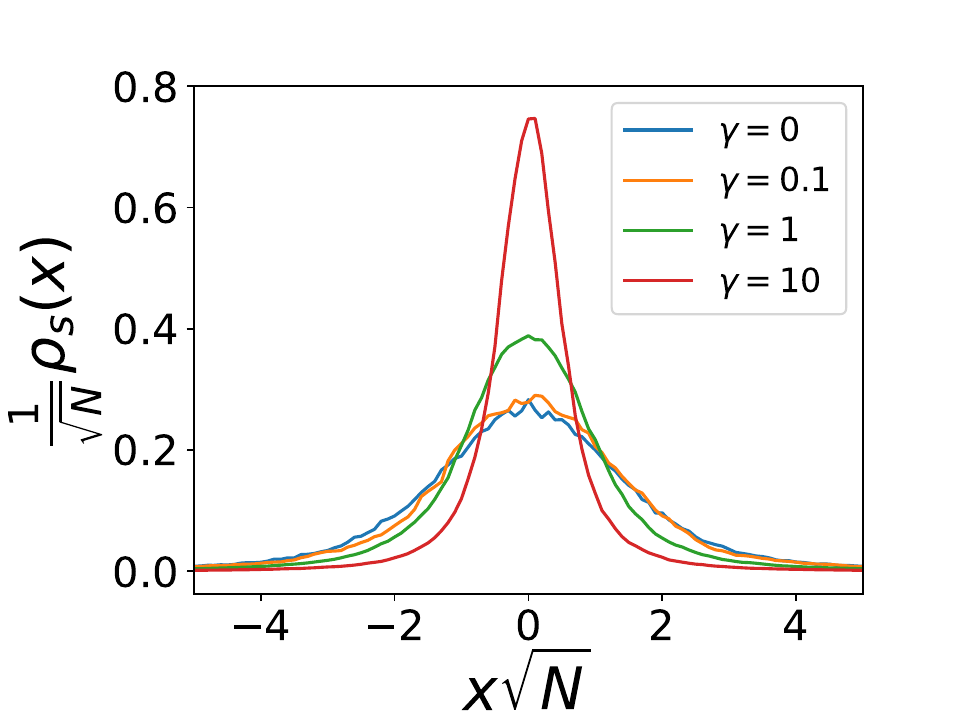}
    \caption{Left: Fraction $r(n)$ of particles in clusters of size $n$, with $r(n)=n p(n)/\sum_{m=1}^N m p(m)$, in the limiting model $g=0^+$ for $N=10000$ and  different values of $\gamma$ (averaged over $10^6$ realisations for $\gamma=0$ and over a time $10^5$ for $\gamma>0$). $r(n)$ is independent of $n$ for $\gamma=0$ and decays exponentially for $\gamma>0$. Center: Rescaled particle density $\rho_s(x/\sqrt{N})/\sqrt{N}$ for different values of $N$ in the limiting model $g=0^+$ for $\gamma=1$. With this rescaling all the plots collapse on the same curve, which is compatible with \eqref{phi}. Right: Rescaled particle density for $N=10000$ and different values of $\gamma$.}
    \label{g0figs}
\end{figure}

The second interesting regime is $\frac{g}{v_0^2} \ll 1/N$. In the limit $\frac{g}{v_0^2} \to 0^+$ the interactions still play an important role since they forbid crossing of the particles. 
One finds that a reliable effective model in that limit can be defined as follows. When particles meet they form a point like cluster. The instantaneous velocity of each cluster is the mean velocity of all the particles in the cluster. A cluster is characterized by the ordered list of the velocities of the particles which have joined. For $\gamma>0$ each particle can change its velocity, which may result in breaking of the cluster in pieces, according to precise rules 
(see \cite{SM} for details). For small $\gamma$ the particles tend to form large clusters, see Fig. \ref{spacetime_g0},
as observed in some RTP lattice models \cite{SG2014,slowman,Dandekar2020}.
We have determined numerically the distribution $p(n)$ of sizes $n$ of these clusters. For $\gamma=0$, $p(n)$ decays as $\frac{1}{n}$, for $n \leq N$, while for $\gamma>0$ the leading behavior is exponential in $n$ with a rate depending on $\gamma$, see Fig.~\ref{g0figs}. 
Finally, our numerics are consistent with the scaling form in Eq. \eqref{phi} for the total density,
with a scaling function $\phi$ which depends on $\gamma$, see Fig.~\ref{g0figs}. Interestingly, this scaling function seems to exhibit a power law tail $\phi(x) \propto 1/x^3$ for large $x$ \cite{SM}. Here also, we find that $\rho_d(x)$ vanishes at large $N$. 


To understand better the above regimes for model II and in particular the appearance of a semi-circle
for $\frac{g}{v_0^2}=O(1)$ it is useful to study the fluctuations of the positions $x_i$ of the particles.
We consider the limit $\gamma=0^+$ such that at large time the system is with equal probability near the
fixed point corresponding to any of the $2^N$ values of $\vec \sigma$. 
We perform an expansion for small $v_0^2/g$ of $\delta x_i=x_i-x_i^{\rm eq}$, where $x_i^{\rm eq}$ denotes
the equilibrium positions for $v_0=0$. One can diagonalize the Hessian matrix ${\cal H}_{ij}=\partial_{x_i} \partial_{x_i} V_{\vec \sigma}(\vec x)|_{\vec x=\vec x^{\rm eq}}$ associated
to the potential in \eqref{potential}, either approximately as in \cite{bouchaud_book},
or exactly \cite{Hermite1}. To linear order one has $ \delta \vec x= v_0\,{\cal H} \vec \sigma$,
and using $\langle \sigma_i \sigma_j \rangle = \delta_{ij}$ we find the following estimates at large $N$ 
\cite{UsFuture}
\be 
{\rm Var}(\delta x_i)  \sim \frac{v_0^2}{N g \rho_i^2} ~ , ~
{\rm Var} (\delta x_i - \delta x_{i+n}) 
\sim \frac{v_0^2 n}{g^2 \rho_i^4 N^2} \label{var}
\ee 
where here $\rho_i= \rho_s(x_i^{\rm eq})$ is the local mean density, which in 
the bulk is $\rho_i \sim 1/\sqrt{g}$. The first quantity
is dominated by soft large wavelength fluctuations which are cutoff by the quadratic well.
It allows to predict the various regimes as $g/v_0^2$ varies, summarized in Fig \ref{phase_diagrams}. 
If $\delta x_i$ is much smaller than the size of the support $\sim \sqrt{g}$ the
semi-circle density which holds for $v_0=0$ remains unchanged. This yields the
condition $g/v_0^2 \gg 1/N$. For $g/v_0^2=O(1/N)$ a crossover occurs to the
regime $g/v_0^2 \ll 1/N$ where particles tend to aggregate as in Fig.~\ref{spacetime_g0}. These conclusions are compatible with the forms in \eqref{fsc} and \eqref{phi}. Note that for very large $g$, $\delta x_i$ is much smaller than the interparticle distance $1/(N \rho_i)= \sqrt{g}/N$, which explains why for $g/v_0^2 \gg N$
the density exhibits peaks at the interparticle scale {(see Fig. \ref{phase_diagrams} and \cite{SM}}). Since we observe numerically that the fluctuations
decrease as $\gamma$ increases the above discussion remains valid for $\gamma>0$. 
Finally, the second result in \eqref{var} indicates that the variance of  
the number of particles in an interval grows linearly with its size.
This is in contrast with the case of the standard DBM where a similar calculation
leads to logarithmic growth (see \cite{bouchaud_book}).



As a final remark, in the diffusive limit $\gamma \to +\infty$, $v_0 \to +\infty$ and $\frac{v_0^2}{2 \gamma}=D$ fixed, we expect model II 
to converge to a variant of the DBM, where $\beta=2g/(N D) \ll 1$, with an additional hard-core repulsion, which remains to be studied. 

{\bf Conclusion}. We introduced a model of interacting active particles in one dimension, the active Dyson Brownian motion, with two different variants,
and studied the stationary density. While for finite $N$ it presents singularities (corresponding to the fixed points of the $\gamma=0$ dynamics), for 
$N \to +\infty$ those singularities are washed out and the density becomes smooth inside its finite support. We developed an analytical approach to compute it for
model I and found $1/2$ and $3/2$ exponents at the edges. For model II, the particles tend to form clusters, and we found strong evidence that the density takes a semi-circular 
shape in a wide range of parameters (see Fig.~\ref{phase_diagrams}). Our results raise several challenging open questions.
The first would be to obtain analytical results for model II, such as for the stationary density, the gaps, and the
cluster statistics. This requires
a better understanding of the failure of the DK equation for model II, which in turn
could provide a new insight into the study of 
active particle systems through hydrodynamic equations. Second, the effect of additional passive noise ($T>0$) 
should be important for model II since for $T>2g$ it enables particle crossings. 
Third, the $g=0^+$ effective model clearly deserves further investigation. 
Finally an intriguing question is whether there exists a matrix model associated to the active DBM.
\\

{\it Acknowledgments:} 
We thank David S. Dean and Satya N. Majumdar for collaborations on related topics. 
We thank LPTMS (Orsay) and Coll{\`e}ge de France for hospitality.
We thank the Erwin Schrödinger Institute (ESI) of the University of Vienna for the hospitality during the workshop {\it Large deviations, extremes and anomalous transport in non-equilibrium systems} in October 2022.

{}

\newpage

.

\newpage

.

\newpage

\begin{widetext}

\setcounter{secnumdepth}{2}

\begin{large}
\begin{center}

Supplementary Material for\\  {\it Interacting, running and tumbling: the active Dyson Brownian motion}

\end{center}
\end{large}

\tableofcontents

\bigskip

We give the principal details of the calculations described in the main text of the Letter. 
We display additional analytical and numerical results which support the findings presented in the Letter. 

\bigskip

\section{Definition of the models}
The two models studied in this paper (model I and model II) consist of $N$ interacting particles whose positions $x_i(t)$ evolve according to the stochastic equation
\bea \label{modelsupmat} 
 \dot x_i(t) &=& - \lambda x_i(t) +  \frac{2 }{N} \sum_{j \neq i} 
\frac{g_{\sigma_i(t), \sigma_j(t)}}{x_i(t)-x_j(t)}
+  v_0 \sigma_i(t) + \sqrt{\frac{2 T}{N}} \xi_i(t) \quad, \quad i = 1,2, \cdots, N\\
g_{\sigma,\sigma'}&=&
\begin{cases}
& g \, \delta_{\sigma,\sigma'} \quad \quad \quad \quad \;\, \rm{(model \; I)} \;,\\ 
& g \quad, \quad \forall (\sigma, \sigma') \quad  \rm{(model \; II)} \;.
\end{cases}
\eea
In Eq. (\ref{modelsupmat}) the variables $\sigma_i(t)=\pm 1$ are independent telegraphic noises which switch sign with a constant rate $\gamma$, and the $\xi_i(t)$ are independent standard white noises. We take $g>0$ so that the interaction between the particles is repulsive. Each particle is subject to an external potential $V(x)=\frac{\lambda}{2} x^2$ (in the rest of the paper we set $\lambda=1$). In this paper we restrict our study of model I and II to the case $T=0$.

For $v_0=0$ (but $T \neq 0$) model II corresponds to the stationary version of the well-known Dyson Brownian motion (DBM), with $\beta=\frac{2g}{T}$ (see e.g., \cite{bouchaud_book} Section 9.4 setting $T=1$ there). In this case the stationary measure for the joint distribution of the positions of the $N$ particles is given by
\beq \label{pdf_beta}
{P}_{\rm joint}(x_1,...,x_N) = \frac{1}{Z_N} e^{- \frac{N \lambda}{2 T} \sum_i x_i^2} \prod_{i<j}|x_i-x_j|^\beta \;,
\eeq
where $Z_N$ is a normalization constant. This joint probability density function (\ref{pdf_beta}) coincides with the joint distribution of eigenvalues for the $\beta$-ensemble of random matrices \cite{forrester_book}. It is well known that in this case the 
one-particle density converges in the $N \to \infty$ limit to the Wigner semi-circle law
\be \label{semicircle}
\rho_{sc}(x) = \frac{\lambda}{g} \frac{\sqrt{(\frac{4g}{\lambda}-x^2)_+}}{2\pi} \;,
\ee 
where we used the notation $(x)_+=x$ if $x>0$ and $0$ otherwise. The limiting density thus has a finite support $[-2\sqrt{g/\lambda}, +2\sqrt{g/\lambda}]$. 

For the Dyson Brownian motion at $T=0$, as well as for the model II at $v_0=0$ and $T=0$,
the equilibrium positions of the particles for any finite $N$ are given by the zeros of the rescaled Hermite polynomial $H_N \left(\sqrt{\frac{N\lambda}{2g}}x \right)$ \cite{mehta_book,Hermite2} (see below for a quick derivation). In the limit $N\to+\infty$ the density of these zeros converges to the Wigner semi-circle \eqref{semicircle}. For the DBM the density remains the same semi-circle for any $T=O(1)$, i.e. the thermal noise scales as $O(1/\sqrt{N})$ and
its effect is weak. Note that the average characteristic polynomial remains equal to the same Hermite polynomial, see e.g. Section 6.1 in \cite{bouchaud_book}). 
However if one takes instead $T=O(N)$, i.e. $\beta = O(1/N)$ (i.e. the passive noise term is of order $O(1)$), 
the particle density is not anymore a semi-circle and extends on the whole real axis \cite{BouchaudGuionnet}. In model II at $T=0$ and $v_0>0$ the noise is instead purely active, but scales as $O(1)$. Hence it is not obvious whether the stationary particle density will be a semi-circle or not. This question
is discussed below and in the main text. Surprisingly, our numerical simulations for model II suggest that the density seems to remain a semi-circle in the limit of large $N$. This is at variance with model I where the stationary density is never a semi-circle. In the diffusive limit discussed in the text, model I converges again to a DBM but in the high temperature regime $\beta = O(1/N)$. For the same reason we expect model II to converge to a DBM with $\beta = O(1/N)$, but with the additional constraint that the particles cannot cross. For $\beta<1$ this constraint changes the statistics of the process
\cite{Allez13,Lepingle07}, which remains to be studied in the present context.


{\bf Non interacting case}. Finally, let us briefly mention some existing results concerning the non-interacting case ($g=0$). In this case the density was computed exactly and is very different from an equilibrium Boltzmann distribution: it has finite support $[-\frac{v_0}{\lambda},\frac{v_0}{\lambda}]$ and exhibits singularities at the edges of the support (see e.g. \cite{DKM19})
\bea \label{rho_inde_app}
&&\rho_{s}(x) = A \left(1 - \left(\frac{\lambda x}{v_0}\right)^2 \right)^{\frac{\gamma}{\lambda}-1} \quad, \quad \quad \quad \quad \; -\frac{v_0}{\lambda} \leq x \leq + \frac{v_0}{\lambda} \;, \\
&& \rho_{\pm}(x) = \frac{A}{2} \, \left(1 \pm \frac{\lambda x}{v_0} \right)^{\frac{\gamma}{\lambda}} \left(1 \mp \frac{\lambda x}{v_0} \right)^{\frac{\gamma}{\lambda}-1}  \quad, \quad -\frac{v_0}{\lambda} \leq x \leq + \frac{v_0}{\lambda} \;,
\eea
with $A = (\lambda/v_0)\frac{\Gamma(1/2+\gamma/\lambda)}{\sqrt{\pi} \Gamma(\gamma/\lambda)}$. 
The difference of 1 between the exponents at the left and right edges of $\rho_\pm(x)$ is actually more general, and valid even
for a large class of external forces $f(x)$ -- such that the support is a single interval $[x_-,x_+]$, i.e. $f(x_\pm)=0$, with the additional assumption $f'(x_\pm) \neq 0$. 
One predicts that
\be \label{ratio}
\frac{\rho_+(x)}{\rho_-(x)} \simeq_{x \to x_+} \frac{1}{x_+-x} \quad , \quad \frac{\rho_+(x)}{\rho_-(x)} \simeq_{x \to x_-} \, (x-x_-)  \;.
\ee 
Indeed, replacing the harmonic force $-\lambda x$ with an arbitrary external force $f(x)$ (still without interactions) the expression in \eqref{rho_inde_app} becomes \cite{DKM19,LMS2020} 
\begin{equation} \label{rho_inde_general}
\rho_\pm(x) = \frac{B}{v_0(v_0 \pm f(x))} \exp\left[2\gamma \int_0^x dy \frac{f(y)}{v_0^2-f^2(y)} \right] \;,
\end{equation}
with $B$ a normalization constant. Here $f(x)$ is assumed to derive from a potential with a single minimum, which we choose to be at $x=0$, and this solution is valid on the interval $[x_-,x_+]$ where $x_\pm$ is the point closest to $x=0$ such that $f(x_\pm)=\mp v_0$. From \eqref{rho_inde_general} we immediately see that if the derivative of $f(x)$ does not vanish at $x=x_\pm$, the exponent with which $\rho_+(x)$ vanishes at $x=x_+$ will be smaller by 1 compared to the exponent of $\rho_-(x)$ (and conversely at $x=x_-$). We will observe something similar in our model in the presence of interactions, see
Fig. \ref{singularity_finiteN_rhopm} below.



\section{Finite $N$}
\label{finiteN_sec}
In this section we study the dynamics of the system for finite $N$. One starts with the case $\gamma=0$ where the particles do not
change their internal state. In that case there are equilibrium configurations which are fixed points of the dynamics. We will study these
fixed points in detail. Next we will consider $\gamma>0$ and see that the above fixed points play an important role to understand the
form of the stationary measure for $\gamma>0$. In particular, the support of the stationary measure can be determined by studying the fixed
point corresponding to a state where all particles have the same velocity.

For $\gamma=0$, for each given $\vec \sigma = (\sigma_1,...,\sigma_N)$ there is a set of fixed points in the space $\vec x=(x_1,...,x_N)$
which are by definition all the solutions of (we have set $\lambda=1$) 
\beq
\dot x_i = \sigma_i v_0 - x_i + \frac{2}{N}\sum_{j\neq i}\frac{g_{\sigma_i,\sigma_j}}{x_i-x_j}=0 \ \ \ {\rm for} \ i=1, \ \ldots \ , N \;. 
\label{eqFixedSigma}
\eeq
Note that the center of mass $\bar x= \frac{1}{N} \sum_i x_i$ satisfies the simple equation
\be \label{c_of_m}
\dot{\bar x} = v_0 \frac{1}{N} \sum_i \sigma_i - \bar x \;.
\ee 
Hence at any of the fixed point one has $\bar x = v_0 \frac{1}{N} \sum_i \sigma_i$. 
Each of these fixed points corresponds to a stationary point of the following potential 
\begin{equation} \label{potential2} 
    V_{\vec \sigma}(\vec x) = \frac{1}{2} \sum_i x_i^2 -\frac{2}{N}\sum_{i<j} g_{\sigma_i,\sigma_j} \log |x_i-x_j| - v_0 \sum_i \sigma_i x_i
\end{equation}
i.e. $\partial_t \vec x = - \vec{\nabla} V_{\vec \sigma}(\vec x)=0$. The potential and the set of fixed points
are invariant by a global permutation $\tau$ of the particles $(x_i,\sigma_i) \to (x_{\tau(i)},\sigma_{\tau(i)})$.
We will show below that all fixed points are stable, i.e. {\it attractive} for the dynamics. 
\\

\subsection{$N=2$ case}
\label{N2section}
Let us illustrate the structure of fixed points for two particles $N=2$. We consider the two cases:

(i) Model II, with $g_{\sigma,\sigma'}=g$. The equations for the fixed points are:
\bea
v_0 \sigma_1 - x_1 + \frac{g}{x_1-x_2}=0 \quad , \quad  v_0 \sigma_2 - x_2 + \frac{g}{x_2-x_1}=0
\eea
which are easily solved introducing $x=x_1+x_2= v_0 (\sigma_1+\sigma_2)$ and $y=x_1-x_2$
which obeys $v_0(\sigma_1-\sigma_2)-y+ \frac{2 g}{y}=0$. For each pair of values $(\sigma_1,\sigma_2)$ we get 2 fixed points: $x=\pm 2 v_0$ and $y=\pm \sqrt{2g}$ or $x=0$ and $y=\pm v_0 \pm \sqrt{v_0^2+2g}$. Taking into account the exchange symmetry between the two particles, this corresponds to 4 different situations, see Fig. \ref{TableN2} : 
\begin{itemize}
    \item both particles are in a state of velocity $+v_0$ (resp. $- v_0$) and have positions $x_{1,2}=v_0 \pm \sqrt{\frac{g}{2}}$ (resp. $-v_0 \mp \sqrt{\frac{g}{2}}$) 
    \item the particles want to move away from each other and have positions $x_{1,2}= \pm \frac{1}{2} (\sqrt{v_0^2+2g}+v_0)$ 
    \item the particles want to move towards each other and have positions $x_{1,2}= \pm \frac{1}{2} (\sqrt{v_0^2+2g}-v_0)$
\end{itemize}
Note that in model II since the particles never cross each other, for each pair $(\sigma_1,\sigma_2)$ the fixed point reached by the dynamics 
is determined by the initial ordering of the particles.
\\

(ii) Model I, with $g_{\sigma,\sigma'}=g \delta_{\sigma,\sigma'}$. The equations are
\be v_0\sigma_1 - x_1 + g\frac{\delta_{\sigma_1,\sigma_2}}{x_1-x_2}=0 \quad , \quad  v_0\sigma_2 - x_2 + g\frac{\delta_{\sigma_1,\sigma_2}}{x_2-x_1}=0
\ee
and the fixed points correspond to 3 different situations, see Fig. \ref{TableN2}:
\begin{itemize}
    \item both particles want to move to the right (resp. left) and have positions $x_{1,2}=v_0 \pm \sqrt{\frac{g}{2}}$ (resp. $-v_0 \mp \sqrt{\frac{g}{2}}$) 
    \item the particles want to move away from each other and have positions $x_i=v_0\sigma_i$. 
\end{itemize}
In model I, if the particles are in states with opposite velocities, the fixed point which is reached by the dynamics is independent of the
initial condition, see Fig. \ref{TableN2} (bottom left). 

\begin{figure}[t!]
    \centering
    \includegraphics[width=0.8\linewidth,trim={0 1.8cm 0 0},clip]{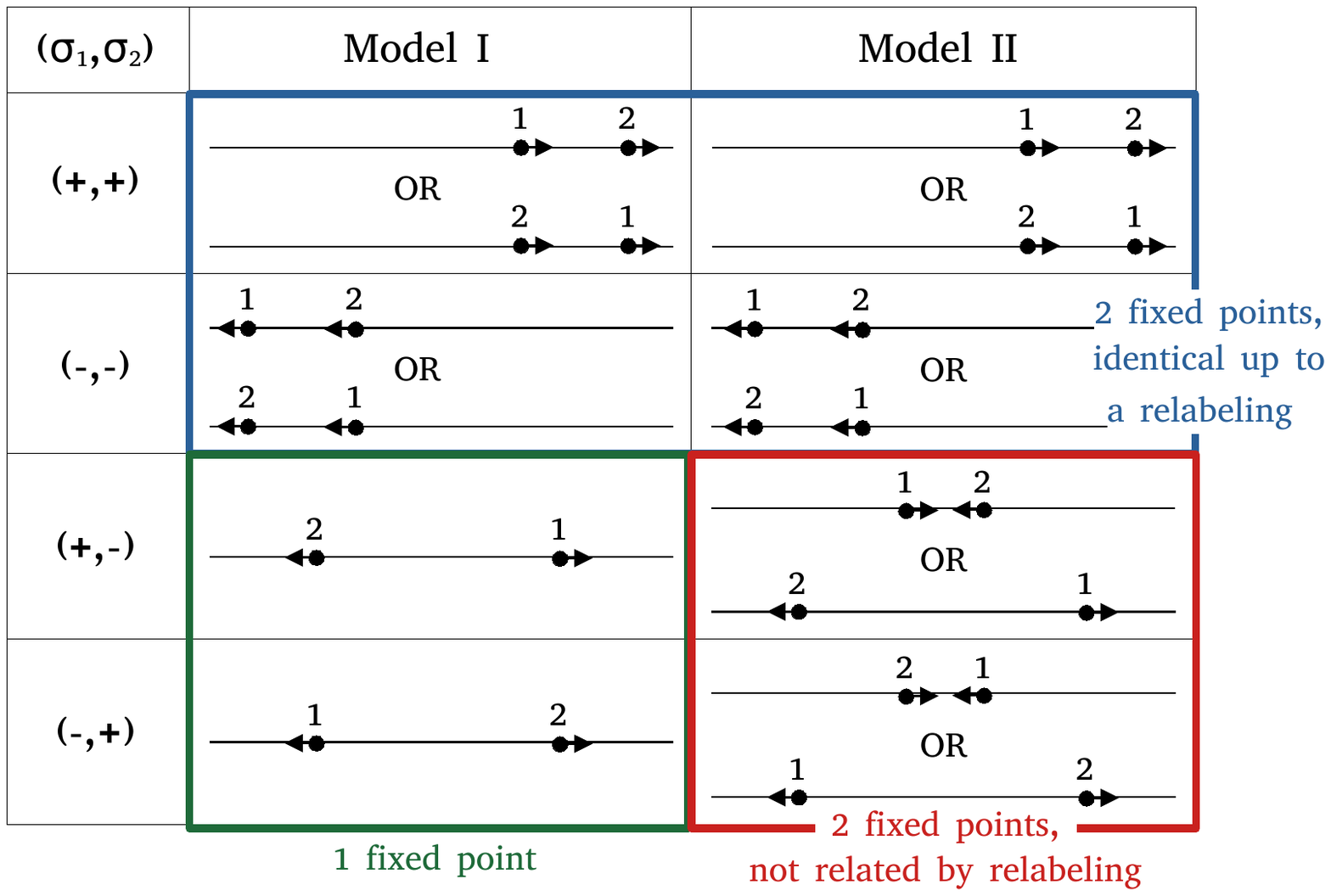}
    \caption{Schematic description of the fixed points for $N=2$ in model I and II together with the corresponding values
    of $(\sigma_1,\sigma_2)$ (first column). There is a total of 6 fixed points in model I and 8 in model II, which reduces to $N+1=3$ (for model I) and 
    $2^N=4$ (for model II) up to a permutation of the labels. The notation "OR" means that the fixed point reached under the $\gamma=0$ dynamics 
    depends on the initial condition. Indeed, the ordering of the particles is conserved in model II, and in model I for particles in the same
    internal state.}
    \label{TableN2}
\end{figure}

\subsection{Fixed point for $N$ particles in the same state and support of the density}
\label{AppSupport}

As explained in the main text, and anticipating a bit on the following sections, 
the fixed points which correspond to all $N$ particles being in the same state (all velocities being $+v_0$ or all
being $-v_0$) allow to determine the edges of the support of the density for $\gamma>0$. The upper edge $x_{+,N}$ for finite $N$ is given by $x_{+,N} = \max_{1\leq i\leq N} x^{eq}_i$ where the $x^{eq}_i$ are the solutions of the following set of equations valid for both models (setting all $\sigma_i=+1$ in \eqref{eqFixedSigma}) 
\beq
v_0 - x_i + \frac{2g}{N}\sum_{j\neq i}\frac{1}{x_i-x_j}=0 \ \ \ {\rm for} \ i=1, \ ... \ , N
\eeq
Performing the change of variable $x_i=v_0 + \sqrt{\frac{2g}{N}}y_i$ we obtain :
\beq
-y_i + \sum_{j\neq i}\frac{1}{y_i-y_j}=0 \ \ \ {\rm for} \ i=1, \ ... \ , N
\label{eq_support}
\eeq
The solution of this set of equations is well known to be given by the zeros of the Hermite polynomials, $H_N(y)=0$ \cite{Hermite1,Hermite2}. We briefly recall the proof here for the sake of completeness.

We write equation (\ref{eq_support}) under the form :
\beq
-y_i\prod_{k \neq i}(y_i-y_k) + \sum_{j\neq i}\prod_{k \neq i, j}(y_i-y_k)=0 \ \ \ {\rm for} \ i=1, \ ... \ , N
\eeq
Introducing the polynomial $P(y) = \prod_{l=1}^N (y-y_l)$, this can be written :
\beq
-y_i P'(y_i) + \frac{1}{2} P''(y_i)=0 \ \ \ {\rm for} \ i=1, \ ... \ , N
\eeq
The polynomial $P''(y)-2yP'(y)$ is of degree $N$ and has $N$ common roots with $P$, therefore it is proportional to $P$. Looking at the coefficient of $y^N$ to find the proportionality factor, we obtain
\beq
P''(y)-2yP'(y)+2NP(y)=0
\eeq
for which the solutions are of the form $C_1 H_N(y)$ (where $C_1$ is a constant) plus a non-polynomial term, which has to be zero in this case.

Similarly the lower edge is obtained from the fixed point with all $\sigma_i=-1$, which is simply obtained from
the above by changing $v_0 \to - v_0$. Let us denote $y=\zeta_k^N$ the $k^{th}$ largest zero of the Hermite polynomial $H_N(y)$. We thus 
obtain that the two edges of the support of the density 
are 
\be 
x_{\pm,N} = \pm \left( v_0 + \sqrt{\frac{2 g}{N}} \zeta_1^N \right) 
\label{Hermite_support}
\ee 
a result valid for both models I and II.
\\

{\bf Large $N$ asymptotics of the support}. In the limit $N\rightarrow\infty$ with fixed $k=O(1)$ the asymptotics of the $k^{th}$ largest zero of 
$H_N(y)$ is known to be given by \cite{Hermite_asymptotics} 
\begin{eqnarray}
\zeta_k^N = \sqrt{2N+1} + 2^{-1/3} (2N+1)^{-1/6} a_k + O(N^{-5/6})
\label{asympt_zeta}
\end{eqnarray}
where $a_k$ is the $k^{th}$ zero of the Airy function, which for large $k$ is given by $a_k = - (\frac{3 \pi}{8} (4k-1))^{2/3} + O(k^{-4/3})$. Taking $k=1$ and using \eqref{Hermite_support} we get
\beq
x_{\pm,N} = \pm \left[v_0 + 2\sqrt{g} \left( 1 +\frac{a_1}{2} N^{-{2/3}} + \frac{1}{4}N^{-1} + O(N^{-4/3}) \right)\right]
\label{asymptot_support}
\eeq


More generally, at finite $N$, in the case of model II where particles cannot cross, this computation also gives us the support of the stationary measure for each particle individually. Indeed, consider the particle located at the $k^{th}$ position starting from the right and denote $x_k$ its position. One expects that $x_k$ cannot be larger (resp. smaller) than its equilibrium value corresponding to the state where all the particles have $\sigma=+1$ (resp. $-1$). According to the computation above, this means that $x_k$ is always included in the interval $[-v_0+\sqrt{\frac{2g}{N}}\zeta_k^N, v_0+\sqrt{\frac{2g}{N}}\zeta_k^N]$. This is at variance with model I where every particle has the same support $[x_{-,N},x_{+,N}]$.

\subsection{Stability of the fixed points and their determination in the general case}
Let us now look at the stability of the fixed points. For now, let us consider $\vec \sigma$ to be given and fixed. 
For both models, the Hessian takes a simple form for any configuration $\vec{x}$:
\begin{eqnarray}
&&\mathcal{H}_{\vec \sigma}(\vec{x}) = \left(\frac{\partial^2 V_{\vec \sigma}(\vec{x})}{\partial x_i \partial x_j} \right)_{1 \leq i,j \leq N} = I + A \\
&&A_{ij} = A_{ji} = -\frac{2}{N} \frac{g_{\sigma_i,\sigma_j}}{(x_i-x_j)^2} \leq 0 \quad {\rm for} \quad i \neq j \label{off_diag} \\
&&A_{ii} = -\sum_{j \neq i} A_{ij}   \label{def_diag}
\end{eqnarray}
{ (note that $\mathcal{H}_{\vec \sigma}(\vec{x})$ does not depend on $\vec{\sigma}$ in model II).} From the relations (\ref{off_diag}) and (\ref{def_diag}) we see that the matrix $A$ is a symmetric diagonally dominant real matrix (i.e. $|A_{ii}| \geq \sum_{j\neq i} |A_{ij}|$) with non-negative diagonal entries (see e.g. \cite{wolfram}). For such matrices, a classical result of linear algebra states that $A$ is positive semi-definite. Therefore all the eigenvalues of $\mathcal{H}_{\vec{\sigma}}(\vec x)$ are larger or equal to 1 (and 1 is an eigenvalue associated to the eigenvector $(1, \ ... \ ,1)${, which describes the relaxation of the center of mass}). Actually it turns out that the matrix $\mathcal{H}_{\vec \sigma}(\vec{x})$ can be diagonalised exactly (see Ref. \cite{Hermite1}) and its eigenvalues are simply the integers from 1 to $N$.
%
%
Thus the eigenvalues of ${\cal H}_{\vec{\sigma}}(\vec x)$ are always strictly positive for any configuration $\vec x$.
This implies that all the fixed points are attractive for the dynamics. In addition, 
the potential $V_{\vec \sigma}(\vec{x})$ is strictly convex on any subspace such that the interaction term does not diverge.
For model II each such subspace corresponds to a given ordering of the particles. For model I 
each such subspace corresponds to a given ordering of the particles with $\sigma_i=+1$ and a given ordering of the particles with $\sigma_i=-1$.
Hence for each of these subspaces, in both models, from convexity there is a unique fixed point, which is attractive. 
In other words the basin of attraction of that fixed point is the subspace in question. For each $\vec \sigma$
the number of subspaces is $N!$ for model II and $N_+! N_-!$ for model I, where $N_\pm$ is the number of $\sigma_i=\pm$. For $N=2$ this
is illustrated in Fig.~\ref{TableN2}. 

One can now ask what is the total number of fixed points as $\vec \sigma$ is varied. 
In model I, we see that all vectors $\vec \sigma$ with the same $N_+$ will give the same fixed points up to a global permutation $\tau$ of the particles. Therefore if we are interested in the number of distinct fixed points up to a permutation of the particles when varying $\vec \sigma$, there will be only $N+1$ of them in model I. In contrast, in model II each value of $\vec \sigma$ will lead to a different fixed point a priori, so that there are generically $2^N$ fixed points in this case.


Although for model II it is difficult to compute all the fixed points for arbitrary $N$, it is possible to do so in model~I, which we now focus on. Indeed in this case there is no coupling between $+$ and $-$ particles and they can be treated as two independent sets of particles with positions $\vec{x}^\pm$ subject to a potential
\begin{equation} \label{potential2} 
    V^\pm(\vec{x}^\pm) = \frac{1}{2} \sum_{i} (x_i^\pm)^2 -\frac{2g}{N}\sum_{i<j} \log |x_i^\pm-x_j^\pm| \mp v_0 \sum_{i} x_i^\pm
\end{equation}
This potential has a form similar to the one studied in \ref{AppSupport} (since it describes particles which all have the same sign) and therefore the results from this section can be applied here independently to both sub-systems with $N_{\pm}$ particles. 
Note that in both sub-systems, the interaction strength remains $2g/N$ [see Eq. (\ref{potential2})] and not $2g/N_{\pm}$. 
Hence the equilibrium positions can be written as $x_i^\pm=\pm v_0 + \sqrt{\frac{2g}{N}}y_i^\pm$ where the $y_i^\pm$'s are the zeroes of the Hermite polynomials $H_{N_+}$ and $H_{N_-}$. 
For finite $N$, all $+$ particles are therefore included in the interval$[-v_0-2\sqrt{\frac{2 g}{N}} \zeta_1^{N_+},-v_0+2\sqrt{\frac{2 g}{N}} \zeta_1^{N_+}]$ and all $-$ particles in the interval $[v_0-2\sqrt{\frac{2 g}{N}} \zeta_1^{N_-},v_0+2\sqrt{\frac{2 g}{N}} \zeta_1^{N_-}]$. Therefore, 
in the $N \to +\infty$ limit, using the asymptotic behavior of $\zeta_1^{N_{\pm}}$ from Eq. (\ref{asympt_zeta}), one finds that 
the density will thus be the sum of two semi-circles of support $[v_0-2\sqrt{gp_+},v_0+2\sqrt{gp_+}]$ and $[-v_0-2\sqrt{gp_-},-v_0+2\sqrt{gp_-}]$ where $p_\sigma = N_\sigma/N$ is the fraction of particles with sign $\sigma$.

\subsection{Effect of $\gamma>0$}

\subsubsection{1. Simple argument in the generic case} 
\label{subsec:simple} 

\begin{figure}[h!]
    \centering
    \includegraphics[width=0.45\linewidth]{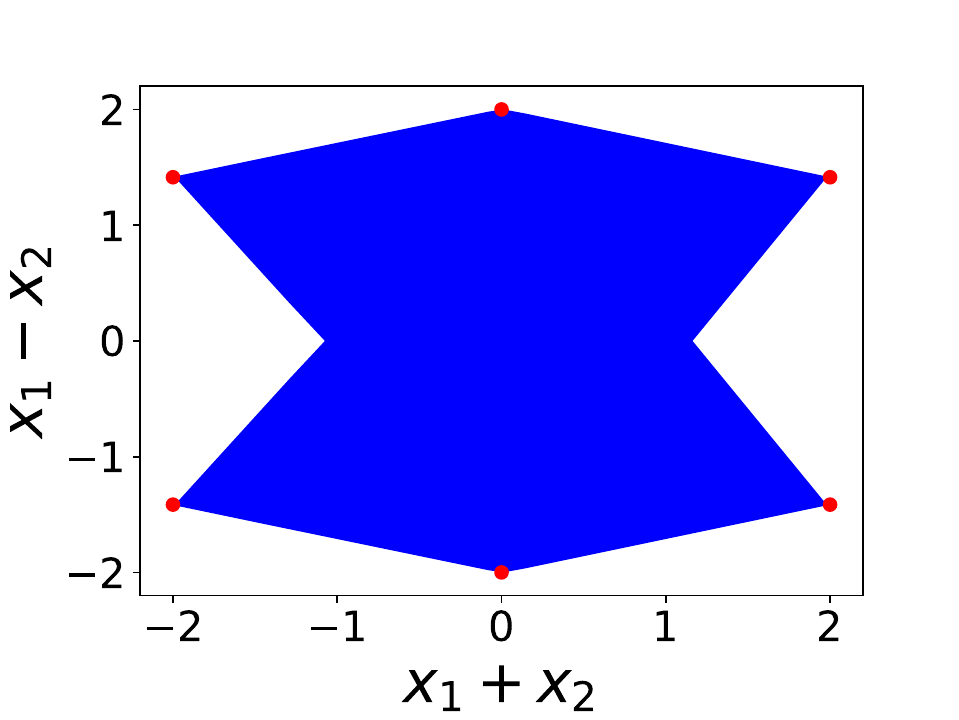}
    \includegraphics[width=0.45\linewidth]{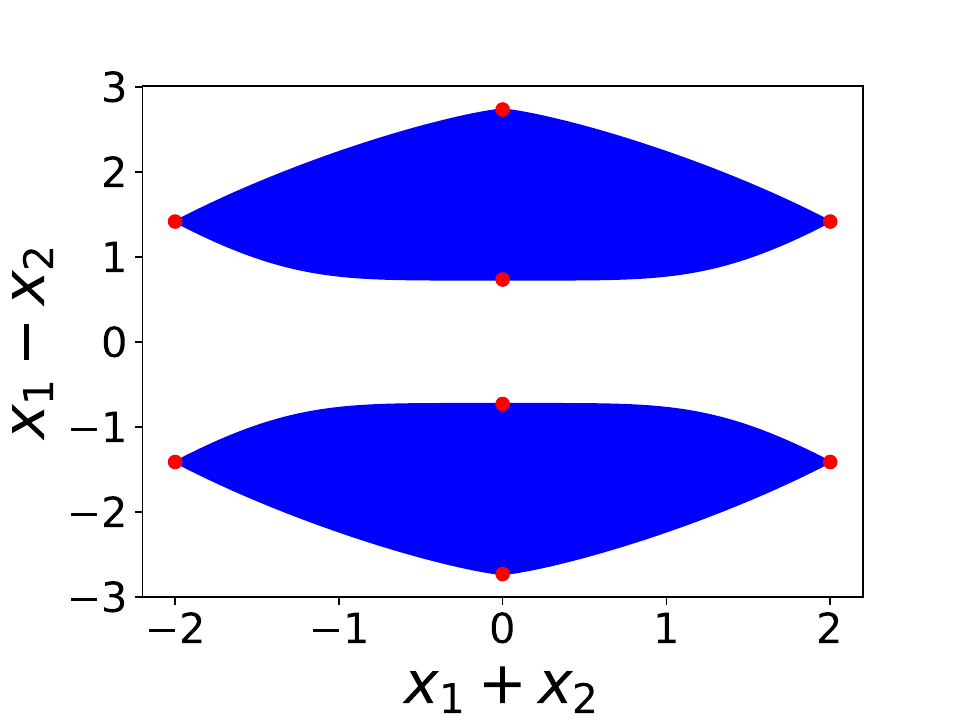}
    \caption{Support of the stationary joint distribution $P_{\rm stat}(x_1,x_2)$ for $\gamma>0$ in the $(x_1+x_2,x_1-x_2)$ plane for $N=2$,
    for model I (left) and model II (right). The figures were obtained by simulating the dynamics and saving the successive positions of both particles. Since on each half plane $x_1-x_2 \gtreqless 0$ the set of points obtained is convex, it is enough to plot the convex hull of this set of points on each half plane. 
    The red dots are the fixed points which are listed in Fig \eqref{TableN2}
    and whose coordinates are given in the text in Section \eqref{N2section}.
    Since the support is independent of $\gamma$ we used $\gamma=0.01$ (small values of $\gamma$ allow particles to spend more time close to the fixed points so that shorter simulations are required). All other parameters have been set to 1. In model I the two particles can cross hence there is a unique ergodic component, while there are two components in model II.}
    \label{x1vsx2}
\end{figure}

The next question is how do these fixed points manifest in the stationary joint distribution $P_{\rm stat}(\vec x)$, and in the stationary particle density
$\rho_s(x)$ for $\gamma>0$ ? For $N=1$ these two quantities are identical and given in \eqref{rho_inde_app}. For any $N$, 
each time one particle switches sign, the system flows towards the corresponding fixed point until the next change of sign, at which point the potential will suddenly change leading to a different fixed point. The smaller the value of $\gamma$, the more time the particles will spend near the fixed point and the more we expect them to be visible as singular points in the joint distribution and in the density. It is important to note that the support of both the joint distribution
and the density is independent of $\gamma$. For $N=2$ this support is plotted in the space $(x_1,x_2)$ in Fig. \ref{x1vsx2} for both models I and II. The positions of the fixed points are visible in the figure as corners. The density itself $\rho_s(x)$ exhibits non-analyticities which we now analyze.

Let us denote by $\vec x^*= (x_1^*,\dots,x_N^*)$ any one of those fixed points (we consider here both model I and II). 
For $N=1$, we know that there are only two fixed points at $x^*=x_1^*=\pm v_0$ which correspond to the edges of the support,
see \eqref{rho_inde_app} setting $\lambda=1$. From the exact formula, we see that the density near $x^*$ behaves as $C_\pm |x-x^*|^{\gamma-1}$,
where $C_\pm$ refers to the amplitude of the singularity on either side of $x^*$. Since in this case $x^*$ is also
an edge, one of the two amplitudes vanishes. This behavior can actually be recovered from a simple heuristic argument, which we will generalize below to $N>1$. 
Let us consider for instance the upper edge $x^*=x_+=v_0$. When $\sigma$ is fixed to $1$ the position of the particle satisfies the equation $\dot{x} = -x + v_0$, or after a change of variable $y=x-x_+$:
\begin{equation}
    \dot{y} = -y \quad \Rightarrow \quad y(t) = y_0 e^{-t}
\end{equation}
The probability for a particle to still have $\sigma=+1$ after a time $t$ is proportional to $e^{-\gamma t}$. Combining those two information, we can write 
the stationary probability density $p(y)$ of $y$ as
\begin{equation} 
    p(y) = \frac{p(t)}{|\frac{dy}{dt}|} \propto \frac{e^{-\gamma t}}{|y|} \propto |y|^{\gamma-1}
\end{equation}
which is indeed the correct result for the exponent of the singularity. 

This argument can be generalised to arbitrary $N$ (and $g>0$). Let us assume that all the $\sigma_i$ remain fixed for a time $t \gg 1/\lambda =1$. 
Let us consider the vicinity of a fixed point and $\vec{y} = \vec{x} - \vec{x}^*$, where $\vec{x} \in \mathbb{R}^N$ is the vector of all particle positions. 
In the limit of large time the convergence to the fixed point is dominated by the smallest
eigenvalue of the Hessian $H_{\sigma}(\{x_i\})$, which is equal to unity, and corresponds to an eigenvector 
$(1,\dots,1)$ [see the discussion below Eq. (\ref{def_diag})]. Hence one has $|\vec y| \propto e^{-t}$
for $t$ large enough. Since the probability that all the $N$ particles keep the same $\sigma$ for a time $t$ decays as $e^{-N \gamma t}$, the stationary joint probability density $p(\vec y)$ of $\vec y$ behaves as 
$p(\vec y) \propto |\vec y|^{N\gamma-1}$. Since the dominant eigenmode corresponds to a global translation of all the particles,
it is clear that the total density will inherit the same singularity near any point $x^*=x_i^*$ 
with $i=1,\dots,N$ (i.e. which corresponds to any coordinate of the fixed point). 
Note that $x^*$ can be either one of the two edges of the density, $x^*=x_{\pm,N}$, or a point inside the support. There are thus two main cases:\\

(i) In the case $\gamma<1/N$ the density
diverges at $x^*$ as $\rho_s(x) \sim C_\pm |x-x^*|^{N\gamma-1}$. This holds both for edge points or for internal points.
Note that for general $\lambda$ the exponent is $N\gamma/\lambda-1$. \\

(ii) In the case $\gamma>1/N$ the density has a smooth part
and a singularity of the form $\rho_s(x) - \rho_s(x^*) \sim C_\pm |x-x^*|^{N\gamma-1}$. 
Note that if $x^*$ is at the edge of the support of the density, $x^*=x_\pm$, then one can show that $\rho_s(x^*)=0$, i.e.
the density vanishes at the edge. \\

In both cases, for edge points and the internal points which correspond to the same fixed point in phase space as the edge point,
one of the two amplitudes vanishes. For the other internal points both can be non zero (they can be different). 

In the limit $N \to +\infty$ the singularity in the density becomes weaker and weaker. 
Hence the density becomes vanishingly small in the vicinity of the edge. 
This explains the fact that the support of the density for $N \to +\infty$ (which is studied in the text and below) is strictly smaller than the limit for
$N \to +\infty$ of the support for finite $N$ given in \eqref{asymptot_support}. 

Finally note that we have studied here the singularities of $\rho_s(x)$. One can also study the singularities
of $\rho_\pm(x)$ for finite $N$. In particular, concerning the edges $x_{\pm,N}$, we 
have observed that the difference of $1$ in the exponents discussed for the non interacting case in \eqref{ratio},
seems to hold also in presence of interactions. This is shown in Fig. \ref{singularity_finiteN_rhopm}
for model II and $N=2$, but we expect it to be more general.

\begin{figure}[h!]
    \centering
    \includegraphics[width=0.32\linewidth]{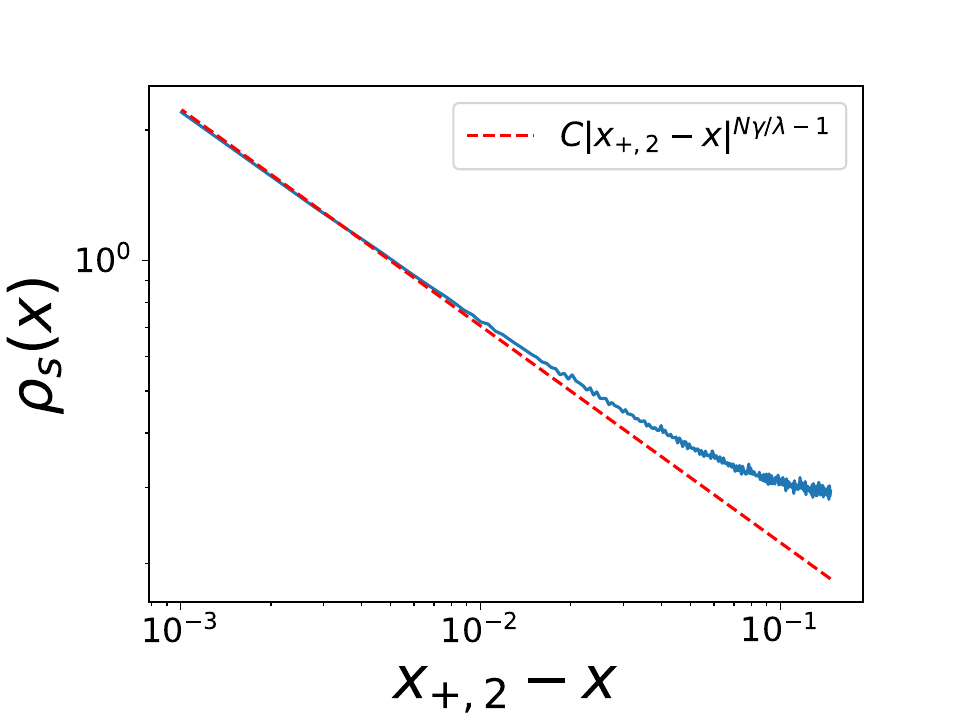}
    \includegraphics[width=0.32\linewidth]{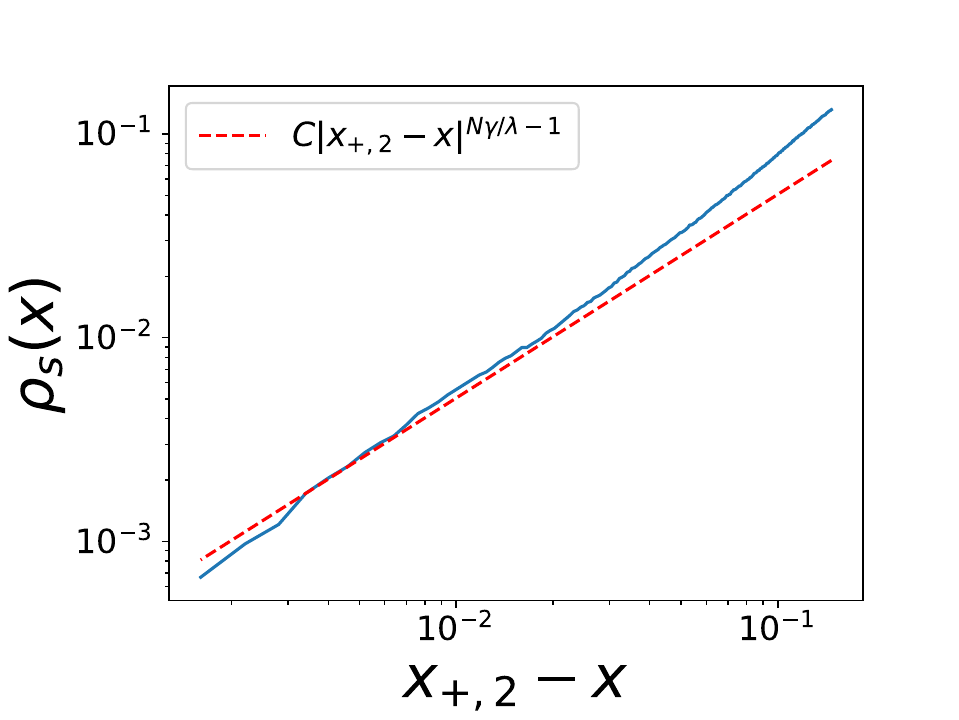}
    \includegraphics[width=0.32\linewidth]{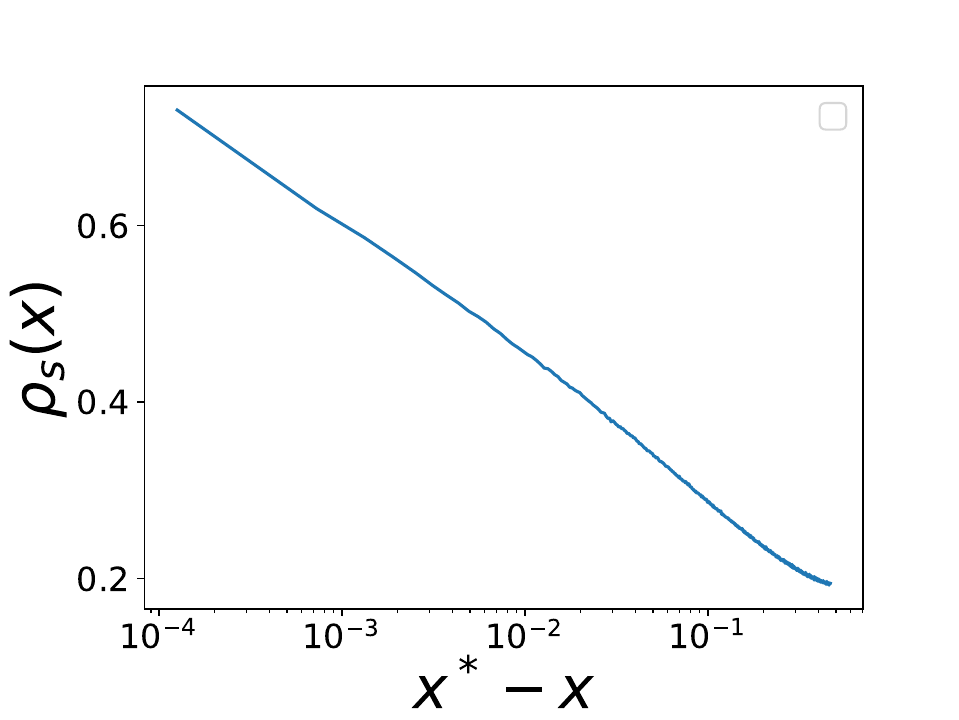}
    \caption{Comparison of simulations with the predictions of the singularity exponent $N \gamma/\lambda-1$ for the density for two particles $N=2$ for model II with $v_0=1$, $g=1$, $\lambda=1$, in log-log scale (we get similar results for model I). Right: Density of particles near the right edge of the support for $\gamma=0.25$ ($N \gamma/\lambda-1=-0.5$). Center: Same plot for $\gamma=1$ ($N \gamma/\lambda-1=1$). Right: Density of particles near the second largest fixed point (at the left of the singularity) for $\gamma=0.5$ ($N \gamma/\lambda-1=0$), in log-linear scale. We can see the log divergence in this case.}
    \label{N2_vary_gamma_logscale}
\end{figure}

\begin{figure}
    \centering
    \includegraphics[width=0.32\linewidth]{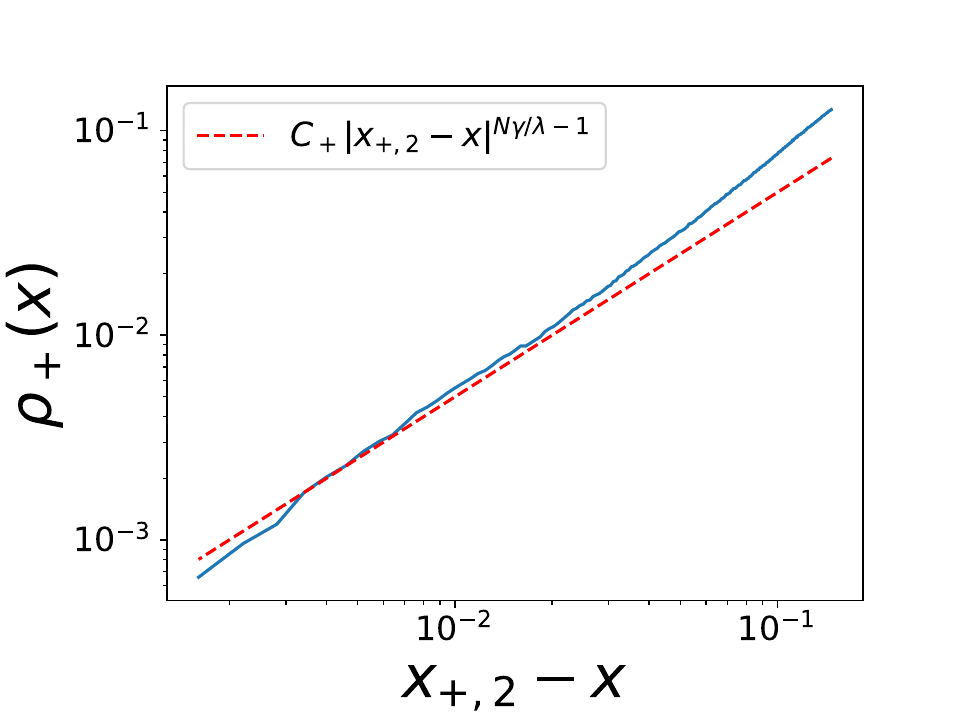}
    \includegraphics[width=0.32\linewidth]{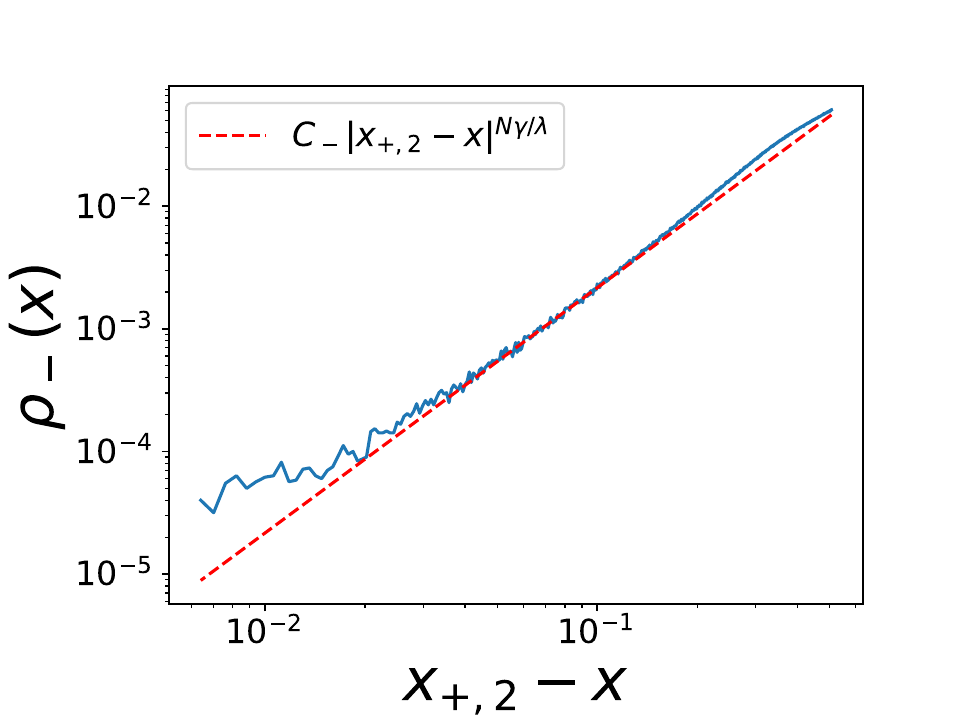}
    \caption{Density of particles $\rho_+(x)$ (left panel) and $\rho_-(x)$ (right panel) near the right edge $x_{+,2}$ of the support for model II with $N=2$ particles with $\gamma=1$, $v_0=1$, $g=1$ and $\lambda=1$, in log-log scale. We observe a difference of 1 between the exponents.}
    \label{singularity_finiteN_rhopm}
\end{figure}

\subsubsection{2. More refined argument, and the marginal case $N \gamma=1$}

The argument above is valid for $x^*$ at the edge of the support as well as for internal points $x^*$ corresponding to the same fixed point in phase space as one of the edge points (in general there are $2N$ such points in both model I and II). Those points can only be reached by the corresponding particle ($x^*=x_i$) if $\vec{\sigma}(t)$ remains constant for an infinite time, which is an underlying hypothesis of the reasoning above. Numerically we observe that for those singularities the argument above predicts the correct behavior for any $\gamma$ (see Fig. \ref{N2_vary_gamma_logscale}). However, there are other cases.
Consider now a singular point $x^*$ in the bulk together with the corresponding fixed point $\vec x^*$ which has a generic $\vec \sigma^*$ (not 
all $\sigma^*_i$'s being equal). During the evolution, the system $(\vec x(t),\vec \sigma(t))$ can "by chance" be in the vicinity
$\vec x(t) \approx \vec x^*$ while $\vec \sigma(t) \neq \vec \sigma^*$. In this case, it turns out that the result above still applies, except in the case $N\gamma=1$ where we observe a logarithmic divergence (see Fig.\ref{N2_vary_gamma_logscale} right panel). This phenomenology is similar to what has been observed in the case of a single particle with 3 states $\sigma=-1$, $0$ or $1$ \cite{3statesBasu}. Explaining this log divergence requires a slightly refined argument which we now describe.


To understand better the behavior of the density near fixed points, we consider a simplified model of model I and II near a fixed point
with a fixed set $\vec \sigma$. 
Consider dynamics of the center of mass
$\bar x(t)= \frac{1}{N} \sum_i x_i(t)$. Let us denote $x(t)$ its deviation from its value at the fixed point,
i.e. $x(t)= \bar x(t) -  \frac{v_0}{N} \sum_i \sigma_i$. It satisfies the equation of motion (\ref{c_of_m})
\beq
\frac{dx}{dt} = - x \;.
\eeq
This can be considered as an effective single particle model by adding some injection and absorption processes, which
allow to study the density of particles on a finite interval $x \in [0,x_0]$. The effective Fokker-Planck equation
together with its stationarity condition for the PDF $P_{\vec \sigma}(x)$ reads
\beq
\frac{\partial P_{\vec \sigma^*}(x)}{\partial t} = \frac{\partial}{\partial x}(xP_{\vec \sigma^*}(x)) - N\gamma P_{\vec \sigma^*}(x) + r = 0 \;.
\label{eq_source}
\eeq
In the right hand side of Eq. (\ref{eq_source}), 
the term proportional to $- N \gamma$ takes into account the flips from $\vec \sigma^*$ to any another configuration $\vec \sigma$.
The constant term $r$  takes into account the fact that particles with $\vec \sigma \neq \vec \sigma^*$ which are in the interval $[0,x_0]$ can switch to $\vec \sigma=\vec\sigma^*$.
Since we consider a small interval and we do not expect $P_{\vec\sigma \neq \vec \sigma^*}(x)$ to vary strongly in this interval (contrary to $P_{\vec\sigma^*}(x)$), we will assume $P_{\vec\sigma \neq \vec\sigma^*}(x)=cst \equiv p_0$ on the interval. Thus we just need to add to the Fokker-Planck equation a constant source term $r=N \gamma p_0$: this explains the third term in the rhs of (\ref{eq_source}). Note that in the case described in Section \ref{subsec:simple} {\red 1} where all the $\sigma_i^*$'s are equal, e.g. to describe the edges of the support, one has $r=0$. 
In addition, particles which are already in the state $\vec \sigma^*$ can enter the interval $[0,x_0]$ at $x_0$. This fixes the density at $x_0$ to a certain value $q_0$.
Thus the solution of Eq. \eqref{eq_source} 
reads
\bea
P_{\vec \sigma^*}(x)&&=\frac{r}{N\gamma-1} + \left(q_0-\frac{r}{N\gamma-1}\right) \left(\frac{x}{x_0}\right)^{N\gamma-1} \quad, \quad \begin{rm}for\end{rm} \ N\gamma \neq 1 \\
&&= q_0-r\ln\frac{x}{x_0} \quad , \quad \begin{rm}for\end{rm} \ N\gamma = 1
\eea
Using the fact that $p_0<\frac{1}{x_0}$ it is easy to check that both expressions are positive for any values of $\gamma$ and for any $x$ in $[0,x_0]$.
For $r=0$ this refined argument recovers the exponent $x^{N\gamma-1}$ for the singularity of the density $\rho_s(x)$
given in Section \ref{subsec:simple} {\red 1}.
In addition, this model predicts a logarithmic divergence of the density $\rho_s(x)$ for $N\gamma=1$ when $r\neq 0$ (i.e. at the internal points).
Such a logarithmic divergence in the bulk was also found from an exact solution in a 3-state model.
In the case of the 3-states model, $x$ represents the position of a particle near $x^*=0$. 
For $1<\gamma<3$ we get a maximum with a cusp if $r$ is sufficiently large (i.e. if the density of particles with $\sigma \neq \sigma_0$ is large enough), as in \cite{3statesBasu} (see Fig. 2 there). However, this simplified model does not reproduce the quadratic behavior found in \cite{3statesBasu} for $\gamma>3$.

\section{Dean-Kawasaki versus Fokker-Planck equation}

\subsection{Dean-Kawasaki equation}
Here we show how to extend the Dean-Kawasaki approach \cite {Dean,Kawa} to derive the evolution equations for the densities $\rho_\pm(x, t)$ defined in \eqref{def_rho} 
and in \eqref{defrho2} in presence of the active noise. We start from the general equation :
\beq
\dot x_i(t) = - W'(x_i(t)) -  \sum_{j\neq i} V_{\sigma_i,\sigma_j}'(x_i(t)-x_j(t))  +v_0\sigma_i(t) + \sqrt{\frac{2 T}{N}} \xi_i(t)
\label{dx}
\eeq
where the $\sigma_i(t)$ are independent telegraphic noises and the $\xi_i(t)$ are independent Gaussian white noises. 
For the model of interest here \eqref{modelsupmat} one has 
\be
W(x) = \frac{\lambda x^2}{2} \quad , \quad V_{\sigma, \sigma'} (x) = - \frac{2 g_{\sigma,\sigma'}}{N} \log|x| \;,
\ee 
where, as in the text, we will set $\lambda=1$. 

Consider an arbitrary function $f$. We first introduce :
\bea \label{defrho2} 
\rho_\sigma(x,t) = \frac{1}{N} \sum_i \delta(x_i(t)-x) \delta_{\sigma_i(t),\sigma} \quad , \quad F_\sigma(\vec x(t)) = \frac{1}{N} \sum_i f(x_i(t))\delta_{\sigma_i(t),\sigma} = \int dx f(x) \rho_\sigma(x,t)
\eea

Then :
\beq
\frac{d F_\sigma(\vec x(t))}{dt} = \frac{1}{N} \sum_i \delta_{\sigma_i(t),\sigma} f'(x_i(t)) \dot x_i(t) + \frac{T}{N^2} \sum_i \delta_{\sigma_i(t),\sigma} f''(x_i(t)) +  \frac{1}{N} \sum_i f(x_i(t)) \frac{d\delta_{\sigma_i(t),\sigma}}{dt}
\eeq

We can write $\delta_{\sigma_i(t),\sigma}=\frac{\sigma\sigma_i(t)+1}{2}$ so that $\frac{d\delta_{\sigma_i(t),\sigma}}{dt}=\frac{\sigma}{2} \frac{d\sigma_i(t)}{dt}$. Thus we get using eq. (\ref{dx}) :
\beq
\begin{aligned}\frac{d F_\sigma(\vec x(t))}{dt} = && \frac{1}{N}  \sum_i \delta_{\sigma_i(t),\sigma} f'(x_i(t)) [- W'(x_i(t)) - \sum_{j \neq i} V_{\sigma_i,\sigma_j}'(x_i(t)-x_j(t)) + v_0\sigma + \sqrt{\frac{2 T}{N}}  \xi_i(t)] \\ && + \frac{T}{N^2} \sum_i \delta_{\sigma_i(t) ,\sigma} f''(x_i(t)) + \frac{1}{N} \frac{\sigma}{2}\sum_i f(x_i(t)) \frac{d\sigma_i(t)}{dt}  \end{aligned}
\label{Dean_discret}
\eeq
We now need to distinguish model I and model II. In model I the interactions are limited to particles with the same $\sigma$, so that we write $V_{\sigma_i,\sigma_j}(x_i-x_j)=\tilde V(x_i-x_j) \delta_{\sigma_i,\sigma_j}$, where here we will consider later $\tilde V (x) = - \frac{2 g}{N} \log|x|$.
In model II all particles interact together and we simply have $V_{\sigma_i,\sigma_j}(x_i-x_j)=\tilde V(x_i-x_j)$. Let us first consider for simplicity,
as in \cite{Dean}, the case where $\tilde V'(0)=0$. In that case we obtain
\beq
\begin{aligned} \frac{d F_\sigma(\vec x(t))}{dt} =&& \int dx \rho_\sigma(x,t) [v_0\sigma f'(x) - f'(x) W'(x) - f'(x) \int dy N\tilde\rho(y,t;\sigma) \tilde V'(x-y) + \frac{T}{N} f''(x)] \\&&  + \frac{1}{N} \sum_i \delta_{\sigma_i(t),\sigma} f'(x_i(t)) \sqrt{\frac{2 T}{N}}  \xi_i(t) + \frac{1}{N} \frac{\sigma}{2}\sum_i f(x_i(t)) \frac{d\sigma_i(t)}{dt} \end{aligned}
\eeq
where $\tilde{\rho}(y,t;\sigma)=\rho_\sigma(y,t)$ in model I and $\tilde{\rho}(y,t;\sigma)=\rho_s(y,t)=\rho_+(y,t) + \rho_-(y,t)$ in model II. After integration by part we obtain
\bea
\int dx f(x) \partial_t  \rho_\sigma(x,t)
= \int dx f(x) \left( \partial_x  ( \rho_\sigma(x,t)  [-v_0\sigma + W'(x) + \int dy N \tilde\rho(y,t;\sigma) \tilde V'(x-y) ]) \right. \label{integralDean} \\
\left. + \frac{T}{N} \partial_x^2 \rho_\sigma(x,t) + \frac{1}{N} \hat\zeta_\sigma(x,t) - \frac{1}{N} \partial_x \Xi_\sigma(x,t) \right) \nonumber
\eea 
The last two terms correspond respectively to an active noise $\hat \zeta_\sigma$ (originating from the telegraphic noises)
and a passive noise $\Xi_\sigma$ (originating from the thermal white noises). Let us examine these two terms.

The passive noise term $\Xi_\sigma$ is Gaussian and reads
\be
\Xi_\sigma(x,t) = \sqrt{\frac{2 T}{N}}  \sum_i \delta_{\sigma_i(t),\sigma} \delta(x_i(t)-x) \xi_i(t)
\ee
Hence it is fully determined by its covariance which is (here we use $\overline{\dots}$ indifferently for averages over the thermal 
and telegraphic noise) 
\bea
\overline{ \Xi_\sigma(x,t) \Xi_{\sigma'}(x',t')} && = 
2 T \delta_{\sigma,\sigma'} \rho_\sigma(x,t) \delta(x-x') \delta(t-t')
\eea
since $\delta_{\sigma_i(t),\sigma} \delta_{\sigma_i(t),\sigma'} =\delta_{\sigma,\sigma'} \delta_{\sigma_i(t),\sigma}$. Hence we can write :
\beq
\Xi_\sigma(x,t) = \sqrt{2 T \rho_\sigma(x,t)} \  \eta_\sigma(x,t) \quad , \quad \overline{ \eta_\sigma(x,t) \eta_{\sigma'}(x',t')}= \delta_{\sigma,\sigma'} \delta(x-x')\delta(t-t')
\eeq
where $\eta_\pm$ are two independent unit Gaussian white noises. 

The active noise term reads
\beq
\hat\zeta_\sigma(x,t)=\frac{\sigma}{2}\sum_i \delta(x_i(t)-x) \frac{d\sigma_i(t)}{dt}
\eeq
To deal with the term $\frac{d\sigma_i(t)}{dt}$, we discretize time into small intervals $dt$. In the time interval $[t,t+dt]$, $\frac{d\sigma_i(t)}{dt}=-\frac{2\sigma_i(t)}{dt}$ with probability $\gamma dt$ and $0$ otherwise. Thus $\overline{\frac{d\sigma_i(t)}{dt}}=-2\gamma\sigma_i(t)$. Separating the mean from the fluctuations we get :
\bea
\hat\zeta_\sigma(x,t)&&=-\gamma\sum_i \sigma\sigma_i(t)\delta(x_i(t)-x) + \sigma \sqrt{N} \zeta(x,t) \\
&&=-\gamma\sum_i \delta_{\sigma_i(t),\sigma}\delta(x_i(t)-x) +\gamma\sum_i \delta_{\sigma_i(t),-\sigma}\delta(x_i(t)-x) + \sigma \sqrt{N} \zeta(x,t) \\
&&=-\gamma N\rho_\sigma(x,t) +\gamma N\rho_{-\sigma}(x,t)  + \sigma \sqrt{N} \zeta(x,t)
\eea
using that $\sigma \sigma_i=\delta_{\sigma_i,\sigma} - \delta_{\sigma_i , - \sigma}$ and where we have defined (the factor $\sqrt{N}$ is conveniently chosen so that the fluctuating part $\zeta$ is of order unity, see below) 
\beq
\zeta(x,t)=\frac{1}{2\sqrt{N}}\sum_i \delta(x_i(t)-x) r_i(t) \quad , \quad r_i(t)=\frac{d\sigma_i(t)}{dt} - \overline{\frac{d\sigma_i(t)}{dt}}
\label{active_noise}
\eeq
which we will study in detail below. In particular we will show that it becomes Gaussian at large $N$ with a covariance function
that we compute explicitly. 

Let us return to the equation (\ref{integralDean}). Using that it holds for any $f$ we obtain the stochastic evolution equation for
the densities
\bea \label{dean1} 
\partial_t \rho_\sigma(x,t) =&& \partial_x [ \rho_\sigma(x,t) (- v_0 \sigma + W'(x) +
N \int dy \tilde \rho(y,t;\sigma) \tilde V(x-y) )] - \gamma \rho_{\sigma}(x,t) + \gamma \rho_{-\sigma}(x,t)
\\
&& + \frac{T}{N} \partial^2_x \rho_\sigma(x,t)  
+ \frac{1}{N} \partial_x [ \sqrt{ 2 T \rho_\sigma(x,t) } \eta_\sigma(x,t) ] + \frac{\sigma}{\sqrt{N}}\zeta(x,t) \nonumber
\eea 
We now want to specialize this equation to model I and II, with $W'(x)= x$ and $\tilde V'(x_i-x_j) = -\frac{2g}{N} \frac{1}{x_i-x_j}$. The difficulty comes from the fact that in this case $\tilde V'(x)$ diverges at $x=0$. This leads to a self-interaction term that we should remove. It is possible to treat correctly this interaction term for model I. To this aim we
go back to the discrete sum in \eqref{Dean_discret} 

\bea \label{78} 
-\frac{1}{N}  \sum_i \delta_{\sigma_i(t),\sigma} f'(x_i(t)) \sum_{j \neq i} V_{\sigma_i,\sigma_j}'(x_i(t)-x_j(t)) &=& \frac{2g}{N^2}  \sum_i \delta_{\sigma_i(t),\sigma} f'(x_i(t)) \sum_{j \neq i} \frac{\delta_{\sigma_i(t),\sigma_j(t)}}{x_i(t)-x_j(t)} \\
&=& \frac{g}{N^2} \sum_i \sum_{j \neq i} 
\frac{f'(x_i(t)) - f'(x_j(t))}{x_i(t) - x_j(t)} \delta_{\sigma_i(t), \sigma} \delta_{\sigma_j(t), \sigma} \;. \label{79} 
\eea  
Following Rogers-Shi \cite{RogersShi}, we rewrite it as
\bea
&& \frac{g}{N^2} \sum_{i j} \frac{f'(x_i) - f'(x_j)}{x_i - x_j} \delta_{\sigma_i, \sigma} \delta_{\sigma_j, \sigma} - \frac{g}{N^2} \sum_i f''(x_i) \delta_{\sigma_i, \sigma}
= g \int dx dy \frac{f'(x) - f'(y)}{x - y}
\rho_\sigma(x,t)\rho_\sigma(y,t) - \frac{g}{N} \int dx f''(x) \rho_\sigma(x,t) \nonumber \\
&& = -2g \fint dx f(x) \partial_x [\rho_\sigma(x,t) \int \frac{dy}{x - y} \rho_\sigma(y,t)] - \frac{g}{N} \int dx f(x) \partial_x^2 \rho_\sigma(x,t) \;.
\eea

Introducing $\beta=\frac{2g}{T}$ this leads us to the Dean-Kawasaki equation for model I:
\bea \label{dean2}
\partial_t \rho_\sigma(x,t) =&& \partial_x [ \rho_\sigma(x,t) (- v_0 \sigma  + x  -  2 g
\fint dy \frac{1}{x-y} \rho_\sigma(y,t) )] - \gamma \rho_{\sigma}(x,t) + \gamma \rho_{-\sigma}(x,t) \label{Dean_model12}
\\
&& + \frac{T}{N} (1 -  \frac{\beta}{2})  \partial^2_x \rho_\sigma(x,t)  
+ \frac{1}{N} \partial_x [ \sqrt{ 2 T \rho_\sigma(x,t) } \eta_\sigma(x,t) ] + \frac{\sigma}{\sqrt{N}} \zeta(x,t) \;. \nonumber
\eea 

It is tempting to try to perform the same manipulations for model II. If we forget the subtlety about the self interaction term,
we arrive as in \eqref{dean1} to the same equation \eqref{dean2} replacing $\fint dy \frac{1}{x-y} \rho_\sigma(y,t)$
by $\fint dy \frac{1}{x-y} (\rho_+(y,t)+ \rho_-(y,t))$. However one cannot show this convincingly, since we find that the method
used above for model I fails. Indeed in \eqref{78} one must replace $\delta_{\sigma_i(t),\sigma_j(t)}$ by $1$. If we attempt to
symmetrize as in \eqref{79}, one finds the combination $\delta_{\sigma_i,\sigma} f'(x_i) - \delta_{\sigma_j,\sigma} f'(x_j)$
instead of $f'(x_i)-f'(x_j)$ in the numerator. As a result one fails to extract the self-interaction term. Although this
may seem only a technical difficulty, the study in the next section shows that Eq. \eqref{dean1} fails for model II 
in a more fundamental way. As we will again discuss below, note that if we sum the two equations for $\rho_+(x,t)$
and $\rho_-(x,t)$ the above symmetrization works, and leads to a correct equation. However this equation
is not closed. 
\\

{\bf More details on the active noise}. We now argue that the active noise term $\zeta(x,t)$ defined in \eqref{active_noise} is of order $O(1)$ at large $N$. 
Discretizing time as before :
\bea
r_i(t)=\frac{d\sigma_i(t)}{dt} - \overline{\frac{d\sigma_i(t)}{dt}} &=& -\frac{2\sigma_i(t)}{dt}+2\gamma\sigma_i(t) \quad \rm{with \ probability \ \gamma dt} \\
&=& 2\gamma \sigma_i(t) \hspace*{1.8cm} \quad \rm{with \ probability \ 1-\gamma dt}
\eea

Since $\zeta(x,t)$ has zero average, we need to compute its covariance to obtain some information on its order in $N$ 
\beq
\overline{\zeta(x,t)\zeta(x',t')}=\frac{1}{4N}\sum_{i,j}\delta(x_i(t)-x)\delta(x_j(t')-x')\overline{r_i(t)r_j(t')} \;.
\eeq
In the case where $i \neq j$ or $t \neq t'$, $r_i(t)$ and $r_j(t')$ are uncorrelated (and with zero average) given $\sigma_i(t)$ and $\sigma_j(t')$, so that we can write 
\bea
\overline{r_i(t)r_j(t')} &&
= 0 \;.
\eea
In the case $i=j$ and $t=t'$ we find 
\bea
\overline{r_i(t)^2} &&= \overline{(-\frac{2\sigma_i(t)}{dt}+2\gamma\sigma_i(t))^2} \gamma dt + 4\gamma^2 (1-\gamma dt)
=\frac{4\gamma}{dt}-4\gamma^2 +O(dt)
\eea
In the general case we can write 
\beq
\overline{r_i(t)r_j(t')} = (\frac{4\gamma}{dt}-4\gamma^2)\delta_{ij}\delta_{t,t'} +O(dt) \;.
\eeq
Taking the limit $dt\rightarrow 0$, we replace $\frac{\delta_{t,t'}}{dt}$ by $\delta(t-t')$ and we obtain (as in standard calculations
for the Brownian motion) 
\beq
\overline{r_i(t)r_j(t')} = 4\gamma\delta_{ij}\delta(t-t') \;.
\eeq
This heuristics derivation yields the covariance function for $\zeta(x,t)$ 
\bea
\overline{\zeta(x,t)\zeta(x',t')}&&=\frac{\gamma}{N}\sum_{i}\delta(x_i(t)-x)\delta(x-x')\delta(t-t') 
= \gamma\delta(x-x')\delta(t-t')(\rho_+(x,t)+\rho_-(x,t)) \;,
\eea
which is thus of order $O(1)$. We can also examine the higher cumulants of $\zeta(x,t)$. We find:
\bea
&& \overline{r_i(t_1)r_j(t_2)r_k(t_3)}^c = 0 \\
&& \begin{aligned}\overline{r_i(t_1)r_j(t_2)r_k(t_3)r_l(t_4)}^c =&& 16\gamma\delta_{ij}\delta_{ik}\delta_{il}\delta(t_1-t_2)\delta(t_1-t_3)\delta(t_1-t_4) 
\end{aligned}
\eea
which leads the fourth cumulant of $\zeta(x,t)$ as
\bea
&& \overline{\zeta(x_1,t_1)\zeta(x_2,t_2)\zeta(x_3,t_3)}^c = 0 \\
&& \overline{\zeta(x_1,t_1)\zeta(x_2,t_2)\zeta(x_3,t_3)\zeta(x_4,t_4)}^c =
\frac{\gamma}{N} \delta(x_1-x_2)\delta(x_1-x_3)\delta(x_1-x_4) \delta(t_1-t_2)\delta(t_1-t_3)\delta(t_1-t_4) \rho_s(x_1,t_1) 
\eea
where $\rho_s=\rho_++\rho_-$. This suggests that at large $N$ the active noise becomes Gaussian. 
\\

\subsection{Fokker-Planck equation, large $N$ limit and stationary state}
\label{FP}

\subsubsection{Derivation of the Fokker-Planck equation}

Another useful approach is to define a probability density in the 2$N$-dimensional phase space of the model $(\vec x,\vec \sigma)$, ${\cal P}_t(\vec x, \vec \sigma) = {\cal P}(x_1,...,x_N;\sigma_1,...,\sigma_N)$. In the absence of thermal noise it satisfies the Fokker-Planck equation (for both models)
\be 
\partial_t {\cal P}_t(\vec x, \vec \sigma) = \sum_k \partial_{x_k} [ ( - v_0 \sigma_k + x_k - \frac{2}{N} \sum_{l\neq k} \frac{g_{\sigma_k,\sigma_l}}{x_k-x_l}) {\cal P}_t ] - N \gamma {\cal P}_t + \gamma \sum_k \tau_k^1 {\cal P}_t
\label{FP2ND}
\ee 
where $\tau_k^1 {\cal P}_t(\vec x, \vec \sigma)= {\cal P}_t(\vec x, \sigma_1,\dots,-\sigma_k,\dots,\sigma_N)$. 
The initial condition which has been implicitly chosen in the previous section -- see Eq. (\ref{defrho2}) -- (and from which we derive the Dean-Kawasaki equation) reads
\be 
{\cal P}_{t=0}(\vec x, \vec \sigma)= \prod_i \delta(x_i-x_i(0)) \delta_{\sigma_i,\sigma_i(0)} 
\ee 
with a fixed set of $\vec x(0)$ and $\vec \sigma(0)$. However within the present method more general initial conditions
can be considered. Let us define
\beq 
p_\sigma(x,t) = \langle \rho_\sigma(x,t) \rangle_{{\cal P}_t } = \langle \frac{1}{N} \sum_i \delta_{\sigma,\sigma_i} \delta(x-x_i) \rangle_{{\cal P}_t }
= \sum_{\vec \sigma} \int d\vec x \frac{1}{N} \sum_i \delta_{\sigma,\sigma_i} \delta(x-x_i) 
{\cal P}_t(\vec x, \vec \sigma)
\eeq
Note that contrarily to the empirical density $\rho_\sigma(x,t)$ of the Dean-Kawasaki equation which for finite $N$ is a stochastic function, 
$p_\sigma(x,t)$ is a (deterministic) probability density for any value of $N$ (with $\sum_\sigma \int dx p_\sigma(x,t) = 1$). As for the empirical density we will denote $p_s=p_++p_-$ and $p_d=p_+-p_-$. We also need to introduce the two-point density function involving one particle of sign $\sigma$ and one particle of sign $\epsilon$, $p_{\sigma, \epsilon}^{(2)}(x,y,t)$, which is normalized such that $\sum_{\sigma,\epsilon} \int dx dy p_{\sigma,\epsilon}^{(2)} (x,y,t) = 1$, namely
\beq
p^{(2)}_{\sigma,\epsilon}(x,y,t) = \sum_{\vec \sigma} \int d\vec x \frac{1}{N(N-1)}  \sum_{k \neq i}  \delta_{\sigma,\sigma_i} \delta_{\epsilon,\sigma_j} \delta(x-x_i) \delta(y-x_k) {\cal P}_t(\vec x, \vec \sigma) \;,
\eeq
as well as $p^{(2)}_{\sigma,s}(x,y,t) = p^{(2)}_{\sigma,+}(x,y,t) + p^{(2)}_{\sigma,-}(x,y,t)$. Multiplying \eqref{FP2ND} by $\frac{1}{N} \delta_{\sigma,\sigma_i} \delta(x-x_i)$, summing over all particles $i$ as well as over all configurations $\vec{\sigma}$, and integrating over all components of $\vec{x}$ we can obtain an equation for $p_\sigma(x,t)$. The first two terms on the left-hand side become
\be 
\sum_{\vec \sigma} \int d\vec x \frac{1}{N} \sum_i \delta_{\sigma,\sigma_i} \delta(x-x_i) \sum_k \partial_{x_k} [ ( - v_0 \sigma_k + x_k ) {\cal P}(\vec x, \vec \sigma) ] = \partial_x [(- v_0 \sigma + x) p_\sigma(x,t)] \;,
\ee
which is obtained after integrating by part and using that $\partial_{x_k} \delta(x-x_i) = - \delta_{ik} \partial_x \delta(x-x_i)$. The last term in (\ref{FP2ND}) can be rewritten
\bea 
\gamma \sum_{\vec \sigma} \int d\vec x \frac{1}{N} \sum_i \delta_{\sigma,\sigma_i} \delta(x-x_i) \sum_k  {\cal P}_t(\vec x, \sigma_1,\dots,-\sigma_k,\dots,\sigma_N) =&& \ \gamma \sum_{\vec \sigma} \int d\vec x \frac{1}{N} \sum_i \delta_{-\sigma,\sigma_i} \delta(x-x_i) {\cal P}_t(\vec x, \vec \sigma) \nonumber \\
&& + \ \gamma \sum_{\vec \sigma} \int d\vec x \frac{1}{N} \sum_i \sum_{k \neq i} \delta_{\sigma,\sigma_i} \delta(x-x_i) 
   {\cal P}_t(\vec x, \vec \sigma) \nonumber\\
 =&& \ \gamma p_{-\sigma}(x,t)  + (N-1) \gamma p_{\sigma}(x,t) 
\eea 
The result above combines with the term $- \gamma N {\cal P}_t$ in Eq. (\ref{FP2ND}) to give $\gamma p_{-\sigma}(x,t)  - \gamma p_{\sigma}(x,t)$. 
Finally the interaction term in (\ref{FP2ND}) gives
\bea 
&&- \frac{2 g}{N} \sum_{\vec \sigma} \int \frac{d\vec x}{N} \sum_i \delta_{\sigma,\sigma_i} \delta(x-x_i) \sum_k \partial_{x_k} \sum_{ l\neq k} \frac{\tilde\delta_{\sigma_k,\sigma_l}}{x_k-x_l} {\cal P}_t(\vec x, \vec \sigma) = - \frac{2 g}{N} \partial_x \sum_{\vec \sigma} \int \frac{d\vec x}{N} \sum_i \sum_{l \neq i} \delta_{\sigma,\sigma_i} \delta(x-x_i) \frac{\tilde\delta_{\sigma_i,\sigma_l}}{x_i-x_l} 
{\cal P}_t(\vec x, \vec \sigma) \nonumber \\
&& 
\eea 
where $\tilde\delta_{\sigma_k,\sigma_l}=\delta_{\sigma_k,\sigma_l}$ for model I and $\tilde\delta_{\sigma_k,\sigma_l}=1$ for model II. We then introduce $1=\int dy \ \delta(y-x_l)$ to rewrite this as
\be 
- 2 g \left( 1-\frac{1}{N} \right) \partial_x \int dy \frac{1}{x-y} \tilde p^{(2)}(x,y,t;\sigma)
\ee 
where 
\bea  \label{defptile2}
\tilde p^{(2)}(x,y,t;\sigma) =
\begin{cases}
& p^{(2)}_{\sigma,\sigma}(x,y,t) \quad, \hspace*{4.4cm} \quad {\rm for\; model\; I}   \;, \\
& p^{(2)}_{\sigma,s}(x,y,t) =  p^{(2)}_{\sigma,+}(x,y,t) + p^{(2)}_{\sigma,-}(x,y,t) \quad, \quad {\rm for \; model \; II} \;. 
\end{cases}
\eea
Putting everything together we obtain
\be 
\partial_t p_\sigma(x,t) = \partial_x \left[(- v_0 \sigma + x) p_\sigma(x,t) - 2 g \left( 1-\frac{1}{N} \right) \int dy \frac{1}{x-y} \tilde p^{(2)}(x,y,t;\sigma) \right] + \gamma p_{-\sigma}(x,t)  - \gamma p_{\sigma}(x,t) \;.
\label{eqfromFPfull}
\ee

\subsubsection{Large $N$ limit and the stationary state}

In the limit $N \to +\infty$ one naively expects $\rho_\sigma(x,t)$ to converge to its average $p_\sigma(x,t)$. Thus equation \eqref{eqfromFPfull} should be the same as the Dean-Kawasaki equation \eqref{Dean_model12} for $N \to +\infty$. For this to be valid one needs the following decoupling condition to hold in this limit 
for any couple $(\sigma,\epsilon)$
\begin{equation}
    p^{(2)}_{\sigma,\epsilon}(x,y,t) \to p_\sigma(x,t) p_\epsilon(y,t)
    \label{approx}
\end{equation}
We have checked numerically that this condition indeed holds for model I in the stationary state. 
This is shown in Fig. \ref{check_full_finiteN} where, in the left panel, the different terms of \eqref{eqfromFPfull} are plotted for model I. The interaction term is computed in two different ways: either using the exact expression where $p^{(2)}_{\sigma,\sigma}(x,y)$ is evaluated from the simulations, or by approximating $p^{(2)}_{\sigma,\sigma}(x,y) \simeq p_\sigma(x)p_\sigma(y)$ (i.e., neglecting the correlations). The results of these two computations are compared and we see that they overlap exactly. Thus \eqref{approx} is a good approximation in this case in the large $N$ limit. This implies that the Dean-Kawasaki equations given in the text
\eqref{Dean_eq} hold for large $N$. 

However this is not necessarily true in general, and in particular we find that it is not true for model II (see below).
In general there can be an additional correlation term:
\begin{eqnarray}
    &&- 2 g \left( 1-\frac{1}{N} \right) \int dy \frac{1}{x-y} \tilde p^{(2)}_c(x,y,t;\sigma) = - 2 g \left( 1-\frac{1}{N} \right) \int dy \frac{1}{x-y} \tilde p^a_c(x,y,t;\sigma) \\
    &&{\rm with} \quad \tilde p^{(2)}_c(x,y,t;\sigma) = \tilde p^{(2)}(x,y,t;\sigma) - p_\sigma(x,t) \tilde p(y,t;\sigma) \\
    &&{\rm and} \quad \tilde p^a_c(x,y,t;\sigma) = \frac{1}{2} \left(\tilde p^{(2)}_c(x,y,t;\sigma) - \tilde p^{(2)}_c(x,2x-y,t;\sigma) \right)
\end{eqnarray}
where $\tilde p(y,t;\sigma)=p_\sigma(y,t)$ in model I and $p_s(y,t)$ in model II. Note that only the antisymmetric part of the correlations $\tilde p^a_c(x,y,t;\sigma)$ contribute to the integral. For the rest of this discussion we will focus on the stationary state, so that we drop the time dependence.

The case of model II is quite different from model I. To understand this it is useful to write down the stationary equations for $p_s$ and $p_d$ in the case of model II
(these equations are exact) 
\begin{eqnarray}
0 &=& - v_0 p_d(x) + x p_s(x) - 2 g \left( 1-\frac{1}{N} \right) \int dy \frac{1}{x-y} p^{(2)}_{s,s}(x,y) \label{psfull} \\
0 &=& \partial_x [- v_0 p_s(x) + x p_d(x) - 2 g \left( 1-\frac{1}{N} \right) \int dy \frac{1}{x-y} p^{(2)}_{d,s}(x,y)] - 2 \gamma p_d(x)
\label{pdfull} \;,
\end{eqnarray}
where $p_{s,s}^{(2)}(x,y) = p_{+,s}^{(2)}(x,y)+p_{-,s}^{(2)}(x,y)$ and $p_{d,s}^{(2)}(x,y) = p_{+,s}^{(2)}(x,y)-p_{-,s}^{(2)}(x,y)$. The different terms of these two equations are plotted in Fig.~\ref{check_full_finiteN}. Again we compare the interaction term computed by taking into account or neglecting the correlations. For \eqref{psfull}, the two results again overlap perfectly, which means that \eqref{approx} is valid for this equation. Hence the
following equation holds for large $N$ (equivalent to a Dean-Kawasaki equation for the total density)
\begin{equation}
0 = - v_0 p_d(x) + x p_s(x) - 2 g \int dy \frac{1}{x-y} p_s(x) p_s(y) \label{ps_decoupled} \;.
\end{equation}
However the same cannot be said for \eqref{pdfull}. Indeed the first numerical observation is that for model II, $p_d(x) \approx 0$ as
$N \to +\infty$ (see Fig. \ref{plot_rhod}). 
This implies that if one neglects correlations in \eqref{pdfull} one obtains a vanishing interaction term in the limit $N\to+\infty$ (see right panel of Fig. \ref{check_full_finiteN}). However using the exact expression where $p_{d,s}^{(2)}(x,y)$ is evaluated from the numerics one finds that
the interaction term does not vanish at all. This explains the failure of the Dean-Kawasaki equation for this case. Note however that 
\label{ps_decoupled} being valid, the fact that $p_d(x) \approx 0$ immediately implies that the density $p_s=\rho_s$ is a semi-circle
as discussed in the text.

This effect of correlations in model II may be due to the clustering which is often observed for active particles which cannot pass each other. A $+$ (resp. $-$) particle at position $x$ will create an accumulation of $-$ (resp. $+$) particles immediately at its right (resp. left) because they cannot cross. This results in a symmetric contribution to $p_{s,s}^{(2)}$, which is unimportant for the computation of the interaction term, and an antisymmetric contribution to $p_{d,s}^{(2)}$, which may explain the discrepancy observed. In model I this effect is absent because only particles of the same sign interact together.

\begin{figure}[h!]
    \centering
    \includegraphics[width=0.32\linewidth]{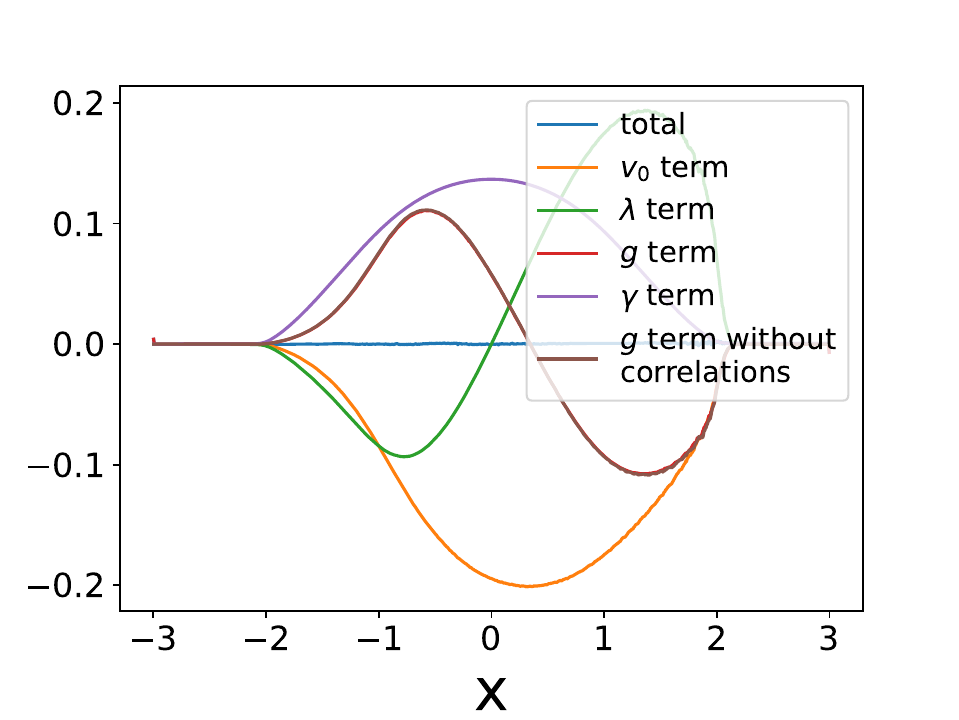}
    \includegraphics[width=0.32\linewidth]{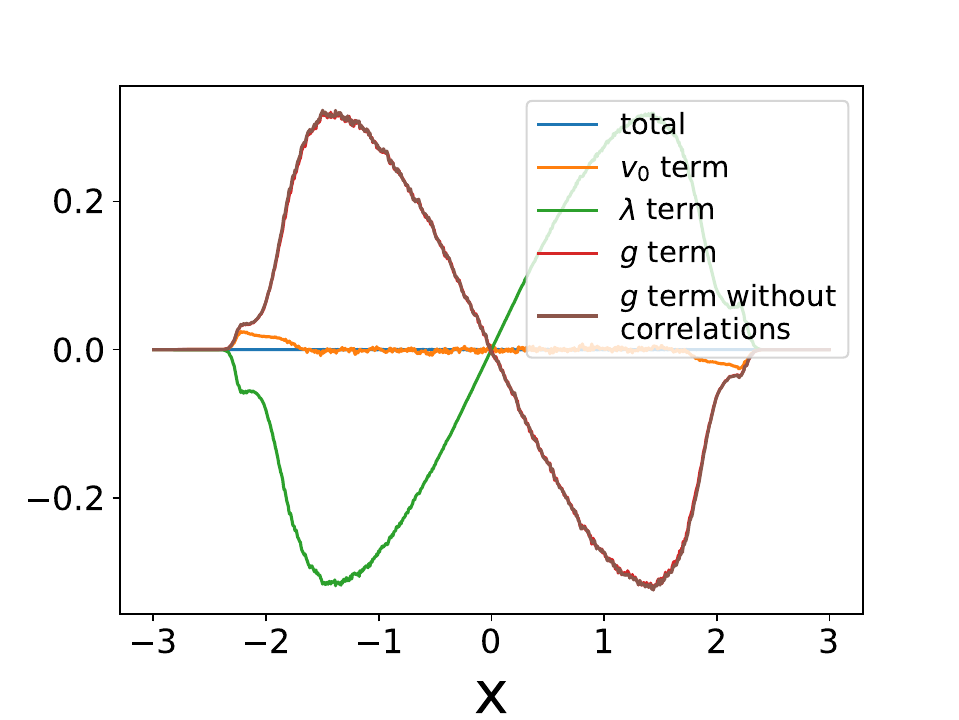}
    \includegraphics[width=0.32\linewidth]{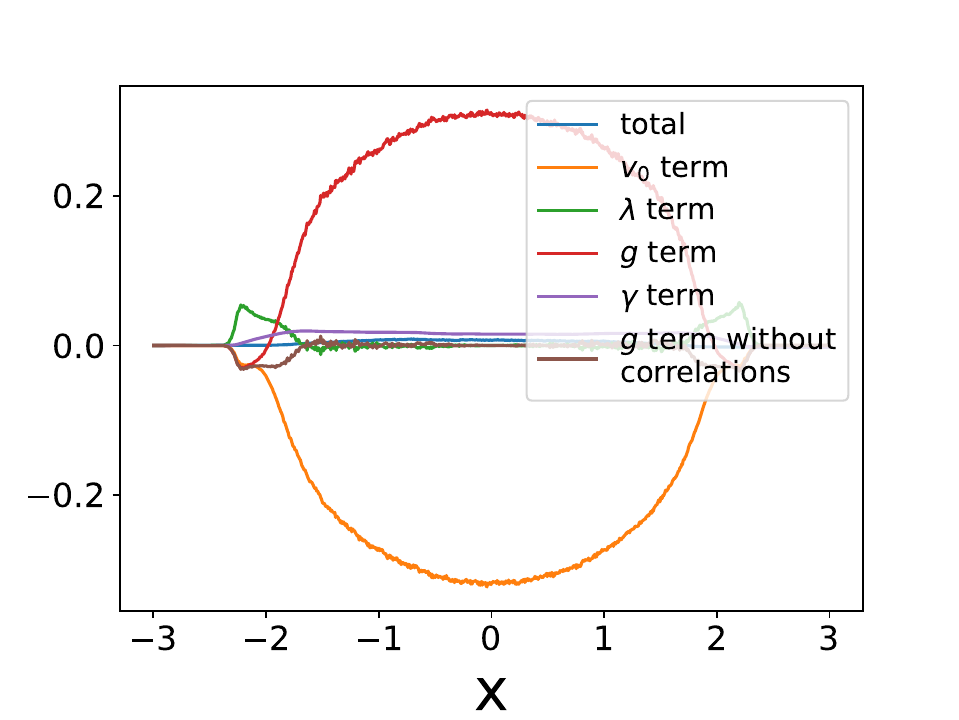}
    \caption{Left: Different terms of the rhs of Eq. \eqref{eqfromFPfull} for model I and their sum, which is zero in both cases as expected, for $N=100$ and all other parameters equal to 1. The interaction term obtained by neglecting correlations is also plotted. It matches perfectly with the true interaction term. Center and Right: Different terms of the rhs of eq. \eqref{psfull} (center) and \eqref{pdfull} (right) for model II and their sum, which is zero in both cases as expected, for the same parameters. The interaction term obtained by neglecting correlations is also plotted. It matches perfectly with the true interaction term in \eqref{psfull} (center) but leads to a completely wrong result in \eqref{pdfull} (right). In all cases the interaction term (including correlations) is used to compute the total sum.}
    \label{check_full_finiteN}
\end{figure}


\subsection{Equation for the resolvent for model I}

In this subsection we show how to use the two approaches discussed above to obtain an equation for the resolvent, which is defined
for $z$ in the complex plane minus the support of the density (which for finite $N$ is a collection of points on the real axis)  
\beq \label{def_resolv}
G_\sigma(z,t) = \int \frac{dx}{z-x} \rho_\sigma(x,t) \;.
\eeq
As defined in (\ref{def_resolv}), $G_\sigma(z,t)$ is a stochastic variable (since, at finite $N$, 
$\rho_\sigma(x,t)$ is a stochastic variable). We will also consider below its average, denoted $\overline{G}_\sigma(z,t)$. 

We first start from the Dean-Kawasaki equation for model I \eqref{Dean_model12} and multiply \eqref{Dean_model12} by $\frac{1}{z - x}$ and integrate over $x$. We then use integrations by parts (the density has finite support so there are no boundary terms) and the identity $(\partial_x + \partial_z) \frac{1}{z - x}=0$ to rewrite the different terms. The second term on the right hand side can be rewritten using the identity
\bea
\int dx \frac{1}{z - x} \partial_x  x \rho_\sigma(x,t)
= \partial_z \int dx \frac{x}{z - x} \rho_\sigma(x,t)
= \partial_z z \int dx \frac{1}{z - x} \rho_\sigma(x,t)
= \partial_z z G_\sigma(z,t) \;,
\eea
while the interaction term yields
\bea
\int dx \frac{1}{z-x} \partial_x [\rho_\sigma(x,t) \int \frac{dy}{x-y} \rho_\sigma(y,t) ]
&=& \partial_z  \int dx dy \frac{1}{(z-x)(x - y)}
\rho_\sigma(x,t) \rho_\sigma(y,t) \\
&=& \frac{1}{2} \partial_z  \int dx dy \frac{1}{(x - y)} [\frac{1}{z-x} - \frac{1}{z-y}] \rho_\sigma(x,t) \rho_\sigma(y,t) \\
&=& \frac{1}{2} \partial_z  \int dx dy \frac{1}{(z-x)(z-y)} \rho_\sigma(x,t) \rho_\sigma(y,t) \\
&=& \frac{1}{2} \partial_z G(z)^2 
\eea 
where in the second step we used the symmetry between $x$ and $y$. This leads to the following exact equation for the resolvent 
(recall that in that equation $G_\sigma$ is thus a stochastic variable) 
\beq
\partial_t G_\sigma = \partial_z (-v_0 \sigma G_\sigma + z G_\sigma - g G_\sigma^2) + \gamma G_{-\sigma} - \gamma G_{\sigma} + \frac{T}{N} (1 -  \frac{\beta}{2})  \partial^2_z G_\sigma + O\left( \frac{1}{\sqrt{N}} \right) \;,
\label{2species_G_apdx}
\eeq
where $O({1}/{\sqrt{N}})$ denotes the active noise term which has zero mean. This equation is thus useful only at large $N$. 

Let us now use the Fokker-Planck approach and, still in the case of model I, derive an equation for the averaged resolvent starting from \eqref{eqfromFPfull}. Defining
\bea
\overline{G}_\sigma(z,t) &=& \int dx \frac{1}{z-x} p_\sigma(x,t) \\
\overline{G}_{\sigma,\sigma}^{(2)}(z,t) &=& \int dx \int dy \frac{1}{z-x} \frac{1}{z-y} p^{(2)}_{\sigma,\sigma}(x,y,t)
\eea
we get from \eqref{eqfromFPfull} 
\beq
 \partial_t \overline{G}_\sigma = \partial_z \left(-v_0 \sigma \overline{G}_\sigma + z \overline{G}_\sigma - g \left( 1 - \frac{1}{N} \right) \overline{G}_{\sigma,\sigma}^{(2)}(z)\right) + \gamma \overline{G}_{-\sigma} - \gamma \overline{G}_{\sigma}
\label{GfromFP}
\eeq
which is valid for any value of $N$. Note that the rewriting of the interaction term is only valid for model I. We did not find a way to obtain an equation for the resolvent in the case of model II.

Finally, in the stationary state one has the symmetry $(x,\sigma) \to (-x,-\sigma)$, ie $\rho_+(x)=\rho_-(-x)$, which implies $G_+(z)=-G_-(-z)$. Using this identity, \eqref{2species_G_apdx} can be rewritten as (in the stationary state, for large $N$)
\beq
0 = \partial_z (-v_0 G_+(z) + z G_+(z) - g G_+^2(z)) - \gamma G_+(-z) - \gamma G_+(z) + O\left( \frac{1}{\sqrt{N}} \right) \;,
\label{2species_G+_apdx}
\eeq
and similarly for \eqref{GfromFP}. This form will be useful in the next section.

\section{Main results for model I}

In this section we present the derivations of the results given in the text, together with additional results. 
We use extensively the approaches introduced in the previous section. 

\subsection{Moments of the density}
\subsubsection{Moments in the stationary state for $N \to +\infty$}
\label{expansion_moments}
In the stationary state, to determine the moments $\langle x^k \rangle_\pm$ of the densities $2 \rho_{\pm}(x)$ (each being thus normalized  
to unity), we can write a large $z$ expansion for $G_s$ and $G_d$ defined in the text under the form :
\bea
&& G_s(z) = G_+(z)+G_-(z) = \frac{1}{z} + \sum_{k=1}^\infty \frac{m^s_k}{z^{k+1}} = \sum_{k=0}^\infty \frac{1}{z^{k+1}} \frac{1}{2} (\langle x^k \rangle_+ + \langle x^k \rangle_-) = \sum_{k=0}^\infty \frac{\langle x^k \rangle}{z^{k+1}} \\
&& G_d(z) = G_+(z)-G_-(z) = \sum_{k=1}^\infty \frac{m^d_k}{z^{k+1}} = \sum_{k=0}^\infty \frac{1}{z^{k+1}} \frac{1}{2} (\langle x^k \rangle_+ - \langle x^k \rangle_-)
\eea
where $\langle \cdot \rangle_\sigma$ denotes an average restricted to particles with spin $\sigma$ and $\langle \cdot \rangle$ the average over all particles. Injecting these expansions into equations (\ref{2species_Gd})-(\ref{eq1integrated}) we can compute recursively the moments for the distributions of the $+$ and $-$ particles in the stationary state. For the first 4 moments this gives :
\bea
&& m^s_1 = 0 \quad ; \quad m^d_1 = \langle x \rangle_+ = - \langle x \rangle_- = \frac{v_0}{1+2\gamma} \\
&& m^s_2 = \langle x^2 \rangle_+ = \langle x^2 \rangle_- = \frac{v_0^2}{1+2\gamma} + \frac{g}{2} \quad ; \quad m^d_2 = 0 \\
&& m^s_3 = 0 \quad ; \quad m^d_3 = \langle x^3 \rangle_+ = - \langle x^3 \rangle_- = \frac{6v_0^3+3 g v_0 (3 + 2\gamma)}{2 (1 + 2\gamma) (3 + 2\gamma)} \\
&& m^s_4 = \langle x^4 \rangle_+ = \langle x^4 \rangle_- = \frac{3v_0^4}{(1+2\gamma)(3+2\gamma)} + \frac{gv_0^2(3+5\gamma)}{(1+2\gamma)^2}+\frac{g^2}{2} \quad ; \quad m^d_4 = 0;
\eea

We see that $\langle x^k \rangle_+ = (-1)^k \langle x^k \rangle_+$, as it should be from the symmetry $(x,\sigma) \to (-x,-\sigma)$. Note that
the odd moments vanish in the diffusive limit $v_0,\gamma \to + \infty$ with $D=\frac{v_0^2}{2 \gamma}$.
These predictions, together with finite $N$ corrections (see Fig. \ref{2species_moments} below), were tested numerically for $N=100$ and a total time $t=10000$ ($10^7$ simulation steps with $dt=0.001$) for different values of the parameters. The relative errors obtained were all between $0.1$ and $0.5\%$.

In general, using the equation for $G_+$ \eqref{2species_G+_apdx} and its expansion in $\frac{1}{z}$ :
\bea
&&G_+(z) = \sum_{k=0}^\infty \frac{\langle x^k \rangle_+}{2z^{k+1}} \quad \Rightarrow \quad G_+^2(z) = \sum_{k=0}^\infty \sum_{l=0}^k \frac{\langle x^l \rangle_+ \langle x^{k-l} \rangle_+}{4z^{k+2}}
\eea
we obtain the following recursion relation (valid for any positive integer $k$ with the convention that the sum is zero for $k<2$) :
\beq \label{recursion_moments}
\langle x^k \rangle_+ = \frac{1}{1+(1+(-1)^{k+1})\frac{\gamma}{k}} (v_0 \langle x^{k-1} \rangle_+ + \frac{g}{2} \sum_{l=0}^{k-2} \langle x^l \rangle_+ \langle x^{k-2-l} \rangle_+)
\eeq

\textbf{Application of \eqref{recursion_moments} to determine the support and the singularity}.
Solving numerically the recursion \eqref{recursion_moments} up to large values of $k$ (using Mathematica) 
gives information on the singularities of $G_\pm(z)=\sum_{k=0}^\infty \frac{\langle x^k \rangle_\pm}{2z^{k+1}}$ (see e.g. \cite{flajolet2009}), and in turn on 
the support and singular behaviors of the densities $\rho_\pm(x)$.
For instance, in the well known example of the semi-circle, we can compute the expansion directly from the explicit expression (with $k=2 p$)
\bea \label{semi-circle2} 
G(z) = \frac{z}{2} - \sqrt{\left(\frac{z}{2}\right)^2-1} = \sum_{p=0}^\infty \frac{1}{p+1} {2 p \choose p} \frac{1}{z^{2p+1}} \;.
\eea
For large $p$ the coefficient of $1/z^{2p+1}$ behaves as $\frac{1}{\sqrt{\pi}} p^{-3/2} 2^{2p}$ (while the coefficient of $1/z^{2p}$ vanishes identically). 
The exponential part tells us the location of the leading singularities, which correspond to the edges of the support ($z=\pm 2$) while the $p^{-3/2}$ indicates a square root singularity. For our model we find a good numerical fit with the form given in the text \eqref{singularity_evenodd}, which
we reproduce here
\beq
\langle x^k \rangle_+ \simeq A k^{-\frac{3}{2}} x_+^k + B k^{-\alpha-1} (-x_+)^k \; \quad , \quad \alpha \approx \frac{3}{2} \;,
\label{singularity_evenodd2}
\eeq 
where the constants $A$, $B$, and the value of the edge $x_+$ are determined numerically.
These fits are shown in Fig. \ref{2species_singularity}.  It is observed that the difference 
between even and odd terms is subdominant at large $k$, with a common $k^{-3/2}$ behavior. 
To determine $\alpha$ numerically it is useful to compute
\be
\kappa_p = \frac{(2p)^{\frac{3}{2}}\langle x^{2p}\rangle_+}{2x_+^{2p}} - \frac{(2p-1)^{\frac{3}{2}}\langle x^{2p-1}\rangle_+}{2x_+^{2p-1}} \sim B(2p)^{\frac{1}{2}-\alpha} + B(2p-1)^{\frac{1}{2}-\alpha} 
\sim 2B(2p)^{\frac{1}{2}-\alpha} \;.
\label{singularity_diff}
\ee
By plotting $\kappa_p$ vs $p$ we are able to obtain the value of $\alpha$ with reasonable accuracy. The results suggest that $\alpha=\frac{3}{2}$ for any set of parameters as long as $\gamma>0$ (however the prefactor $B$ seems to decrease towards zero when $\gamma$ decreases). This is indeed greater than $1$ as we would expect from the simulations. 

From \eqref{singularity_evenodd2} one can obtain the support and the singular behavior of the density $\rho_+(x)$ near its edges.
The first term corresponds to the right edge at $x_+$ with a singular behavior $\rho_+(x) \sim (x_+-x)^{1/2}$. The second
term in \eqref{singularity_evenodd2} corresponds to the left edge at $x_- = -x_+$ with a singular behavior $\rho_+(x) \sim (x-x_-)^{\alpha}$ with $\alpha$
consistent with $\alpha= 3/2$. Note that in the simpler case of an exact semi-circle density, there is
a perfect symmetry of the density near the two edges $\pm 2$, which leads to the cancellation of the odd moments, i.e.
the coefficients of $1/z^{2 p}$ in \eqref{semi-circle2}. Here however, the even-odd effect in \eqref{singularity_evenodd2}
shows that these behaviors are different (they even have different exponents). To check these predictions obtained here by series expansions
and directly for $N=+\infty$, 
we have also performed a direct calculation of the densities $\rho_\pm(x)$ from the numerical solution of the equation of motion,
as explained in Section \ref{simulation_details}. The results are shown in Fig. \ref{2species_singularity_exponents}, where two
distinct behaviors for $\rho_+(x)$ at the two edges $x_+$ (left panel) and $x_-$ (right panel) are observed,
in good agreement with our theoretical predictions. Note that at finite $N$ there are exponential tails 
which extend beyond the infinite $N$ support (as was discussed in Section \ref{finiteN_sec}) which are also visible on the figure
and make the determination of the exponent more delicate. 



\begin{figure}[h!]
    \centering
    \includegraphics[width = 0.9\linewidth]{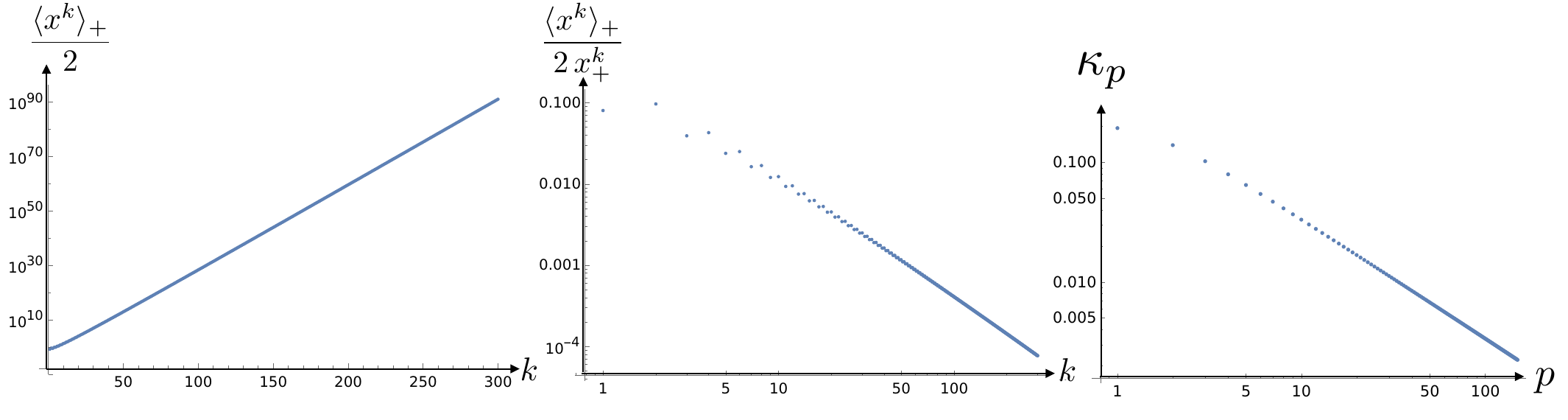}
    \caption{Left and center: coefficients $\langle x_+^k \rangle/2$ of the $\frac{1}{z}$ expansion of $G_+(z)$ as a function of $k$, before (left panel) and after (center panel) renormalizing by $x_+^k$, for $\lambda=1$, $g=1$, $v_0=1$ and $\gamma=1$. In log-lin scale the non-renormalized quantity converges to a line of slope $\ln{x_+}$ for large $k$, while the renormalized one converges to a line of slope $-\frac{3}{2}$ in log-log scale. Right : quantity in eq.(\ref{singularity_diff}) as a function of $p$. The slope is $-1$ in log scale which is compatible with $\alpha=3/2$.
    }
    \label{2species_singularity}
\end{figure}

\begin{figure}[h!]
    \centering
    \includegraphics[width=0.32\linewidth]{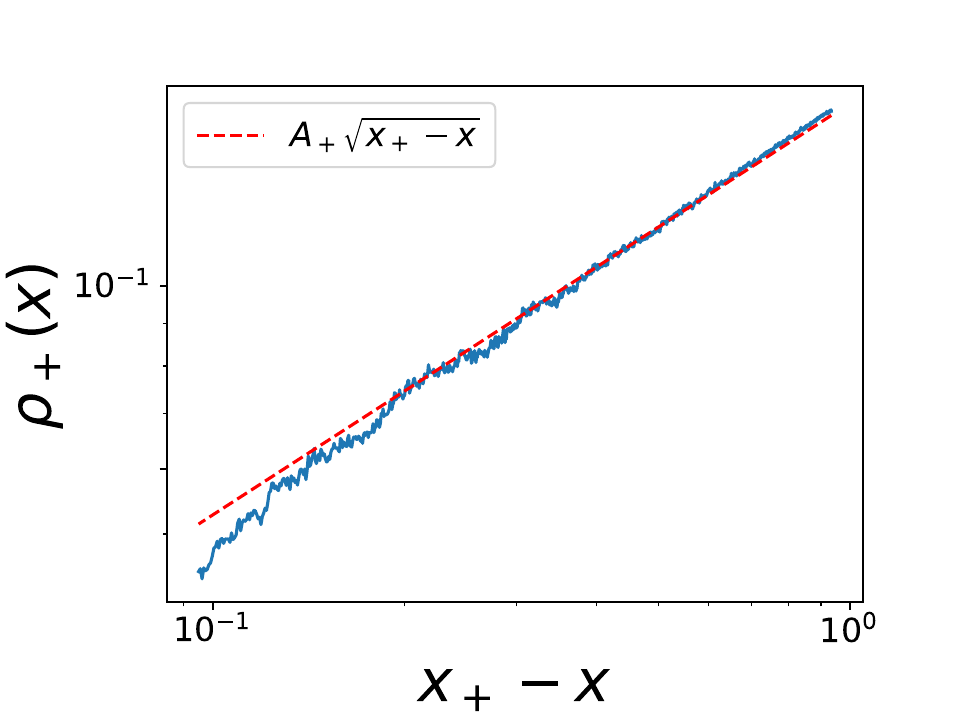}
    \includegraphics[width=0.32\linewidth]{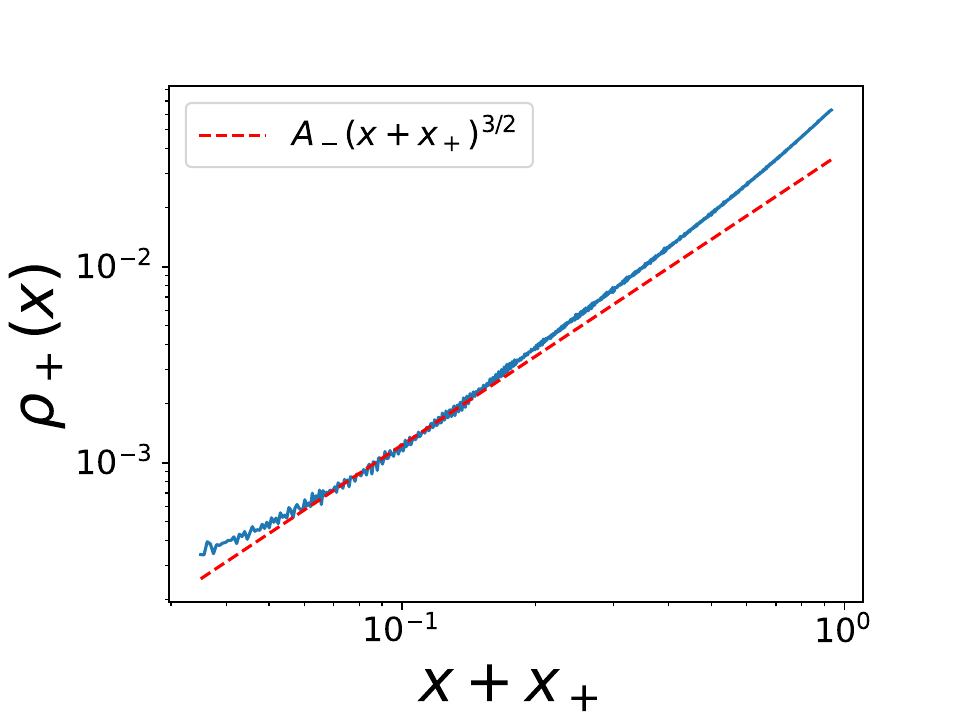}
    \caption{Density $\rho_+(x)$ in the vicinity of the edges $x_+$ (left) and $x_-=-x_+$ (right), for $N=100$ and all parameters equal to $1$. On the left panel, the exponent $1/2$ can clearly be seen far enough from the edge. On the right panel, the exponent $3/2$ near $x_-$ can be observed in a small window between the finite $N$ exponential tail regime (discussed in the text) around $x_-$ and the bulk regime.}
    \label{2species_singularity_exponents}
\end{figure}

We can also mention that the difference of $1$ between the two exponents $1/2$ and $\alpha=3/2$ can also be
naively expected from the fact that the explicit $(-1)^k$ term in the recursion \eqref{recursion_moments}
comes with a $1/k$ coefficient, subleading at large $k$. That may explain 
the factor $\frac{1}{k}$ between the $x_+^k$ term and the $(-x_+)^k$ term in \eqref{singularity_evenodd2}.

Finally, note that $\rho_+(x)$ and $\rho_-(x)$ have the same support as soon as $\gamma >0$ (we assume $\gamma$ to be independent on $N$)
because particles switch sign independently of their position. In addition since one has the symmetry $\rho_+(x)=\rho_-(-x)$ the support of $\rho_\pm(x)$ is always symmetric around $x=0$ (ie $x_-=-x_+$) which is compatible with what we obtained here.



\subsubsection{Moments in the stationary state for finite N}
For $N=1$, the moments can be computed since the distribution is known (see equation \eqref{rho_inde_app}). For the first four moments we get 
\bea
&&\langle x \rangle_+ = - \langle x \rangle_- = \frac{v_0}{1+2\gamma} \\
&& \langle x^2 \rangle_+ = \langle x^2 \rangle_- = \frac{v_0^2}{1+2\gamma} \label{moment2_N1} \\
&& \langle x^3 \rangle_+ = - \langle x^3 \rangle_- = \frac{3v_0^3}{(1+2\gamma)(3+2\gamma)} \label{moment3_N1} \\
&& \langle x^4 \rangle_+ = \langle x^4 \rangle_- = \frac{3v_0^4}{(1+2\gamma)(3+2\gamma)} \label{moment4_N1}
\eea

For intermediate values of $N$, we can compute the moments from equation \eqref{GfromFP} by expanding $\overline{G}_\sigma$ and $\overline{G}_{\sigma,\sigma}^{(2)}$ in $\frac{1}{z}$. For this computation we will directly use the symmetry $(x,\sigma) \to (-x,-\sigma)$ to write them as:
\bea
&& \overline{G}_\sigma(z) = \sum_{k=0}^\infty \sigma^k \frac{\langle x^k \rangle_+}{2z^{k+1}} \\
&& \overline{G}_{\sigma,\sigma}^{(2)}(z) = \sum_{k=0}^\infty \sigma^k \sum_{l=0}^k \frac{\langle x_1^l x_2^{k-l} \rangle_{+,+}}{4z^{k+2}} \;,
\eea
where we have introduced the moments for pairs of $+$ particles. Inserting this in Eq.~\eqref{GfromFP} we can directly obtain the first three moments 
\bea
&& \langle x \rangle_+ = - \langle x \rangle_- = \frac{v_0}{1 + 2\gamma} \label{mom1}\\ 
&& \langle x^2 \rangle_+ = \langle x^2 \rangle_- = \frac{v_0^2}{1+2\gamma} + \frac{g}{2} \left(1-\frac{1}{N}\right) \label{mom2} \\
&& \langle x^3 \rangle_+ = - \langle x^3 \rangle_- = \frac{6v_0^3+(1-\frac{1}{N})3g v_0 (3 + 2\gamma)}{2 (1 + 2\gamma) (3 + 2\gamma)} \;. \label{mom3}
\eea
Interestingly, the average position is independent of $N$. The first three moments as a function of $N$ are compared with their nunerical
determination from the solution of the equation of motion in Fig.~\ref{2species_moments}. The agreement is very good. 
For large $N$ these three moments converge to the predictions for $N=+\infty$. 
Unfortunately, equation \eqref{GfromFP} does not allow to obtain the moments of order 4 of higher. Indeed it leads to
a system of equations which does not close, as it 
involves unknown correlations between the particles. 


\begin{figure}[t]
    \centering
    \includegraphics[width=0.43\linewidth]{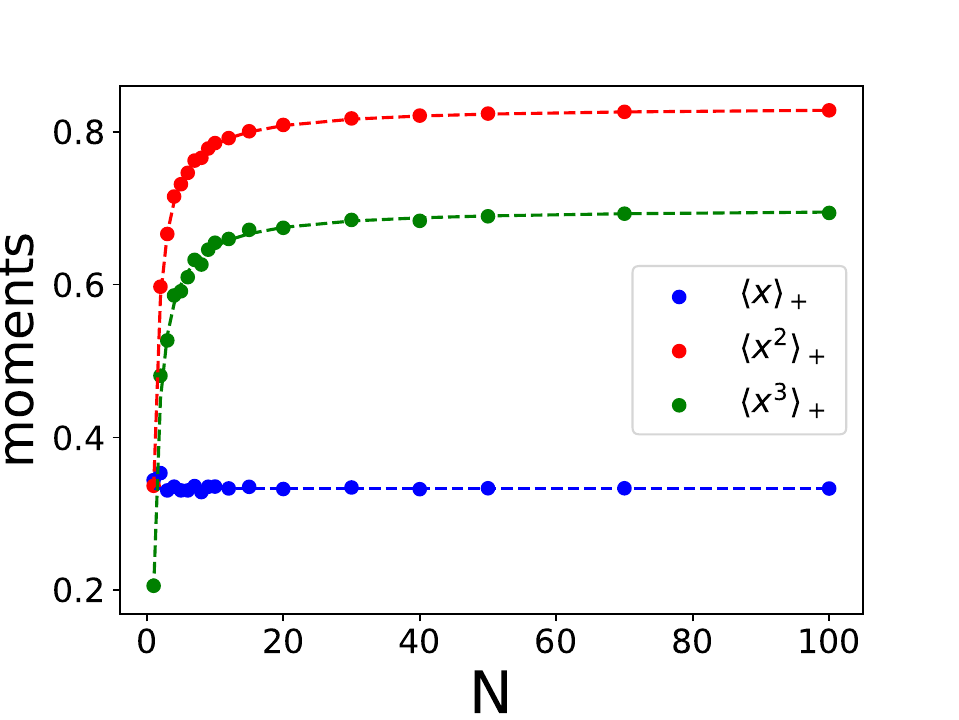}
    \includegraphics[width=0.43\linewidth]{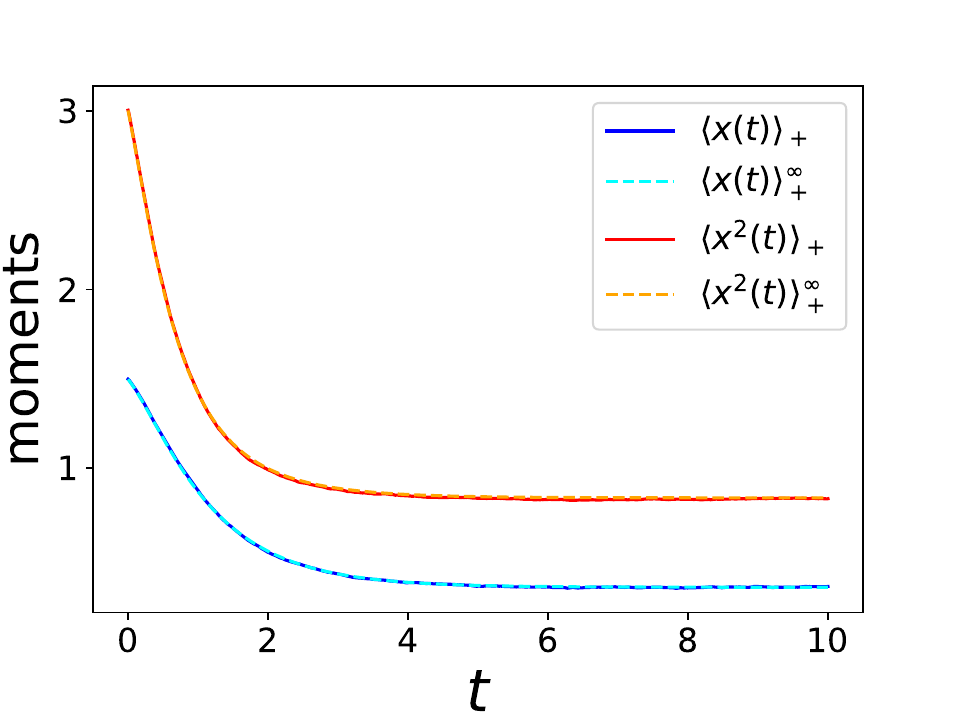}
    \caption{Left: Values of the first three moments 
    of the stationary distribution of the positions of the $+$ particles in model I, as a function of $N$. All the parameters of the model are set to 1. The dots correspond to the results of the numerical simulation (averaged over 100 realisations) while the dashed lines correspond to the predicted values given 
    in Eqs.(\ref{mom1})-(\ref{mom3}). Right: first and second moment of the distribution of the position of the $+$ particles, given by $\langle x^k \rangle_+ = m^s_k + m^d_k$ when the initial proportion of $+$ and $-$ particles are equal, as a function of time. At $t=0$ the particles are spaced equally over the interval $[0,v_0/\lambda+2\sqrt{g}{\lambda}]$. All parameters are set to $1$. The trajectories were obtained by averaging over 1000 realisations with $N=100$. The dashed lines correspond to the predicted evolution for $N=+\infty$.}
    \label{2species_moments}
\end{figure}

\subsubsection{Time evolution of the moments}

We consider now the time evolution of the system towards the stationary state. It can be studied from
the equations obeyed by the time dependent resolvents $G_s(z,t)$ and $G_d(z,t)$
(which can be obtained from \eqref{2species_G_apdx} by taking the sum and differences and neglecting terms subdominant at large $N$)
\bea
&& \partial_t G_s = \partial_z (-v_0 G_d +  z G_s  - \frac{g}{2} (G_s^2+G_d^2)) \\
&& \partial_t G_d = \partial_z (-v_0 G_s +  z G_d  - g G_s G_d) - 2\gamma G_d
\eea
Upon expanding in powers in $1/z$ one obtains the time evolution for the moments $m^{s,d}_k(t) = \int dx x^k \rho_{s,d}(x,t)$ 
for $N \to +\infty$. Focusing on the first two moments, we obtain the linear evolution equations :
\bea
&& \partial_t m^s_1 + m^s_1 = 0 \\
&& \partial_t m^d_1 + (1+2\gamma) m^d_1 = v_0 \\
&& \partial_t m^s_2 + 2 m^s_2 = 2v_0 m^d_1 + g \\
&& \partial_t m^d_2 + 2(1+\gamma) m^d_2 = 2v_0 m^s_1
\eea
which for $\gamma \neq 1/2$ leads to :
\bea
&& m^s_1(t) = m^s_1(0)e^{-t} \\
&& m^d_1(t) = \frac{v_0}{1+2\gamma} + \left(m^d_1(0) - \frac{v_0}{1+2\gamma} \right) e^{-(1 + 2\gamma) t} \\
&& \begin{split} m^s_2(t) =& \frac{v_0^2}{1+2\gamma} + \frac{g}{2} + \frac{2v_0}{1-2\gamma} \left( m^d_1(0) - \frac{v_0}{1+2\gamma} \right) e^{-(1+2\gamma)t} \\
&+ \left(m^s_2(0) - \frac{v_0^2}{1+2\gamma} + \frac{g}{2} - \frac{2v_0}{1-2\gamma} \left( m^d_1(0) - \frac{v_0}{1+2\gamma} \right) \right) e^{-2 t} \quad {\rm if} \ \gamma \neq \frac{1}{2} \\
=& \frac{1}{2} \left( v_0^2 + g \right) + v_0 \left(\frac{v_0}{2} - m^d_1(0) \right) t e^{-2 t} + \left(m^s_2(0) - \frac{1}{2} \left( v_0^2 + g \right) \right) e^{-2 t} \quad {\rm if} \ \gamma=\frac{1}{2} \end{split} \\
&& m^d_2(t) = \frac{2v_0}{1+2\gamma}m^s_1(0) e^{- t} + \left( m^d_2(0) - \frac{2v_0}{1+2\gamma}m^s_1(0) \right) e^{-2(1+\gamma) t}
\eea
The case $\gamma=1/2$ is degenerate and leads to a separate set of equations. These predictions are compared with the simulation results in Fig.~\ref{2species_moments}, showing a perfect agreement. 



\subsubsection{Fluctuations of the moments at finite $N$}

In this section we study a measure of the active noise in the Dean-Kawasaki equation of motion.
The observable that we can easily compute numerically is the absolute value of the difference between the predicted value for the first three stationary moments (including the finite $N$ correction) given in Eqs. (\ref{mom1})-(\ref{mom3}),
and their empirical values computed from the simulations by averaging over some fixed time window. Since the values computed for the moments are exact at all $N$, this corresponds to the fluctuations due to the telegraphic noise. These results seem to confirm that these fluctuations 
scale as $\frac{1}{\sqrt{N}}$, for large $N$.

\begin{figure}[h!]
    \centering
    \includegraphics[width=0.32\linewidth]{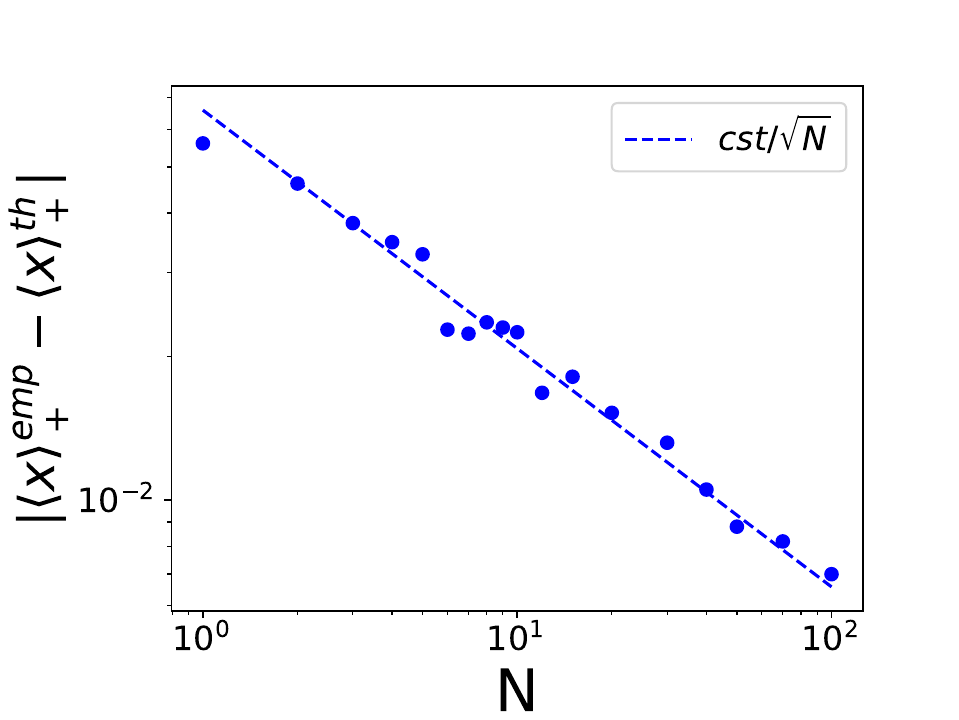}
    \includegraphics[width=0.32\linewidth]{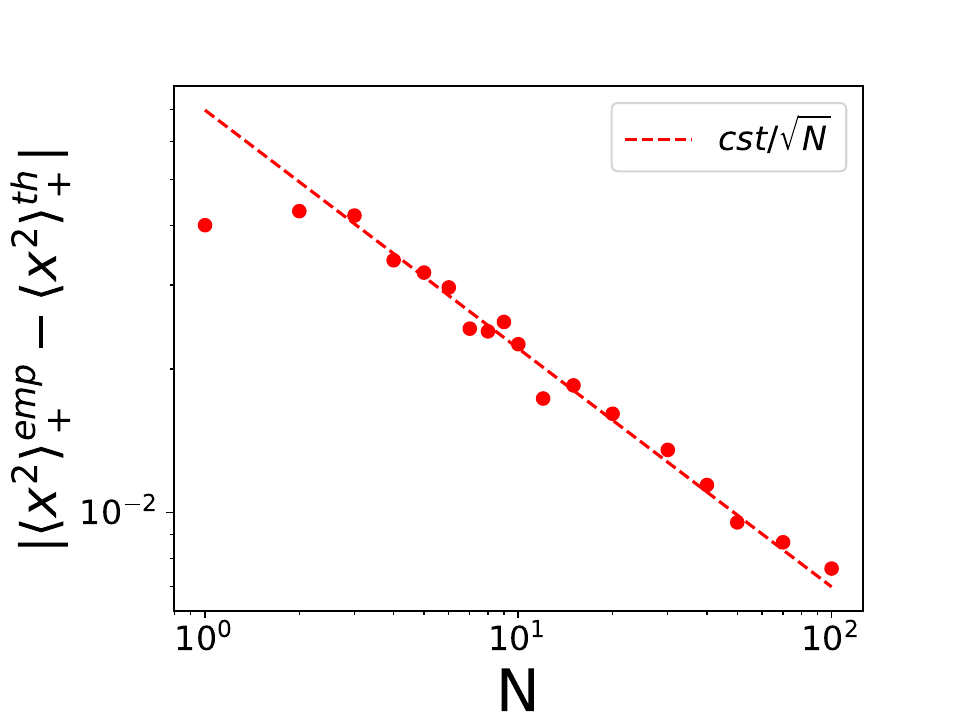}
    \includegraphics[width=0.32\linewidth]{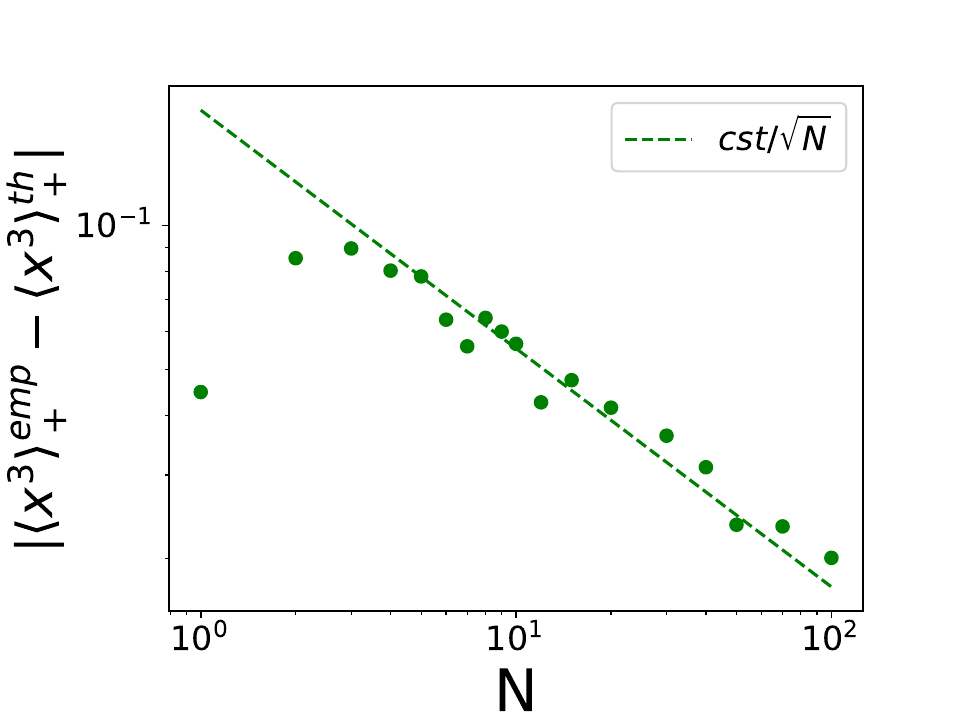}
    \caption{Absolute value of the difference between the predicted value of the first 3 moments of the $+$ particles distribution (including the finite $N$ correction) and the empirical value computed from simulations by averaging over a time $t_{\text{sim}}=100$. After taking the absolute value this was averaged over 10 simulations. The results seem compatible with a $\frac{1}{\sqrt{N}}$ scaling (dashed line).}
    \label{2species_moments_fluctuations}
\end{figure}

\subsection{Small $\gamma$ expansion}
Although we could not solve (\ref{2species_G+_apdx}) explicitly, we were able to obtain a perturbative expansion for the stationary density in the limit of small $\gamma$. Let us start from the integrated version of equation (\ref{2species_G+_apdx}) :
\beq
-v_0 G_+(z) + (z G_+(z) - \frac{1}{2}) - g G_{+}^2(z)) = \gamma \int_{\pm \infty}^z [G_+(z') + G_+(-z')] dz'
\label{integrated_G+}
\eeq
where we used $G_+(z) \sim \frac{1}{2z}$ when $z \to \pm \infty$. The integral in the right hand side represents formally taking
the primitive of a function of the complex variable $z$ (in the upper half complex plane)
which vanishes at infinite $z$.
The bound of the integral can be chosen among $\pm \infty$ arbitrarily. Indeed one has : $\int_{-\infty}^{+\infty} [G_+(z') + G_+(-z')] dz' = \int_{-\infty}^{+\infty} [G_+(z') - G_-(z')] dz' = 0$ by symmetry and using that 
$\int dx \rho_+(x)=\int dx \rho_-(x)$. In the following, we will take $g=1$ for simplicity (the parameter $g$ can then be reintroduced by dimensional analysis). We write $G_+$ as :
\beq \label{expansion_gam}
G_+(z) = g_+^0(z) + \gamma g_+^1(z) + O(\gamma^2)
\eeq
where the $g_+^i$ are of order $1$. $g_+^0$ is the solution of the equation
\beq
-v_0 g_+^0(z) +  z g_+^0(z) - \frac{1}{2} - (g_{+}^0(z))^2 = 0
\eeq
which has the correct behavior at $\pm \infty$ ie :
\beq
g_+^0(z) = \frac{z-v_0}{2}  \left( 1  - \sqrt{1 - \frac{2}{(z-v_0)^2}}  \right) 
\eeq
which as we know corresponds to the semi-circle density centered at $v_0$ with radius $\sqrt{2}$, normalized to $\frac{1}{2}$ :
\beq \label{dens00} 
\rho_+^0(x) = \frac{\sqrt{(2-(x-v_0)^2)_+}}{2\pi}
\eeq

At first order in $\gamma$, we get the equation :
\beq
(z - v_0 - 2 g_+^0(z)) g_+^1(z) = \int_{\pm \infty}^z [g_+^0(z') + g_+^0(-z')] dz'
\eeq
ie :
\beq
g_+^1(z) = - \frac{1}{(z-v_0)\sqrt{1-\frac{2}{(z-v_0)^2}}} \int_{\pm \infty}^z \left[ v_0 + \frac{1}{2}(z'-v_0)\sqrt{1-\frac{2}{(z'-v_0)^2}} - \frac{1}{2}(z'+v_0)\sqrt{1-\frac{2}{(z'+v_0)^2}} \right] dz'
\eeq
It is quite straightforward to see that outside the interval $[-v_0-\sqrt{2}, v_0+\sqrt{2}]$, $g_+^1(z)$ has no imaginary part, so that the density remains zero outside this interval, as we expect. We now focus on the interval $[\max(v_0-\sqrt{2}, -v_0+\sqrt{2}), v_0+\sqrt{2}]$, and choose the bound to be $+\infty$. We want to compute the first order correction to the density :
\beq
\rho_+^1(x) = -\frac{1}{\pi}\lim_{\epsilon \to 0^+} {\rm Im} \ g_+^1(x+i\epsilon)
\eeq
On the interval we consider, there are 2 contributions :
\bea
\rho_+^1(x) &=& \frac{1}{\pi \sqrt{2-(x-v_0)^2}} \left[ \int_x^{v_0+\sqrt{2}} [v_0 - \frac{1}{2}\sqrt{(x'+v_0)^2-2}] dx' + \int_{v_0+\sqrt{2}}^{+\infty} [v_0 + \frac{1}{2} \sqrt{(x'-v_0)^2-2} - \frac{1}{2}\sqrt{(x'+v_0)^2-2}] dx' \right] \nonumber \\
&&\equiv \frac{1}{\pi \sqrt{2-(x-v_0)^2}} (\tilde{I}_1(x) + I_0) \;. \label{order1_rhosup} 
\eea
Computing the first integral yields :
\bea
&&\begin{split} \tilde{I}_1(x) \ =& \ v_0(v_0+\sqrt{2}-x) + \frac{1}{4}[(v_0+x) \sqrt{(v_0+x)^2-2} - 2(2v_0 + \sqrt{2}) \sqrt{v_0(v_0+\sqrt{2})} \ ] \\
&- \frac{1}{2} [\ln (v_0+x+\sqrt{(v_0+x)^2-2} \ ) - \ln (2v_0+\sqrt{2}+2\sqrt{v_0(v_0+\sqrt{2})} \ )] \end{split} \nonumber \\
&& \ \ \ \ \underset{x \to v_0+\sqrt{2}}{=} (v_0 - \sqrt{v_0 (v_0+\sqrt{2})})(v_0 + \sqrt{2} - x) + O((v_0 + \sqrt{2} - x)^2)
\eea
Taking into account the prefactor, this term vanishes as a square root near the edge $v_0+\sqrt{2}$. The second integral gives :
\bea
&&\begin{split} I_0 = & \left[ v_0 x + \frac{1}{4} \left( (x-v_0) \sqrt{(x-v_0)^2-2} - (x+v_0) \sqrt{(x+v_0)^2-2} \right) \right. \\ & \left. - \frac{1}{2} \left( \ln(x-v_0+\sqrt{(x-v_0)^2-2} \ ) - \ln(x+v_0+\sqrt{(x+v_0)^2-2} \ ) \right) \right] _{v_0+\sqrt{2}}^{+\infty} \end{split} \nonumber \\
&& \ \ \ \ = - v_0 (v_0+\sqrt{2}) + (v_0+\frac{1}{\sqrt{2}}) \sqrt{v_0(v_0+\sqrt{2})} - \frac{1}{2} \ln(\sqrt{2}v_0 + 1 + \sqrt{2v_0(v_0+\sqrt{2})} \ )
\eea
(the expression in the brackets vanishes at $+\infty$ and only the contribution from the bound $v_0+\sqrt{2}$ remains). Since this expression is non-zero for any $v_0>0$, this term leads to a divergence with exponent $\frac{1}{2}$ at $v_0+\sqrt{2}$. Overall (using $\frac{1}{\sqrt{2-(x-v_0)^2}} = \frac{1}{2^{3/4} \sqrt{v_0+\sqrt{2}-x}} + \frac{\sqrt{v_0+\sqrt{2}-x}}{8 \times 2^{1/4}} + O((v_0+\sqrt{2}-x)^{3/2})$) we get :
\bea \label{expansion_rho1_edge}
\rho_+^1(x) &=& \frac{1}{\pi \sqrt{2-(x-v_0)^2}} \left( \frac{\ln 2}{4} - v_0 x + \frac{1}{4}(v_0+x) \sqrt{(v_0+x)^2-2} - \frac{1}{2} \ln (v_0+x+\sqrt{(v_0+x)^2-2}) \ \right) \\
&\underset{x \to v_0+\sqrt{2}}{=}& \frac{I_0}{\pi 2^{3/4} \sqrt{v_0+\sqrt{2}-x}} + \frac{1}{\pi}\left( \frac{I_0}{8 \times 2^{1/4}} + \frac{1}{2^{3/4}}(v_0 - \sqrt{v_0 (v_0+\sqrt{2} \ )} \ ) \right) \sqrt{v_0+\sqrt{2}-x} + O((v_0+\sqrt{2}-x)^{3/2}) \nonumber
\eea

The divergence can be interpreted as follows. Assume that near the edge of the support, $\rho_+$ behaves as :
\beq
\rho_+(x) = A\sqrt{x_+-x} + O(z_c-x) \quad {\rm with} \quad x_+ = x_+^0 + \gamma x_+^1 + O(\gamma^2) \quad {\rm and} \quad A = A_0 + \gamma A_1 + O(\gamma^2)
\eeq
In this case we know that $x_+^0=v_0+\sqrt{2}$ and $A_0=\frac{1}{\pi 2^{1/4}}$. Then we have :
\beq
\rho_+(x) = A_0\sqrt{x_+^0-x} + \gamma A_1\sqrt{x_+^0-x} + \gamma \frac{A_0 x_+^1}{2\sqrt{x_+^0-x}}  + O(\gamma^2, \gamma(x_+-x))
\label{double_expansion_gamma_edge}
\eeq
We can identify the terms of order $\frac{1}{\sqrt{x_+^0-x}}$ at first order in $\gamma$ in equations \eqref{expansion_rho1_edge} and \eqref{double_expansion_gamma_edge} to obtain the first order correction for the edge $x_+^{(1)}$ :
\beq
x_+^{(1)} = \sqrt{2}{I_0} = - \sqrt{2} \left( v_0 (v_0+\sqrt{2}) - (v_0+\frac{1}{\sqrt{2}}) \sqrt{v_0^2+\sqrt{2} v_0} + \frac{1}{2} \ln(\sqrt{2}v_0 + 1 + \sqrt{2v_0^2+2\sqrt{2}v_0} \ ) \right)
\eeq
which is a negative, monotonically decreasing function of $v_0$ for any $v_0$. Note that we can also compute $A_1$ but we will not display it here. 

We now focus on the interval $[-v_0-\sqrt{2}, \min(-v_0+\sqrt{2}, v_0-\sqrt{2})]$. Here it is simpler to choose the bound of the integral as $-\infty$. In this case, $\rho_+^1$ only has one contribution :
\bea
\rho_+^1(x) &=& \frac{1}{2\pi \sqrt{(x-v_0)^2-2}} \int_{-v_0-\sqrt{2}}^x \sqrt{2-(x'+v_0)^2} dx' \\
&=& \frac{1}{2\pi \sqrt{(x-v_0)^2-2}} \left( \frac{1}{2} (x+v_0) \sqrt{2-(x+v_0)^2} + \arcsin \left( \frac{x+v_0}{\sqrt{2}} \right) + \frac{\pi}{2} \right)
\label{order1_rhoinf}
\eea
Near the edge $-v_0-\sqrt{2}$, this behaves as :
\beq
\rho_+^1(x) = \frac{(x+v_0+\sqrt{2})^{3/2}}{3\,\pi \, 2^{1/4} \sqrt{v_0^2+\sqrt{2}v_0}} + O((x+v_0+\sqrt{2})^{5/2})
\eeq
Thus we recover the $\frac{3}{2}$ singularity as well as its amplitude, as given in the main text, where the dependence in 
$g$ has been restored.  
\\

\begin{figure}[h!]
    \centering
    \includegraphics[width=0.24\linewidth,trim={1.4cm 17cm 0 1cm},clip]{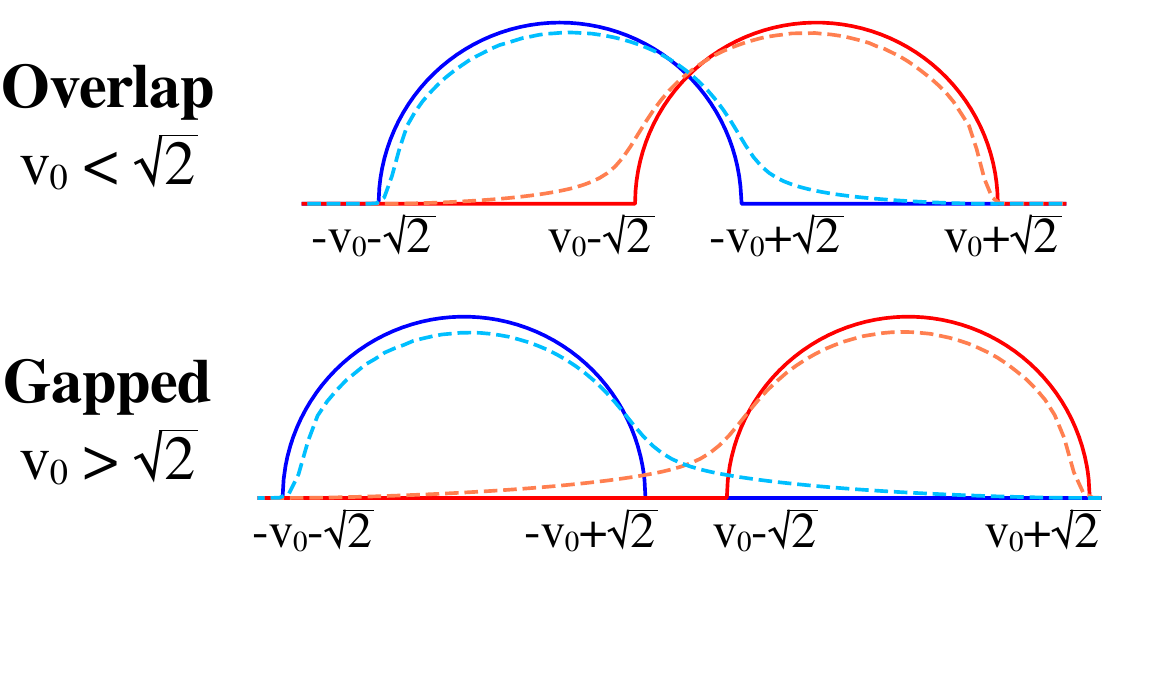}
    \includegraphics[width=0.24\linewidth]{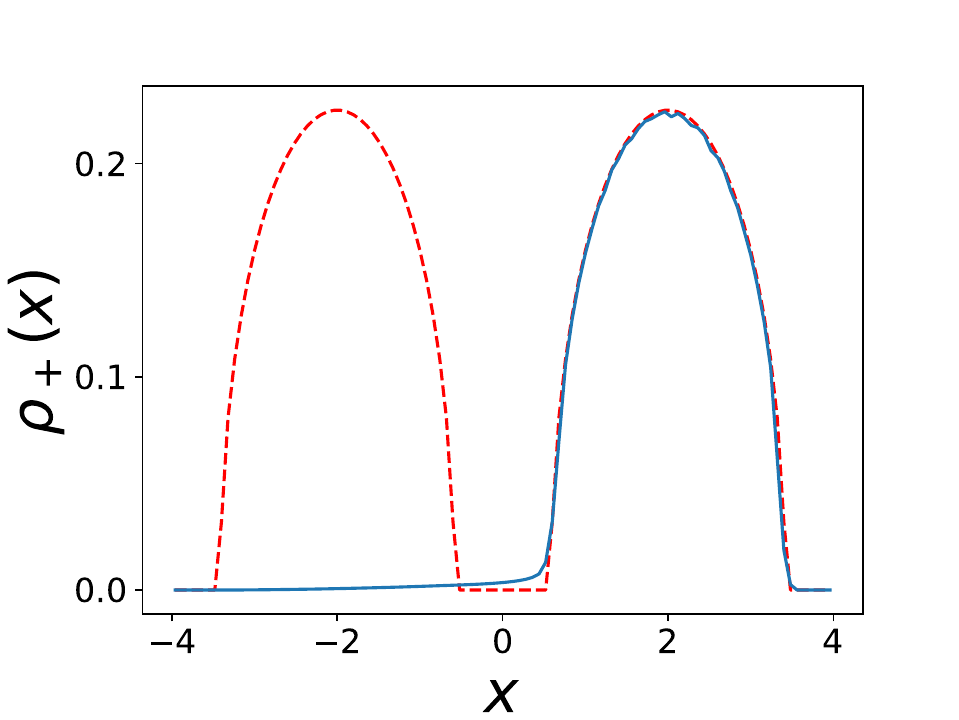}
    \includegraphics[width=0.24\linewidth]{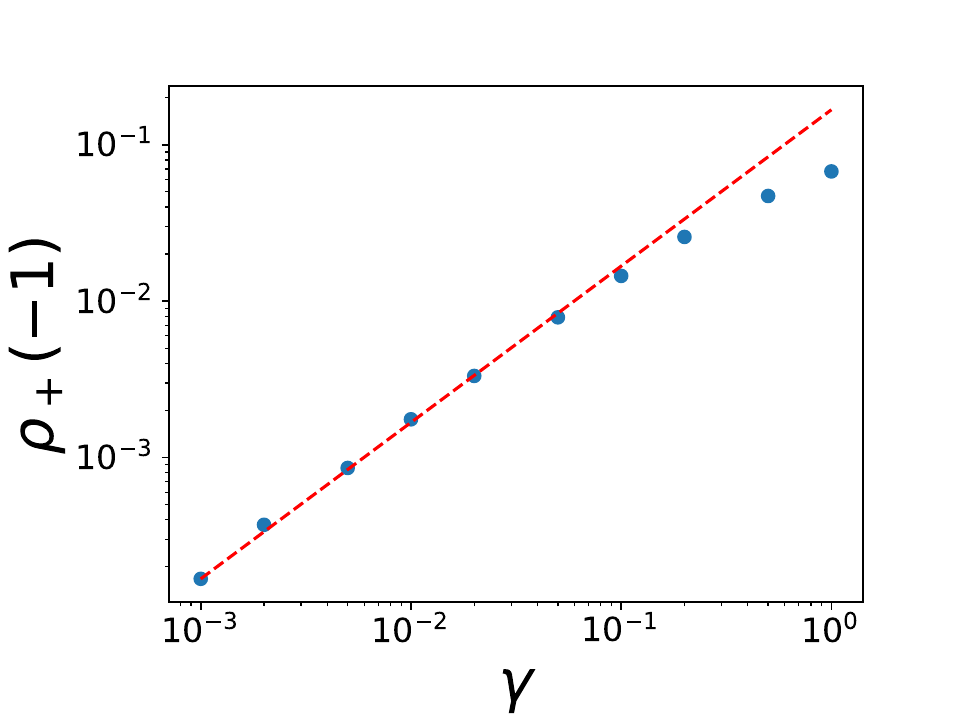}
    \includegraphics[width=0.24\linewidth]{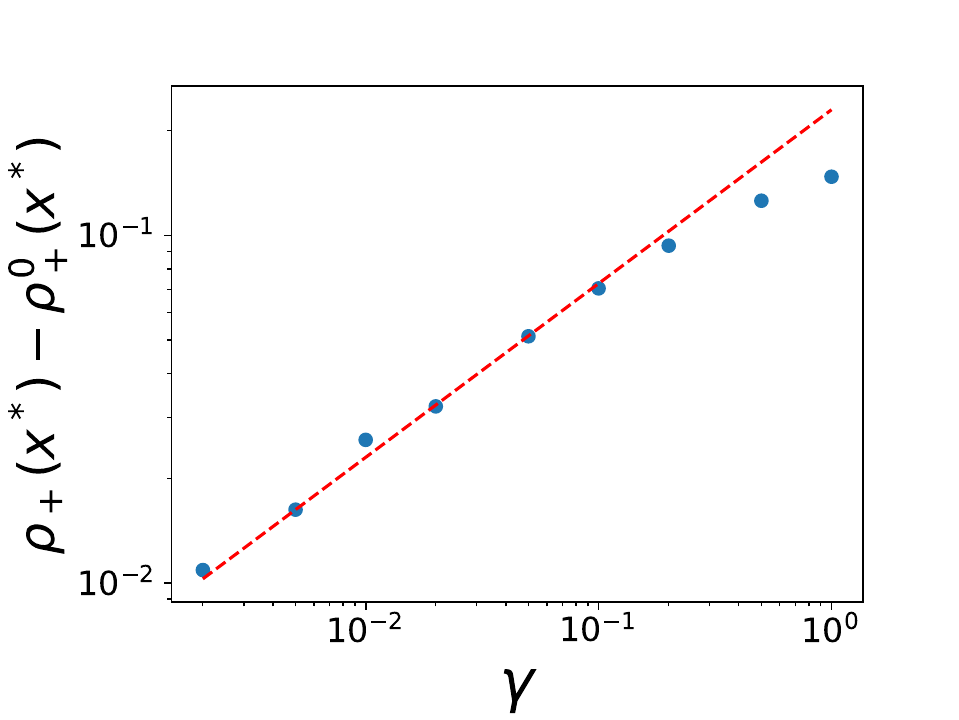}
    \caption{Left : Sketch of the densities $\rho_\pm(x)$ in the two possible cases: overlap ($v_0<\sqrt{2}$) and gapped ($v_0>\sqrt{2}$). Center-left: Numerically measured density $\rho_+(x)$ for $N=100$, $g=1$, $v_0=2$ and $\gamma=0.01$. The dashed line is
    the $\gamma=0$ result for $\rho_s^0(x)$, i.e. two semi-circles. Center-right : $\rho_+(-1)$ as a function of $\gamma$. For small $\gamma$ it is linear in $\gamma$ (the dashed red line as slope $1$). Right : $\rho_+(x^*)$ as a function of $\gamma$, in log scale. The dashed red line as slope $1/2$. This is consistent with the scaling
    form of the boundary layer in \eqref{difference}. }
    \label{2species_gamma0_rhop}
\end{figure}

{\bf Boundary layer form for the density $\rho_+(x)$ at $x= v_0 - \sqrt{2}$}. 

One can perform a similar analysis for the last remaining interval $[- |v_0-\sqrt{2}|, |v_0-\sqrt{2}|]$. We will not
do it in details here. However we will address one point which is still not clear, that is the origin of the divergence 
in the small $\gamma$ expansion in \eqref{order1_rhoinf} at $x^*=v_0-\sqrt{2}$.
Starting from $\gamma=0$ there are two situations either $v_0 < \sqrt{2}$ and there is an overlap
between the two semi-circles, or $v_0 >\sqrt{2}$ and there is a gap between these two
semi-circles. In both cases the small $\gamma$ expansion of $\rho_+(x)$ has a singularity at $x^*=v_0-\sqrt{2}$
as can be seen in \eqref{order1_rhoinf} (and the expansion of $\rho_-(x)$ has a similar singularity at $- x^*=-v_0+\sqrt{2}$). 

For $x$ near $x^*$ we expect a boundary
layer to form at small $\gamma$. To understand better what happens near this point, we thus perform the expansion a bit differently by writing 
\beq \label{eq_bl}
G_+(z) \approx g_+^0(z) + \gamma^\beta f\left(\frac{z-x^*}{\gamma^\nu}\right) \;,
\eeq
where the scaling function $f(\epsilon)$ and the exponents $\beta$ and $\nu$ are determined self-consistently below. By injecting this expansion (\ref{eq_bl}) in Eq.~(\ref{integrated_G+}) one obtains
\beq
\gamma^\beta \sqrt{(z-v_0)^2-2} \ f\left(\frac{z-x^*}{\gamma^\nu}\right) - \gamma^{2\beta} f\left(\frac{z-x^*}{\gamma^\nu}\right)^2 = 
- \gamma \int_z^{+\infty} [G_+(z') + G_+(-z')] dz'
\eeq
Introducing $\epsilon=\frac{z-x^*}{\gamma^{\nu}}$ and keeping only the leading order in $\epsilon$ we get :
\beq
\gamma^{\beta+\frac{\nu}{2}} 2^{3/4} \sqrt{-\epsilon} \ f(\epsilon) - \gamma^{2\beta} [f(\epsilon)]^2 = \gamma C_0 + \gamma^{\beta+1} C_1
\eeq
where $C_0=\int_{x^*}^{+\infty} [g_+^0(z') + g_+^0(-z')] dz'$ and $C_1=\int_{x^*}^{+\infty} [f\left(\frac{z'-x^*}{\gamma^\nu}\right) + f\left(\frac{-z'-x^*}{\gamma^\nu}\right)] dz'$. Since $\beta>0$, the last term on the right hand side can always be neglected at first order in $\gamma$. We then choose $\beta$ and $\nu$ such that the remaining terms are all of the same order. This leads to $\beta=\frac{1}{2}$ and $\nu=1$. The plot in Fig. \ref{2species_gamma0_rhop} shows that the leading correction near $x^*$ is indeed of order $\sqrt{\gamma}$, which explains the divergence when expanding in integer powers of $\gamma$. This leads us to a simple second degree equation :
\beq
f(\epsilon)^2 - 2^{3/4} \sqrt{-\epsilon} f(\epsilon) + C_0 = 0
\eeq
We choose the solution which gives back the $\frac{1}{\sqrt{-\epsilon}}$ behavior for large $\epsilon \to - \infty$ :
\beq
f(\epsilon) = \frac{\sqrt{-\epsilon}}{2^{1/4}} - \sqrt{-C_0-\frac{\epsilon}{\sqrt{2}}}
\eeq
Taking the imaginary part we obtain the density which takes the scaling form for $x- x^*= O(\gamma)$ (inside the boundary layer)
\beq \label{difference} 
\rho_+(x) - \rho_+^0(x) \simeq \sqrt{\gamma} \,  \tilde{\rho}_+\left(\frac{x-x^*}{\gamma}\right) + O(\gamma) \quad , \quad \tilde{\rho}_+(\epsilon)  = -\frac{1}{\pi} {\rm Im} f(\epsilon)
\eeq
where $\rho_+^0(x)$ is the semi-circle density for the $+$ species at $\gamma=0$, see \eqref{dens00}. 
In the case where $v_0>\sqrt{2}$ (the two semi-circles do not overlap), we obtain $C_0=-I_s + i\frac{\pi}{2}$ with $I_s=-I_0-\tilde{I}_1(v_0-\sqrt{2})>0$ $\forall v_0>\sqrt{2}$. When $v_0<\sqrt{2}$, $C_0=\alpha -i I_3$ with $\alpha=v_0(\sqrt{2}-v_0)>0$ and $I_3 = (v_0 - \frac{1}{\sqrt{2}})\sqrt{v_0(\sqrt{2}-v_0)} + \frac{1}{2} \arcsin(\sqrt{2}v_0-1) + \frac{\pi}{4}>0$ $\forall v_0<\sqrt{2}$. This leads to :
\bea
\tilde{\rho}_+(\epsilon) &=& \frac{1}{\pi\sqrt{2}} \sqrt{\sqrt{\left(I_s-\frac{\epsilon}{\sqrt{2}}\right)^2+\frac{\pi^2}{4}} - I_s + \frac{\epsilon}{\sqrt{2}}} - \frac{\sqrt{(\epsilon)_+}}{\pi 2^{1/4}} \quad {\rm for} \ v_0>\sqrt{2} \quad \text{gap case}  \label{rhotilde_gap}\\
&=& \frac{1}{\pi\sqrt{2}} \sqrt{\sqrt{\left(\alpha+\frac{\epsilon}{\sqrt{2}}\right)^2+I_3^2} + \alpha + \frac{\epsilon}{\sqrt{2}}} - \frac{\sqrt{(\epsilon)_+}}{\pi 2^{1/4}} \quad {\rm for} \ v_0<\sqrt{2} \quad \text{overlap case} \label{rhotilde_overlap}
\eea

For $\epsilon \to + \infty$, one can check that 
the first expression behaves as $- \frac{I_s}{2^{3/4} \pi \sqrt{\epsilon}}$ (i.e. with a negative prefactor)
and the second as $\frac{\alpha}{2^{3/4} \pi \sqrt{\epsilon}}$ (i.e with a positive prefactor). 
For $\epsilon \to -\infty$ the first expressions behaves as $\frac{1}{2 \times 2^{3/4}  \sqrt{|\epsilon|}}$ 
and the second as $\frac{I_3}{2^{3/4} \pi \sqrt{|\epsilon|}}$.
The difference $\rho_+(x) - \rho_+^0(x)$ in \eqref{difference} 
should match with $\gamma \rho_+^1(x)$, 
which we recall has a divergence at $x=v_0-\sqrt{2}$, see e.g.  \eqref{order1_rhosup}. More precisely, the above limits should
match $\sqrt{\gamma} \rho_+^1(v_0-\sqrt{2}+\sqrt{\gamma} \epsilon)$ in the limit $\epsilon \to 0$. However, since we only computed $\rho_+^1(x)$ on $[-v_0-\sqrt{2}, -|v_0-\sqrt{2}|]$ and $[|v_0-\sqrt{2}|, v_0+\sqrt{2}]$, we can only check this matching on the right for $v_0>\sqrt{2}$ and on the left for $v_0<\sqrt{2}$. Using equations \eqref{order1_rhosup} and \eqref{order1_rhoinf} respectively, we see that the limits indeed match in these two cases.

We have compared these predictions to a numerical simulation. The plots shown in 
Fig.\ref{gamma_expansion_rho1inf} show that for $\gamma=0.01$ the agreement with the simulations is quite good (i) in the bulk
and (ii) in the boundary layer, both in the case with an overlap ($v_0 < \sqrt{2}$) and in the case with a gap ($v_0 > \sqrt{2}$).
The sizeable difference between the prediction and the simulation just at the right of $x^*$ is probably due to finite $N$ effects.
\\

\begin{figure}[h!]
    \centering
    \includegraphics[width=0.24\linewidth]{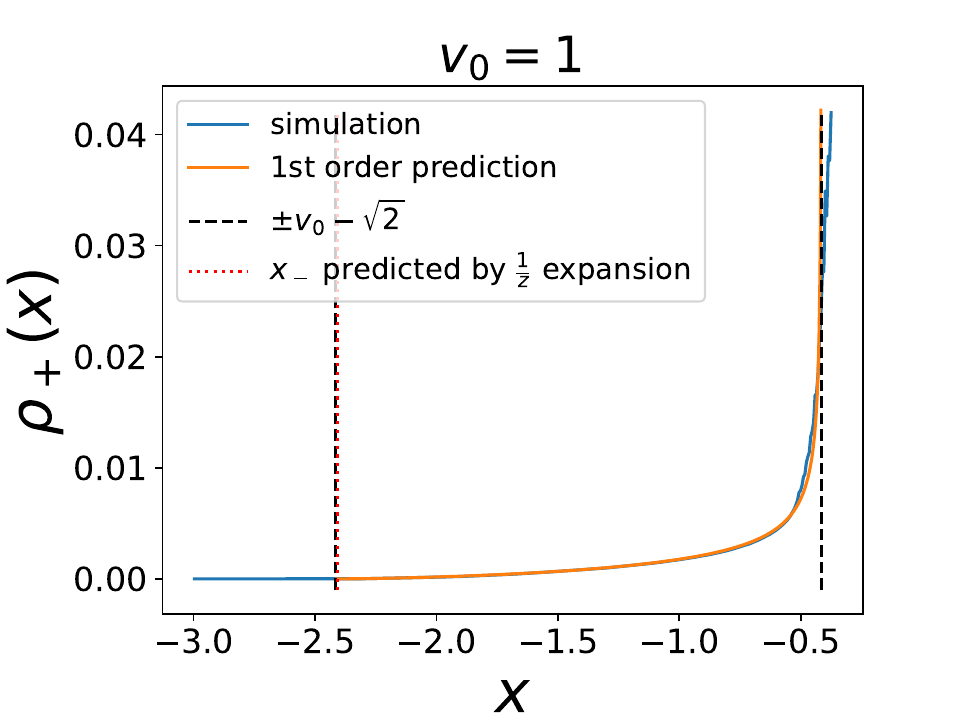}
    \includegraphics[width=0.24\linewidth]{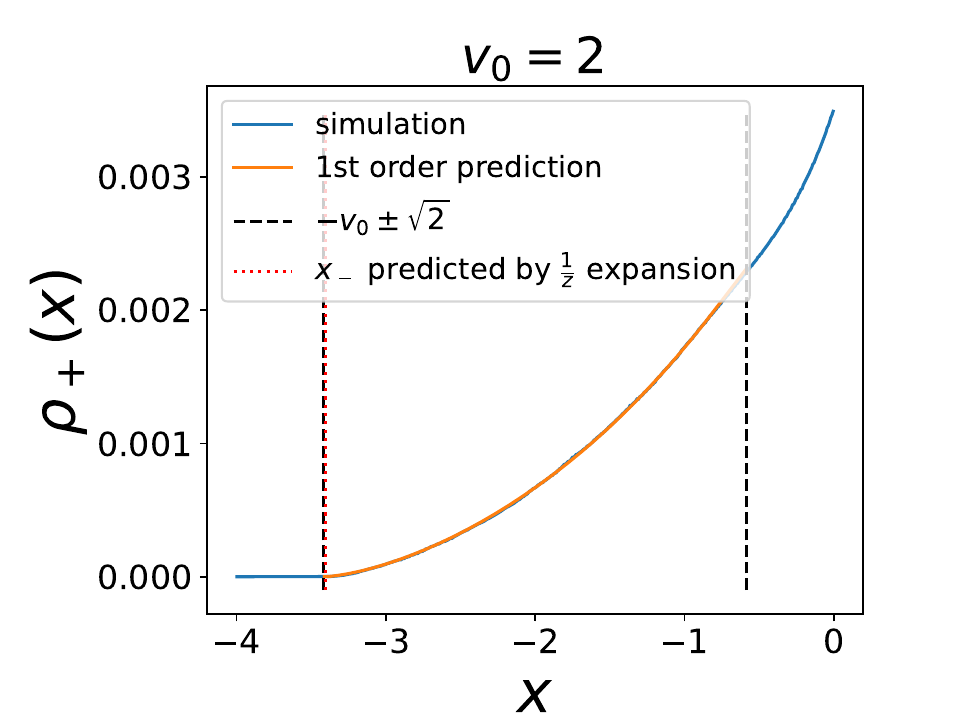}
    \includegraphics[width=0.24\linewidth]{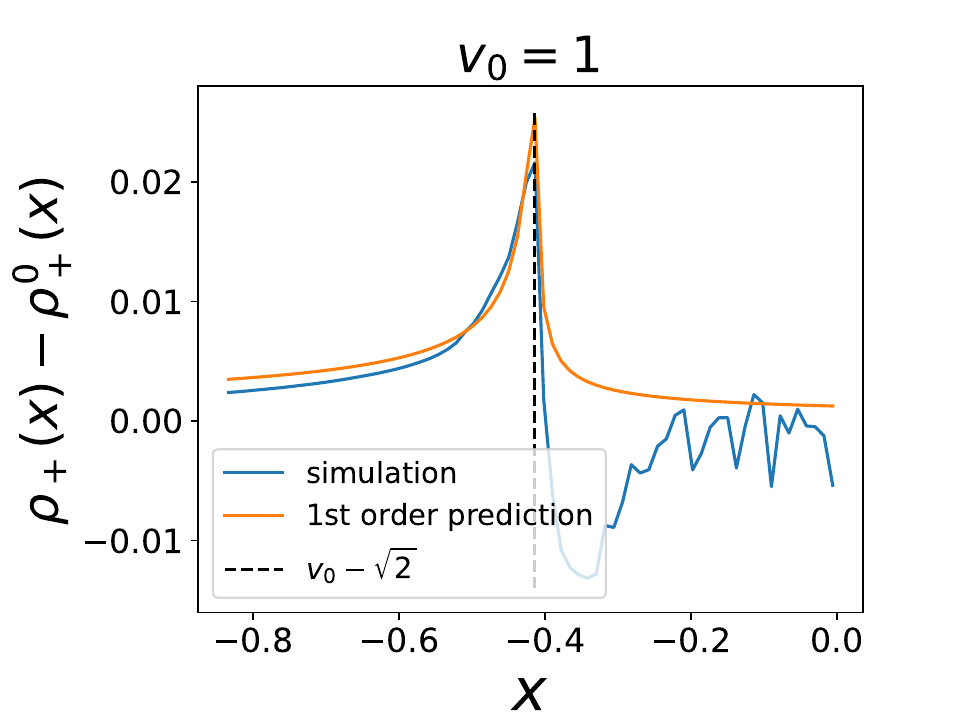}
    \includegraphics[width=0.24\linewidth]{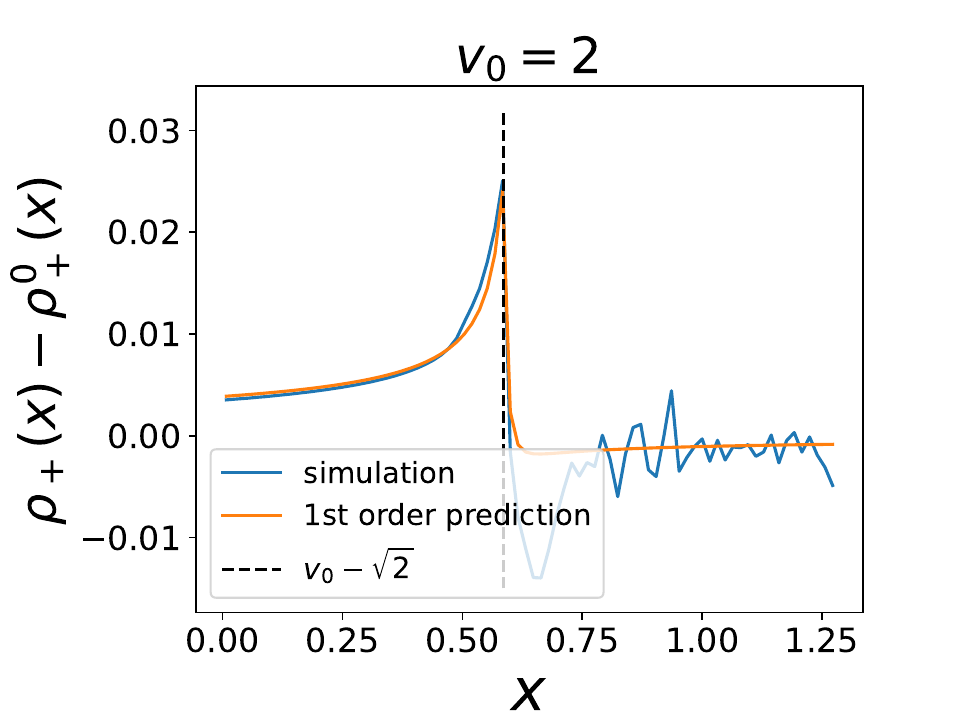}
    \caption{Left : Bulk perturbation theory in $\gamma$, comparison of the first order predictions (order $\gamma$) for the density $\rho_+(x)$ on the interval $[-v_0-\sqrt{2},\min(-v_0+\sqrt{2},v_0-\sqrt{2})]$, see formula \eqref{order1_rhoinf}, 
    with the simulations results for $N=100$, for $\gamma=0.01$ and $v_0=1$ (left) and $2$ (right). Right : Boundary layer 
    form around $x^*$, 
    comparison of the density $\rho_+$ near $x^*=v_0-\sqrt{2}$ after subtracting the 0th order in $\gamma$, with the prediction at order $\sqrt{\gamma}$ given in Eq. 
    \eqref{difference}, for $\gamma=0.01$ and $v_0=1$ (left) when there is an overlap, see \eqref{rhotilde_overlap},
    and $v_0=2$ (right) when there is a gap, see \eqref{rhotilde_gap}.}
    \label{gamma_expansion_rho1inf}
\end{figure}

{\bf Perturbation theory to order $O(\gamma^2)$}. It is possible to carry out the perturbation theory to the next order
in $\gamma$. Since the calculations are quite heavy we will not reproduce them here. Let us give one of the main results.
First we find that the expansion performed above remains fully consistent to the next order. Second, we 
we find that the second order correction $x_+^{(2)}$ of the upper edge reads
\beq \label{edgesecondorder} 
x_+^{(2)} = \sqrt{2}I_2 - I_0 \left(\frac{7}{8\sqrt{2}} I_0 + 2 (v_0 - \sqrt{v_0(v_0+\sqrt{2})} \ )\right)
\eeq
where
\bea
&& I_2 = - \int_{v_0+\sqrt{2}}^{+\infty} dx' \left( \frac{1}{\sqrt{(x'-v_0)^2-2}} + \frac{1}{\sqrt{(x'+v_0)^2-2}} \right) \tilde{I_0}(x')  \\
&&\begin{split} \tilde{I}_0(x) = & -v_0 x - \frac{1}{4} \left( (x-v_0) \sqrt{(x-v_0)^2-2} - (x+v_0) \sqrt{(x+v_0)^2-2} \right) \\ & + \frac{1}{2} \left( \ln(x-v_0+\sqrt{(x-v_0)^2-2} \ ) - \ln(x+v_0+\sqrt{(x+v_0)^2-2} \ ) \right) \end{split}
\eea
with $\tilde{I}_0(v_0+\sqrt{2})=I_0$. The integral $I_2$ has to be computed numerically.

In Fig. \ref{gamma_expansion_zc_order2} we see that taking into account this second order correction $x_+^{(2)}$ significantly improves the agreement with simulations for $\gamma=0.1$, and that the residual error is indeed of the order $O(\gamma^3)$. For $\gamma=0.01$, the numerical errors are too high to see a real improvement.

\begin{figure}[h!]
    \centering
    \includegraphics[width=0.32\linewidth]{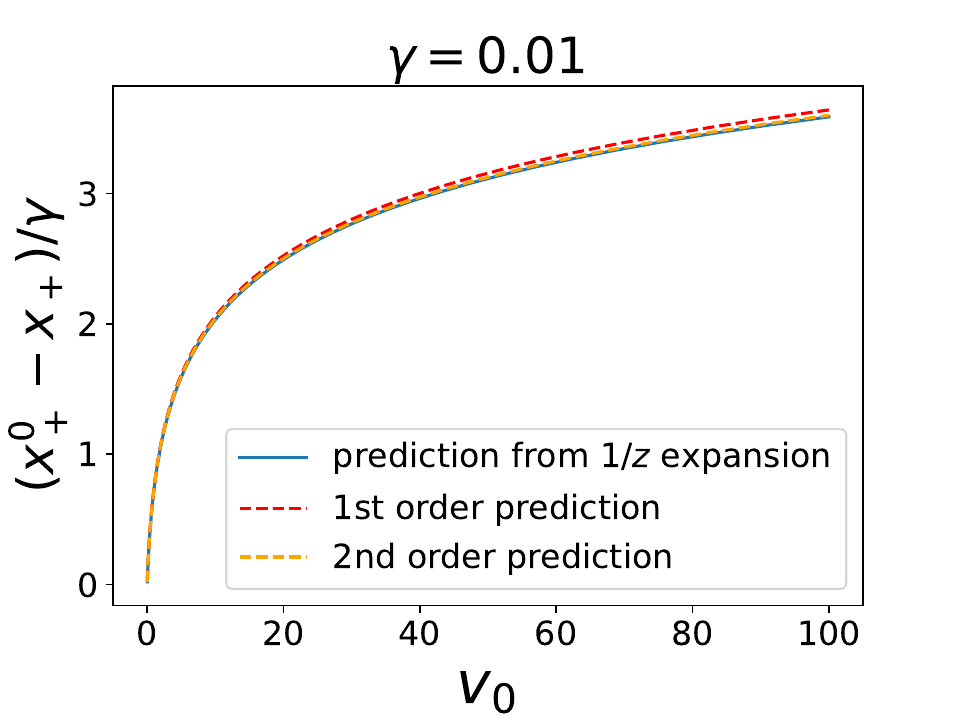}
    \includegraphics[width=0.32\linewidth]{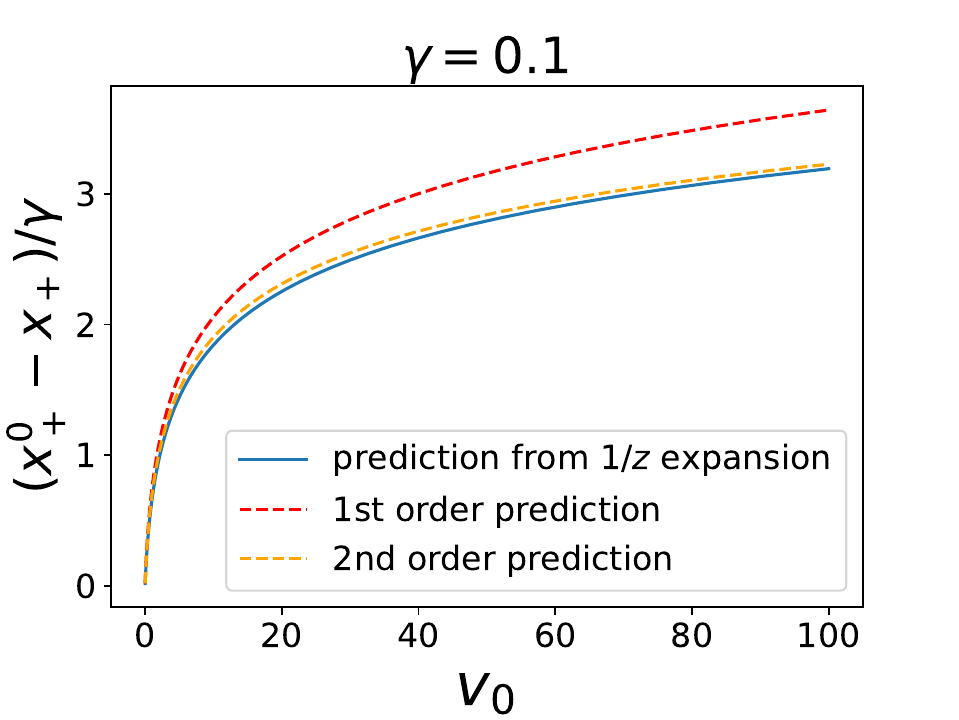}
    \caption{Comparison of the correction to $x_+$ up to second order $O(\gamma^2)$ computed in \eqref{edgesecondorder}
    with the numerical prediction using the $\frac{1}{z}$ expansion, for $\gamma=0.01$ (left) and $0.1$ (right), with $v_0=1$, $g=1$ and $\lambda=1$.}
    \label{gamma_expansion_zc_order2}
\end{figure}

{\bf Remark}: we have shown that $\rho_+(x) \sim (x-x_-)^{3/2}$ both from a perturbation approach for small $\gamma$ as well as, numerically, from the behavior of the moments (see Section \ref{expansion_moments}). Since $\rho_-(x) \sim (x-x_-)^{1/2}$, this exponent $3/2$ is again 
consistent with what seems to be a general property that the difference between the two exponents is $1$, as shown in the noninteracting case in Eq. (\ref{ratio}).

\section{Details of numerical simulations and further results}
\label{simulation_details}

In this section we explain a few details of the numerical methods for model I and II and the limiting model $g=0^+$.
We also present additional numerical results that support the discussions/claims in the text.

\subsection{Model I and II: numerical details}
For model I and II each plot of the density was obtained using a single run of the stochastic dynamics given by Eq. \eqref{modelsupmat}, with $\lambda=1$. We wait for the system to reach the stationary state and then we collect the histogram of positions for all particles over a time window of the order of $10^4$. The time-steps used are typically of the order of $10^{-3}$ for $N=100$. For smaller values of $N$ we use a larger time-step and collect statistics over a larger time window.

\subsection{Model II: further results}

\begin{figure}[h!]
    \centering
    \includegraphics[width=0.45\linewidth]{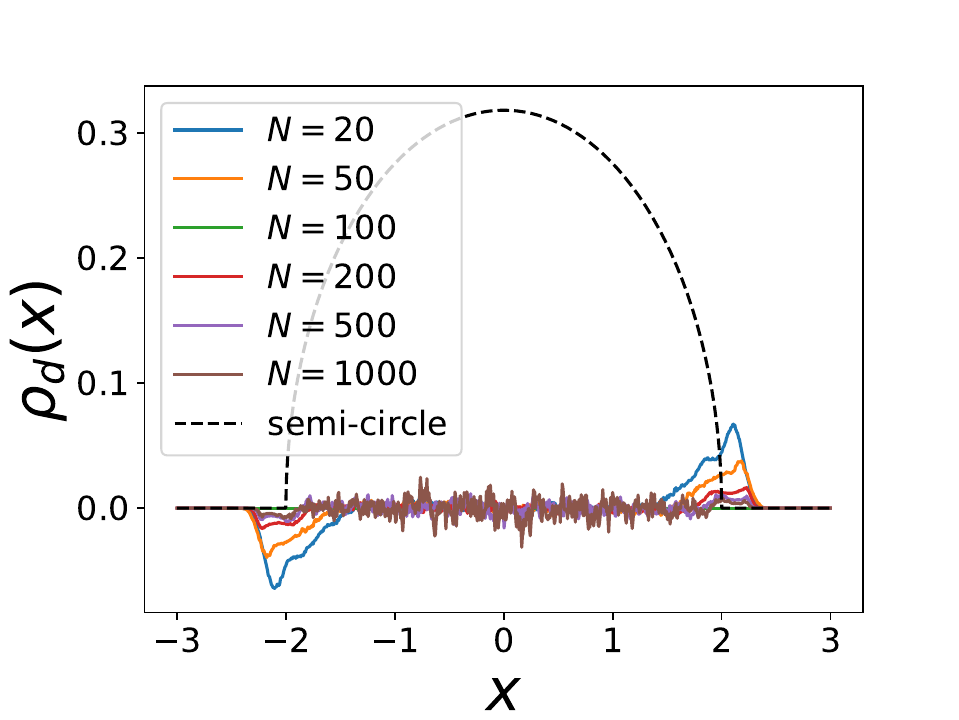}
    \includegraphics[width=0.45\linewidth]{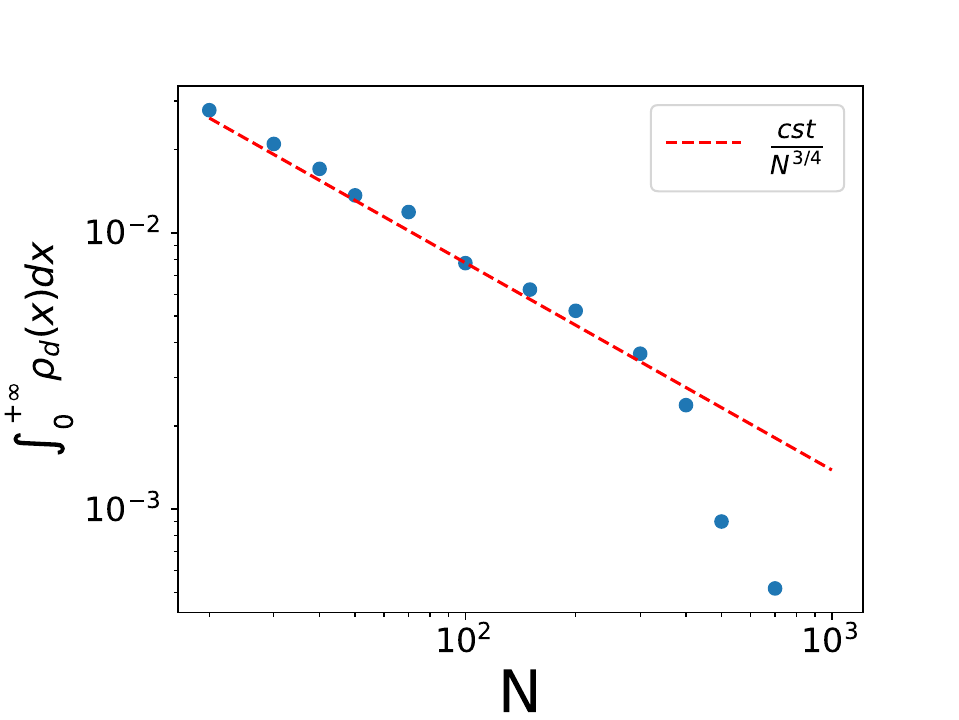}
    \caption{Left: The measured density $\rho_d(x)=\rho_+(x)- \rho_-(x)$ for $\lambda=1$, $v_0=1$, $\gamma=1$ and $g=1$ for different values of $N$.
    Right: the total integral $\int_0^{+\infty} \rho_d(x)\,dx$ as a function of $N$, for the same set of parameters. It appears to decrease as $N^{-3/4}$.}
    \label{plot_rhod}
\end{figure}

Let us now justify the claim in the text that for model II the stationary density $\rho_d(x)$ vanishes at large $N$. Fig.~\ref{plot_rhod} shows the behavior of $\rho_d(x)$ as $N$ increases. For any value of $N$, $\rho_d(x)$ averages to zero in the bulk of the distribution, meaning that the $+$ and $-$ particles are well mixed. This is because in model II the particles cannot cross and change their velocities randomly ($\gamma>0$).
There is however an accumulation of $+$ (resp. $-$) particles at the right (resp. left) edge, which disappears as $N$ is increased. Surprisingly the exponent with which those wings disappear does not seem to match the exponent $1/2$ observed for $\rho_s(x)$ (see Fig. \ref{plot_model2} in the text). Instead it seems to be closer to $3/4$.

\begin{figure}[h!]
    \centering
    \includegraphics[width=0.45\linewidth]{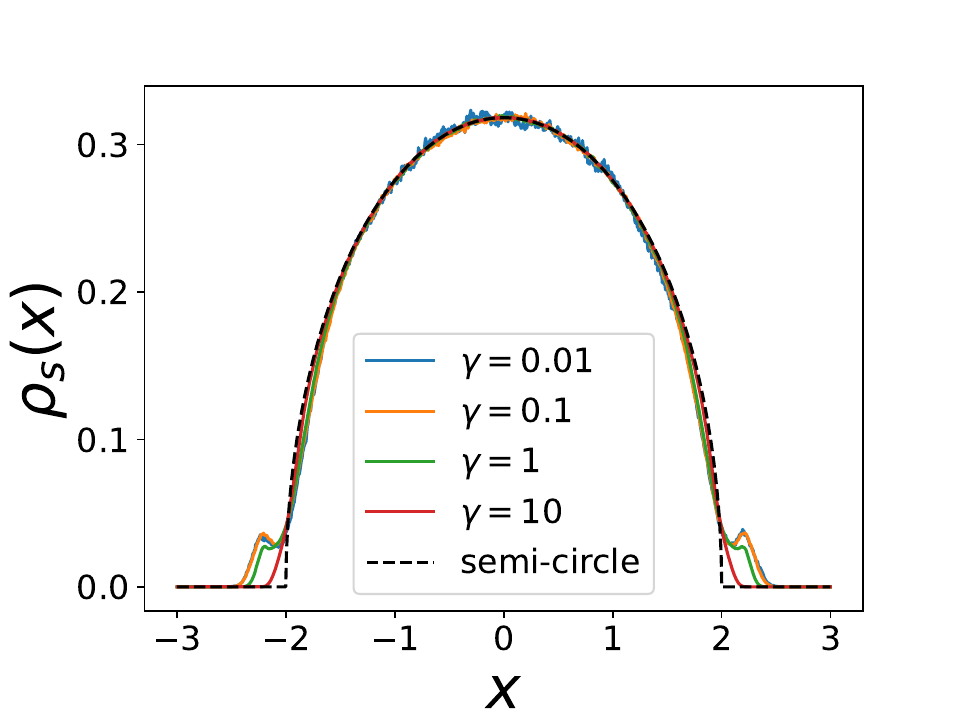}
    \includegraphics[width=0.45\linewidth]{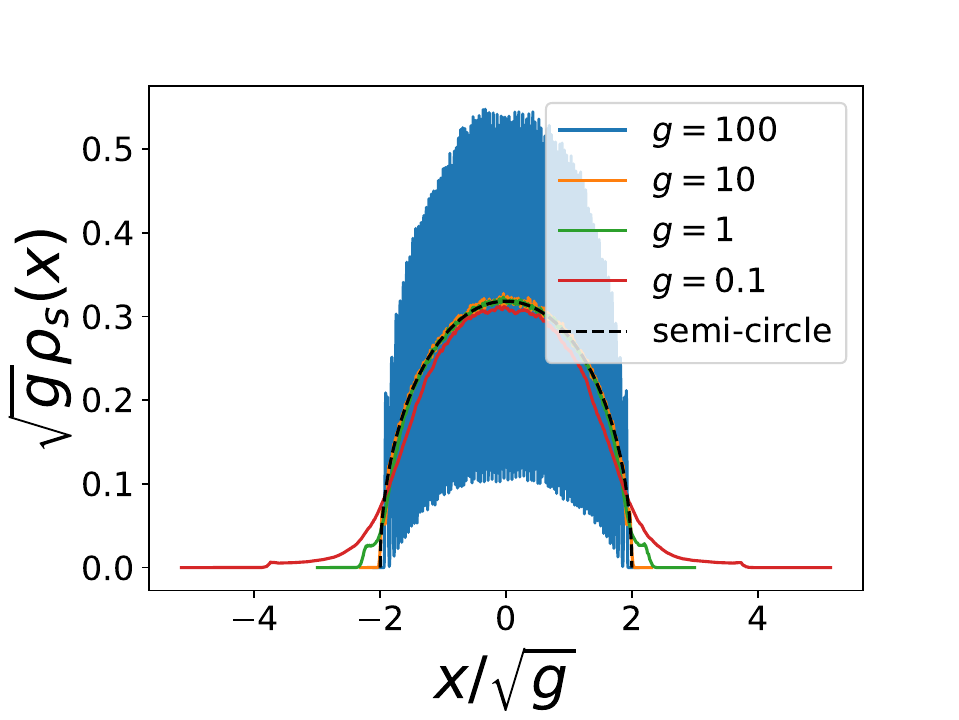}
    \caption{Left: Total density of particles $\rho_s(x)$ for different values of $\gamma$ for $N=100$ and all other parameters set to 1. Increasing $\gamma$ decreases the size of the wings. Right: Scaled density $\sqrt{g} \rho_s(\sqrt{g} x)$ (defined as in \eqref{fsc} in the text) for model II, for different values of $g$ for $N=100$ and all other parameters set to 1. One can see the convergence to a semi-circle in the scaling regime $1/N \ll g \ll N$. 
    Decreasing $g$ increases the size of the wings, while when $g$ is increased much above $Nv_0^2$ the particles become localized leading to oscillations in the density.}
    \label{plot_model2_spikes}
\end{figure}

We have also studied for model II the effect of varying $\gamma$ on the distribution $\rho_s(x)$. 
As mentioned in the text this effect is very weak. This can be seen in
the left panel in Fig. \ref{plot_model2_spikes}. As discussed in the text, for $N\to+\infty$, $\rho_s(x)$ seems to always converge to the semi-circle. 
Therefore it becomes independent of $\gamma$ in this limit. For finite $N$ however we observe a small effect of $\gamma$ on the edges of the distribution: the wings seem to disappear as $\gamma$ is increased.

Finally, we have also studied the effect of varying the interaction parameter $g$ on the distribution $\rho_s(x)$ for model II. 
This was schematically depicted in Fig. \ref{phase_diagrams} in the text where we identified three regimes:  
(i) $g/v_0^2 \leq 1/N$, (ii) $1/N \leq g/v_0^2 \leq N$ and (iii) $N \leq g/v_0^2$. 
Here we show the results of the simulations in Fig. \ref{plot_model2_spikes}
for various values of $g/v_0^2$ in the intermediate regime (ii). The data confirms the
scaling form proposed in the text in \eqref{fsc}, where the scaling function is a semi-circle ($f=f_{sc}$).
Furthermore, we can see that there are wings for intermediate values of $g$, which is
an effect of finite $N$, as can be seen in Fig. \ref{plot_model2} in the text. In addition, as $g$ becomes very large, i.e. in regime (iii), the
density is not smooth anymore and develops spikes. 
This is because the particles become more localized, leading to oscillations in the density.

\subsection{Definition of the limiting model $g=0^+$}
For model II we noticed that decreasing $g$ does not yield back the independent particles model. Indeed even for very small values of $g$,
the interaction term still diverges at coinciding points, which leads to a hard-core repulsion between particles. This has a strong influence even at the level of the one-particle density. In order to better understand this limit we introduced a discrete-time model in which the only type of interaction between particles is a non-intersecting condition. For infinitely small time-steps this should correspond exactly to the limit $g=0^+$ of model II (see Fig. \ref{g0appdx} left panel for a numerical check of this statement). We will take $\lambda=1$ and $v_0=1$ so that the only remaining parameters are the total number of particles $N$ and the time-step $dt$.

For this new model, the dynamics cannot be simulated in the same way as in the case $g \neq 0$, since the interactions between particles cannot be described by a conventional force. Instead we consider that particles which collide form a point-like cluster for which we will describe the dynamics below. It is more convenient to define the whole system as a set of clusters (possibly containing a single particle), each of them characterized by :
\begin{itemize}[noitemsep,nolistsep]
    \item Its position $x$
    \item The vector $\vec{\sigma}$ containing the spins of all the particles in the cluster, ordered from left to right
\end{itemize}
Between two collisions, such a cluster follows the equation (obtained by summing the equations of motion for all the particles in the cluster, which have the same position, and dividing by the size $n$ of the cluster) :
\beq
\dot{x} = - x + \frac{1}{n} \sum_i \sigma_i \;.
\eeq

We start with only clusters of size $1$ distributed uniformly on the interval $[-1,1]$ (for $g=0$ particles are confined to this interval), which we sort according to their position ($x_k<x_{k+1} \ \forall k$). Then at each time-step we perform the following steps in order :
\begin{itemize}[noitemsep,nolistsep]
    \item Flip each spin independently with probability $\gamma dt$
    \item For each cluster containing more than one particle for which at least one spin has flipped, determine if it breaks into several clusters (see below)
    \item For each cluster compute its speed $v=- x + \frac{1}{n} \sum_i \sigma_i$ and its new position $x^{new}=x+v dt$
    \item For $k=1...N_{clusters}$: if $x_k^{new}>x_{k+1}^{new}$ (it means there has been a collision), then:
    \begin{itemize}
        \item compute the collision time $dt_{col}=\frac{x_{k+1}-x_{k}}{v_{k}-v_{k+1}}$
        \item compute the position of the collision $x'=x_k+v_k dt_{col}$
        \item create a new cluster with $\vec{\sigma}$ being the concatenation of the $\vec{\sigma}$'s of the two clusters and :
        \begin{itemize}
            \item $v=-x'+\frac{1}{n} \sum_i \sigma_i$
            \item $x^{new}=x'+v(dt-dt_{col})$
            \item $x=x'-v dt_{col}$ (position of the cluster before the update if it had already existed - useful in case of a new collision at the same step)
        \end{itemize}
    \end{itemize}
    \item Repeat the previous step until there are no collisions
    \item Update the position of all clusters $x\leftarrow x^{new}$
\end{itemize}

To determine if a cluster breaks, we perform the following :
\begin{itemize}[noitemsep,nolistsep]
    \item Decompose the cluster into a list of mini-clusters for which all $+$ particles are on the left and all $-$ particles on the right
    \item Compute for each mini-cluster $k$ its number of particles $n_k$ and its average spin $\bar{\sigma}_k=\frac{\sum_i \sigma_i}{n_k}$
    \item For $k=1...N_{miniclusters}$: if $\bar{\sigma}_k>\bar{\sigma}_{k+1}$, merge mini-clusters $k$ and $k+1$
    \item Repeat until the $\bar{\sigma_k}$'s are ordered\;.
\end{itemize}
The idea is that a cluster breaks if it can be divided into smaller clusters which have individual speeds driving them apart from each other.

Another way to obtain the same decomposition is to use the following characterization: a list of spins $\vec \sigma$ of size $n$ 
forms a cluster iff the running average 
$S_k=\frac{1}{k} \sum_{i=1}^k \sigma_i$ reaches its minimum at the end of the list, i.e. $k_{\min}=n$. Indeed in this case, if we try to divide the list in two at any position, the part on the left will have a larger speed than the part on the right, so they will form a cluster. If the global minimum is reached before, 
i.e. if $k_{\min}<n$, and if cut the list just after this minimum, the part on the left will have a smaller speed than the part on the right, hence will be a genuine cluster, and the two parts will separate. One then repeats the operation for the part on the right. 

We can therefore apply the following algorithm:
\begin{itemize}[noitemsep,nolistsep]
    \item Compute the running average at each position.
    \item Find the global minimum.
    \item The part on the left of this minimum (including the minimum) forms an independent cluster.
    \item Repeat the process removing the independent cluster from the computation of the running average, until the global minimum is at the end of the list.
\end{itemize}

This method is faster than the other one if the clusters do not decompose too often (i.e., when $\gamma$ is not too large), as we have to go through the list only once if there is a single cluster (which is not the case with the previous algorithm).

As for model I and II, the quantities of interest (see below) are computed by running the dynamics and averaging over a large time window. With this code, the simulations are much faster than for small non-zero values of $g$ with the usual code, since we can use larger time-steps (and if there are a lot of large clusters, e.g. for small $\gamma$, there are few positions to update). The results obtained with this model are very similar to what we observe in model II with small values of $g$ (but we can go to larger values of $N$ more easily), see Fig. \ref{g0appdx} left panel. In fact as $g\to 0^+$ the density in model II seems to converge to the density in this effective model. 

Finally, we also studied the double limit $g=0^+$ and $\gamma=0^+$. In this case the results were simply obtained by drawing a random list of $N$ independent spins, each one being $1$ or $-1$ with equal probability, and decomposing this list into clusters as described above. Since 
by definition the clusters cannot cross if there are no spin flips, each
of these clusters then reaches its equilibrium position on time scales much smaller than $1/(N \gamma)$ (where here we assume $\gamma \ll 1/N$). This
equilibrium position is given by
$x_{eq} = \frac{1}{n} \sum_i \sigma_i$,
where the sum is over the spins in the cluster of size $n$. By definition these positions form a strictly increasing sequence.
The quantities of interest are then obtained by averaging over a large number of such realisations. The
argument is again that the clusters will spend much time near their equilibrium positions before a flip occurs.
Hence a random sampling of the spins is appropriate. In all cases the results are very close to what we obtain for small values of $\gamma$ using the dynamics described above, suggesting that the $\gamma \to 0$ limit is well defined for the $g=0^+$ model.

\subsection{Results for the $g=0^+$ model}
\begin{figure}[h!]
    \centering
    \includegraphics[width=0.32\linewidth]{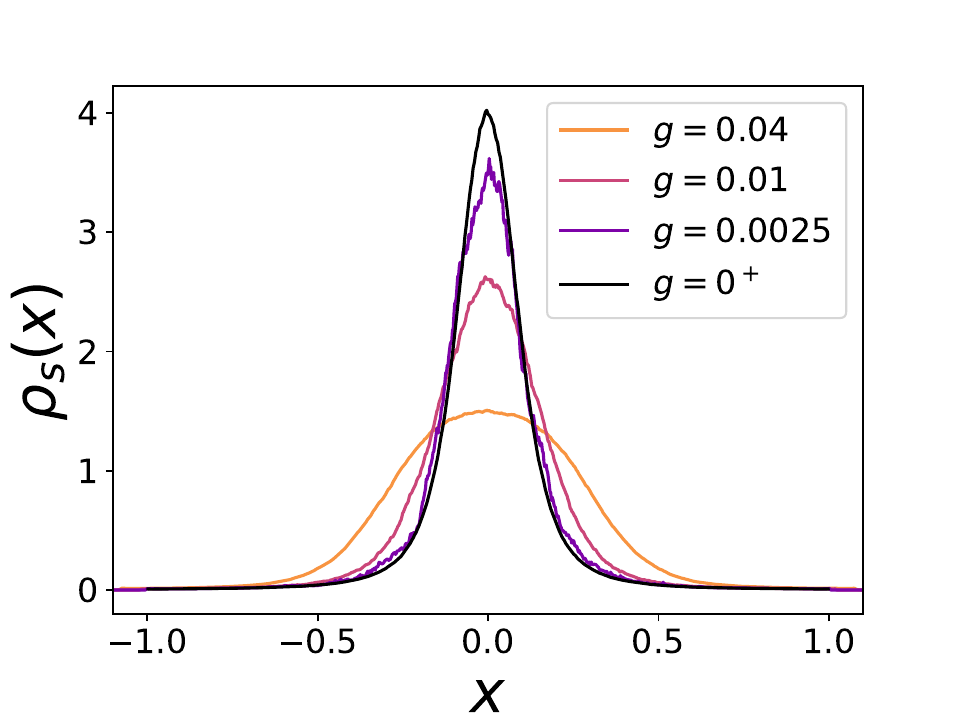}
    \includegraphics[width=0.32\linewidth]{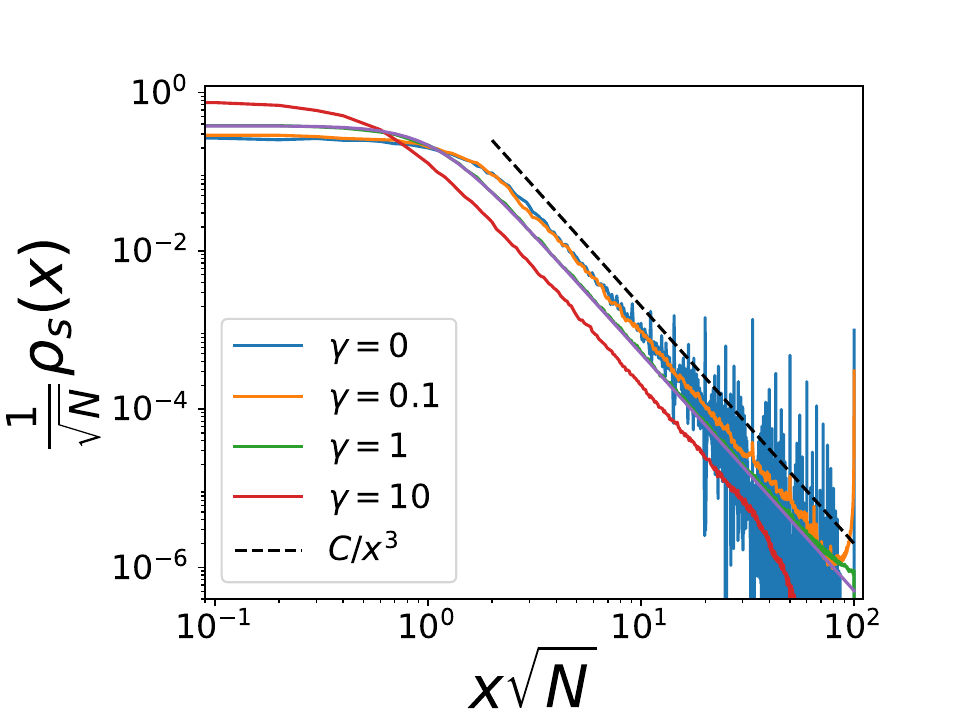}
    \includegraphics[width=0.32\linewidth]{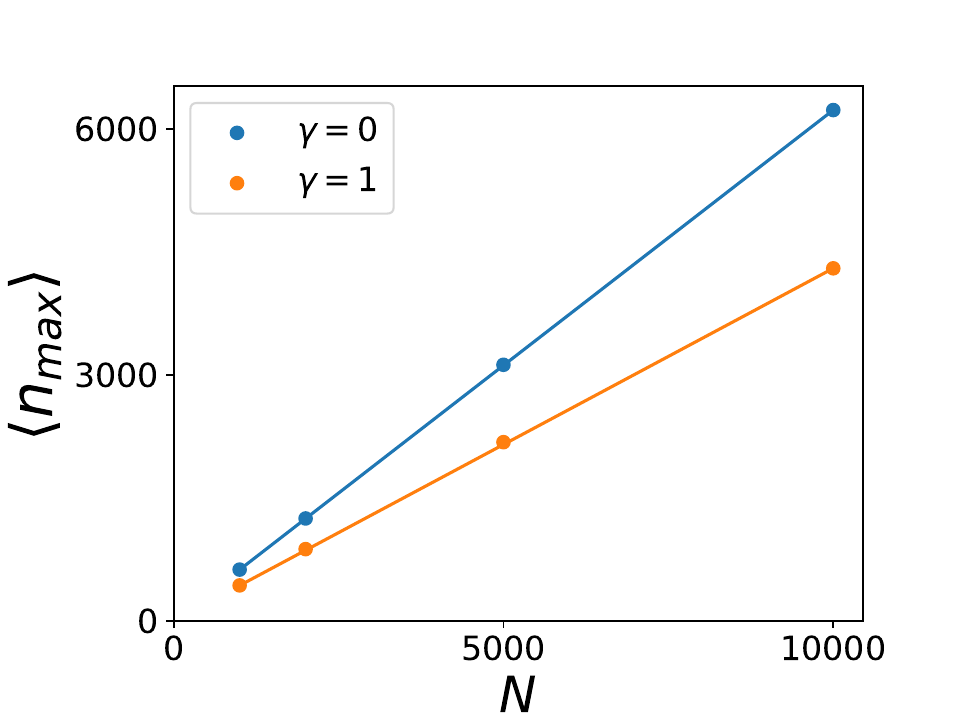}
    \caption{Left: Density $\rho_s(x)$ for small values of $g$, with $N=100$ and all other parameters set to 1. The density converges to the one of the limiting model $g=0^+$ as $g$ is decreased towards zero. Center: Rescaled particle density in the $g=0^+$ model for $N=10000$ and different values of $\gamma$, all other parameters being set to $1$ (same figure as in Fig. \ref{g0figs} right panel given in the main text in log-log scale). For all values of $\gamma$ the tail clearly decays as $x^{-3}$. Right: Average size of the largest cluster as a function of $N$ for $\gamma=0$ and $\gamma=1$. In both cases it behaves linearly in $N$.}
    \label{g0appdx}
\end{figure}

The first observable of interest in the model $g=0^+$ is $\rho_s(x)$. In the text we proposed that it takes a scaling form given in \eqref{phi} with a scaling function $\phi(z)$. This property is tested numerically in Fig. \ref{g0figs}. In the middle panel, it is shown for $\gamma=1$ and
we see that the scaling works very well as a function of $N$ for large $N$. However the right panel in Fig. \ref{g0figs} shows that the
scaling function $\phi(z)$ depends on $\gamma$. We have characterized numerically the tail of this scaling function
$\phi(z)$. It is shown in Fig. \ref{g0figs} (center panel) in log-log scale. It seems that this function has a power law tail
$\phi(z) \sim |z|^{-a}$ with $a \approx 3$ independently of $\gamma$. Note that the density has a step at $x=\pm v_0$ (edges of the support)
of height $=O(1/N)$. 
Furthermore it seems that the scaling function
$\phi(z)$ has a well defined limit $\gamma=0^+$, as can be seen in Fig. \ref{g0figs}. 

Another important observable is the distribution of cluster sizes in the stationary state. 
As discussed in the text, this distribution becomes broader as $\gamma$ increases. 
We first define ${\cal N}_n(t)$ the number of clusters of size $n$ at time $t$ and
the total number of clusters at time $t$, ${\cal N}(t)= \sum_{n=1}^N {\cal N}_n(t) \leq N$.
Then we measure
\be   
p(n) = \frac{ \sum_t {\cal N}_n(t)}{\sum_t {\cal N}(t)} \;,
\ee 
where the summation is over the discrete time steps of the simulation. In the stationary state  we expect that this is also
equal to $p(n)= \langle {\cal N}_n \rangle/\langle{ \cal N}\rangle$, where the brackets denote averages 
with respect to the stationary measure. We have discussed in the text the main features of the distribution $p(n)$,
in particular that it is well fitted by $p(n) \sim 1/n$ when $\gamma \to 0$.
We have also introduced $r(n)=n p(n)/\sum_{m=1}^N m p(m)$ which is the fraction of particles belonging to clusters of size $n$.
One can also define the size of the largest cluster, which we call $n_{\max}$ which for small $\gamma$ is
expected to be of order $O(N)$, since there seems to be a very large cluster where a finite fraction
of the particles condense (see Fig. \ref{spacetime_g0} in the main text). In Fig. \ref{g0appdx} we are showing the average value of
$n_{\rm max}$ measured in the simulation, as a function of the total number $N$ of particles. We see clearly
that this average value is proportional to $N$.

\end{widetext}


\begin{thebibliography}{}




\bibitem{soft} M. C. Marchetti, J. F. Joanny, S. Ramaswamy, T. B. Liverpool, J. Prost, M. Rao, and R. Aditi Simha, {\it Hydrodynamics of soft active matter}, Rev. Mod. Phys. {\bf 85}, 1143  (2013).

\bibitem{BechingerRev} C. Bechinger, R. Di Leonardo, H. L\"{o}wen, C. Reichhardt, G. Volpe, and G. Volpe, {\it Active particles in complex and crowded environments}, Rev. Mod. Phys. {\bf 88}, 045006 (2016).

\bibitem{Ramaswamy2017} S. Ramaswamy, {\it Active Matter}, J. Stat. Mech.  054002, (2017).

\bibitem{Marchetti2018} \'{E}. Fodor, and M. C. Marchetti, {\it The statistical physics of active matter: from self-catalytic colloids to living cells}, Physica A {\bf 504}, 106 (2018). 

\bibitem{Berg2004} H. C. Berg, {\it E. Coli in Motion}, (Springer Verlag, Heidel-
berg, Germany) (2004).

\bibitem{Cates2012} M. E. Cates, {\it Diffusive transport without detailed balance: Does microbiology need statistical physics ?}, Rep. Prog. Phys. {\bf 75}, 042601 (2012).

\bibitem{TailleurCates} 
J. Tailleur, and M. E. Cates, {\it Statistical mechanics of interacting run-and-tumble bacteria}, Phys. Rev. Lett. {\bf 100}, 218103 (2008).
 

\bibitem{W02} G. H. Weiss, {\it Some applications of persistent random walks and the telegrapher's equation}, Physica A {\bf 311}, 381 (2002).

\bibitem{HJ95} P. H\"anggi and P. Jung, {\it Colored Noise in Dynamical Systems}, Adv. Chem. Phys. {\bf 89}, 239 (1995).



\bibitem{ML17} J. Masoliver and K. Lindenberg, {\it Continuous time persistent random walk: a review and some generalizations}, Eur. Phys. J. B {\bf 90}, 1 (2017).

\bibitem{kac74} M. Kac, {\it A stochastic model related to the telegrapher's equation}, Rocky Mountain J. Math. {\bf 4}, 497 (1974).

\bibitem{Orshinger90} E. Orsingher, {\it Probability law, flow function, maximum distribution of wave-governed random motions and their connections with Kirchoff's laws}, Stoch. Process. Their Appl.
{\bf 34}, 49 (1990).


\bibitem{Solon15} A. P. Solon, Y. Fily, A. Baskaran, 
M. E. Cates, Y. Kafri, M. Kardar, and J. Tailleur, {\it Pressure is not a state function for generic active fluids},
Nature Phys. {\bf 11}, 673 (2015).

\bibitem{TDV16} S. C. Takatori, R. De Dier, J. Vermant, and J. F. Brady, {\it Acoustic trapping of active matter}, Nature Comm. {\bf 7}, 10694 (2016).


\bibitem{DKM19} 
A. Dhar, A. Kundu, S. N. Majumdar, S. Sabhapandit and 
G. Schehr, {\it Run-and-tumble particle in one-dimensional confining potentials: Steady-state, relaxation, and first-passage properties}, Phys. Rev. E {\bf 99}, 032132 (2019).



\bibitem{DD19} O. Dauchot and V. D\'emery, {\it Dynamics of a Self-Propelled Particle in a Harmonic Trap}, Phys. Rev. Lett. {\bf 122}, 068002 (2019).


\bibitem{3statesBasu}
U. Basu, S. N. Majumdar, A. Rosso, S. Sabhapandit, and G. Schehr, {\it Exact stationary state of a run-and-tumble particle with three internal states in a harmonic trap}, J. Phys. A: Math. Theor. {\bf 53}, 09LT01 (2020)

\bibitem{LMS2020}
P. Le Doussal, S. N. Majumdar, and G Schehr, {\it Velocity and diffusion constant of an active particle in a one-dimensional force field}, EPL {\bf 130}, 40002 (2020).


\bibitem{slowman}
A. B. Slowman, M. R. Evans, and R. A. Blythe, {\it Jamming and attraction of interacting run-and-tumble random walkers}, Phys. Rev. Lett. {\bf 116}, 218101 (2016).


\bibitem{slowman2}
A. B. Slowman, M. R. Evans, and R. A. Blythe, {\it Exact solution of two interacting run-and-tumble random walkers with finite tumble duration}, J. Phys. A: Math. Theor. {\bf 50}, 375601 (2017).


\bibitem{Active_OU}
D. Martin, J. O'Byrne, M. E. Cates, E. Fodor, C. Nardini, J. Tailleur, F. van Wijland, {\it Statistical mechanics of active Ornstein-Uhlenbeck particles}, Phys. Rev. E {\bf 103}, 032607 (2021)


\bibitem{FHM2014}
Y. Fily, S. Henkes, and M. C. Marchetti, {\it Freezing and phase separation of self-propelled disks}, Soft matter {\bf 10}, 2132 (2014).

\bibitem{FM2012}
Y. Fily, and M. C. Marchetti, {\it Athermal phase separation of self-propelled particles with no alignment}, Phys. Rev. Lett. {\bf 108}, 235702 (2012).

\bibitem{Buttinoni2013}
I. Buttinoni, J. Bialké, F. Kümmel, H. Löwen, C. Bechinger,
and T. Speck, {\it Dynamical Clustering and Phase Separation in Suspensions of Self-Propelled Colloidal Particles}, Phys. Rev. Lett. {\bf 110}, 238301 (2013).


\bibitem{CT2015}
M. Cates, and J. Tailleur, {\it Motility-induced phase separation}, Annu. Rev. Condens. Matter Phys. {\bf 6}, 219 (2015).


\bibitem{BG2021}
C. M. Barriuso Gutiérrez, C. Vanhille-Campos, Francisco Alarcón, I. Pagonabarraga, R. Britoai,  and  C. Valeriani, {\it Collective motion of run-and-tumble repulsive and attractive particles in one-dimensional systems}, Soft Matter {\bf 17}(46) (2021).

\bibitem{CMPT2010}
M. E. Cates, D. Marenduzzo, I. Pagonabarraga, and J. Tailleur, {\it Arrested phase separation in reproducing bacteria creates a generic route to pattern formation}, Proc. Natl. Acad. Sci. U.S.A. {\bf 107}, 11715 (2010).

\bibitem{SG2014}
R. Soto, R. Golestanian, {\it Run-and-tumble dynamics in a crowded environment: Persistent exclusion process for swimmers}, Phys. Review E {\bf 89}, 012706 (2014).

\bibitem{KH2018}
M. Kourbane-Houssene, C. Erignoux, T. Bodineau, and J. Tailleur, {\it Exact Hydrodynamic Description of Active Lattice Gases}, Phys. Rev. Lett. {\bf 120}, 268003 (2018).


\bibitem{Agranov2021}
T. Agranov, S. Ro, Y. Kafri, and V. Lecomte, {\it Exact fluctuating hydrodynamics of
active lattice gases—typical fluctuations}, J. Stat. Mech. 083208 (2021).

\bibitem{Agranov2022}
T. Agranov, S. Ro, Y. Kafri, and V. Lecomte, {\it Macroscopic Fluctuation Theory and Current Fluctuations in Active Lattice Gases}, arXiv:2208.02124.

\bibitem{us_bound_state}
P. Le Doussal, S. N. Majumdar, G. Schehr, {\it Stationary nonequilibrium bound state of a pair of run and tumble particles}, Physical Review E {\bf 104}(4), 044103 (2021).

\bibitem{Maes_bound_state}
P. Dolai, S. Krekels, and C. Maes, {\it Inducing a bound state between active particles}, arXiv:2202.04459 (2022).

\bibitem{nonexistence}
I. Mukherjee, A. Raghu, and P. K. Mohanty, {\it Nonexistence of motility induced phase separation transition in one dimension}, arXiv:2208.05937.


\bibitem{MBE2019}
E. Mallmin, R. A. Blythe, and M. R. Evans, {\it Exact spectral solution of two interacting run-and-tumble particles on a ring lattice}, J. Stat. Mech., 013204 (2019).


\bibitem{KunduGap2020}
A. Das, A. Dhar, and A. Kundu, {\it Gap statistics of two interacting run and tumble particles in one dimension}, J. Phys. A: Math. Theor. {\bf 53}, 345003 (2020).

\bibitem{LMS2019}
P. Le Doussal, S. N. Majumdar, and G. Schehr, {\it Noncrossing run-and-tumble particles on a line}, Phys. Rev. E {\bf 100}, 012113 (2019).


\bibitem{SinghChain2020} 
P. Singh, A. Kundu, {\it Crossover behaviours exhibited by fluctuations and correlations in a chain of active particles}, J. Phys. A: Math. Theor. {\bf 54}, 305001 (2021)


\bibitem{PutBerxVanderzande2019}
S. Put, J. Berx, and C. Vanderzande, {\it Non-Gaussian anomalous dynamics in systems of interacting run-and-tumble particles}, J. Stat. Mech. 123205 (2019).


\bibitem{Metson2022}
M. J. Metson, M. R. Evans, and R. A. Blythe, {\it Tuning attraction and repulsion between active particles through persistence}, arXiv:2207.01317.

\bibitem{MetsonLong}
M. J. Metson, M. R. Evans, and R. A. Blythe, {\it From a microscopic solution to a continuum description of interacting active particles}, arXiv:2207.01321 (2022).


\bibitem{Dandekar2020}
R. Dandekar, S. Chakraborti, and R. Rajesh, {\it Hard core run and tumble particles on a one-dimensional lattice}, Phys. Rev. E {\bf 102}, 062111 (2020).


\bibitem{Thom2011}
A. G. Thompson, J. Tailleur, M. E. Cates, R. A. Blythe, {\it Lattice models of nonequilibrium bacterial dynamics}, J. Stat. Mech. P02029 (2011).


\bibitem{mehta_book}
M. L. Mehta, {\it Random matrices}, Elsevier (2004).

\bibitem{forrester_book}
P. J. Forrester, {\it Log-gases and random matrices}, Princeton University Press (2010). 

\bibitem{bouchaud_book}
M. Potters \& J. P. Bouchaud, {\it A First Course in Random Matrix Theory: For Physicists, Engineers and Data Scientists}, Cambridge University Press (2020).

\bibitem{Dean}
D. S. Dean, {\it Langevin Equation for the density of a system of
interacting Langevin processes}, J. Phys. A: Math. Gen. {\bf 29}, L613 (1996). 

\bibitem{Kawa}
K. Kawasaki, {\it Microscopic analyses of the dynamical density functional equation of dense fluids}, J. Stat. Phys. {\bf 93}, 527 (1998).










\bibitem{Hermite2}
P. Forrester and J. Rogers. {\it Electrostatics and the zeros of the classical polynomials}, SIAM Journal on Mathematical Analysis, {\bf 17}(2):461–468, 1986.


\bibitem{Hermite1}
S. Agarwal, M. Kulkarni, A. Dhar, {\it Some connections between the Classical Calogero-Moser model and the Log Gas}, Journal of Statistical Physics {\bf 176}(3), 2019.

\bibitem{SM}
L. Touzo, P. Le Doussal, G. Schehr, Supplementary Material.



\bibitem{footnotelog} 
For $N \gamma=1$ the singularity is
logarithmic except for fixed points corresponding to all $\sigma_i$ equal. 
It is reminiscent of a similar result obtained for a single RTP with three states \cite{3statesBasu}, as detailed in \cite{SM}.

\bibitem{unicity}
which we have checked numerically.

\bibitem{flajolet2009}
P. Flajolet, R. Sedgewick, {\it Analytic combinatorics}, Cambridge University Press (2009).

\bibitem{foot_infinity} Note that the lower bound of the integral over $z'$ can equivalently chosen to be $\pm \infty$, since $\int_{-\infty}^{+\infty} [G_+(z') + G_+(-z')] dz' = \int_{-\infty}^{+\infty} [G_+(z') - G_-(z')] dz' = 0$ by symmetry. 

\bibitem{foot_frozen}
Here $t$ has been sent to infinity first, i.e. $\gamma t \gg 1$. In the opposite
limit $\gamma t \ll 1$ the variables $\sigma_i$ are frozen and the final state depends on their initial values.
For $N \to +\infty$ the solution of the equations leads to two semi-circles with
different weights

\bibitem{BouchaudGuionnet}
R. Allez, J.P. Bouchaud, A. Guionnet, {\it Invariant $\beta$-ensembles and the
Gauss-Wigner crossover}, Phys. Rev. Lett. {\bf 109}, 094102 (2012).

\bibitem{cuenca}
F. Benaych-Georges, C. Cuenca, V. Gorin, {\it Matrix addition and the Dunkl transform at high temperature}, arXiv:2105.03795.

\bibitem{dumaz}
L. Dumaz, B. Virag, {\it The right tail exponent of the Tracy-Widom $\beta $ distribution}, Annales de l'IHP Probabilités et statistiques, Vol. {\bf 49}, No. 4, pp. 915-933 (2013).

\bibitem{allez_satya}
R. Allez, J. P. Bouchaud, S. N. Majumdar, P. Vivo, {\it Invariant $\beta$-Wishart ensembles, crossover densities and asymptotic corrections to the Marčenko–Pastur law}, Journal of Physics A: Mathematical and Theoretical, {\bf 46}(1), 015001 (2012).

\bibitem{UsFuture}
L.Touzo, P. Le Doussal, G. Schehr, in preparation.

\bibitem{Allez13}
R. Allez, A. Guionnet, {\it A diffusive matrix model for invariant $\beta $-ensembles}, Electronic Journal of Probability, {\bf 18}, 1-30 (2013).

\bibitem{Lepingle07}
E. Cépa, D. L{\'e}pingle, {\it No multiple collisions for mutually repelling Brownian particles}, In Séminaire de Probabilités XL (pp. 241-246). Springer, Berlin, Heidelberg (2007).


\bibitem{Hermite_asymptotics}
\url{https://dlmf.nist.gov/18.16}

\bibitem{wolfram}
\url{https://mathworld.wolfram.com/DiagonallyDominantMatrix.html}


\bibitem{RogersShi} 
L.C.G. Rogers and Z. Shi, {\it Interacting Brownian particules and the Wigner law},
Prob. Th. Rel. Fields {\bf 95}, 555-570 (1993).


\end{thebibliography}
\end{document}